\documentclass[11pt]{article}

\usepackage[table]{xcolor}

\usepackage{titlesec}
\titlespacing*{\section}
  {0pt}           
  {2.1ex plus 0.5ex minus 0.2ex}  
  {1.4ex plus 0.2ex}     

\titlespacing*{\subsection}
  {0pt}            
  {2.0ex plus 0.5ex minus 0.2ex}   
  {1.0ex plus 0.2ex}               

\usepackage{fancyhdr}
\usepackage{extramarks}
\usepackage{amsmath}
\usepackage{amsthm}
\usepackage{amsfonts}
\usepackage{bbold}
\usepackage{tikz}
\usepackage[plain]{algorithm}
\usepackage{algpseudocode}

\usetikzlibrary{automata,positioning}

\usepackage{amsmath}
\usepackage{amssymb}
\usepackage{breqn}
\usepackage{bbm}
\usepackage{mathtools}

\newcommand\sym[1]{\textit{#1}}

\usepackage{mathptmx} 

\usepackage[T1]{fontenc}

\usepackage{tikz}
\usepackage{verbatim}
\usetikzlibrary{positioning}
\usetikzlibrary{snakes}
\usetikzlibrary{calc}
\usetikzlibrary{arrows}
\usetikzlibrary{decorations.markings}
\usetikzlibrary{shapes.misc}
\usetikzlibrary{matrix,shapes,arrows,fit,tikzmark}
\usetikzlibrary{arrows.meta,
                chains,
                positioning}

\usepackage{setspace}
\usepackage[round]{natbib}
\makeatletter
\renewcommand\@biblabel[1]{}
\makeatother

\usepackage[margin=1.0in]{geometry}

\usepackage{pdflscape}

\usepackage[hidelinks]{hyperref}
\usepackage{xcolor,soul}
\sethlcolor{lightgray}

\usepackage{authblk}

\renewcommand{\thefigure}{\arabic{figure}}
\renewcommand{\thetable}{\arabic{table}}

\usepackage{graphicx}
\usepackage{subcaption}

\usepackage{enumitem}
\usepackage{lipsum}
\usepackage[skins,breakable]{tcolorbox}
\newtcolorbox{mybox}[1][colback=gray!15]{breakable,pad at break=1mm,
  oversize,left=1mm,right=1mm,interior hidden,colframe=white,nobeforeafter=,#1}

\usepackage{booktabs} 
\usepackage{adjustbox}
\usepackage{float}

\usepackage[justification=centering]{caption}

\usepackage{longtable}

\usepackage{siunitx} 
\usepackage{array}
\newcolumntype{H}{>{\setbox0=\hbox\bgroup}c<{\egroup}@{}}
\newcolumntype{Z}{>{\setbox0=\hbox\bgroup}c<{\egroup}@{\hspace*{-\tabcolsep}}}

\begin{document}
\title{Better Together? A Field Experiment on \\  Human-Algorithm Interaction in Child Protection\footnote{A version of this paper was previously circulated under the title ``The Impact of Algorithmic Tools on Child Protection: Evidence from a Randomized Controlled Trial.'' We are grateful to Rhema Vaithianathan, Larissa Lorimer, the Centre for Social Data Analytics (CSDA), and the partner agency for their indispensable support and feedback throughout. We also thank Jason Baron, Leah Boustan, Robert Collinson, Janet Currie, Ellora Derenoncourt, Will Dobbie, Joseph Doyle, William Evans, 
Matthew Lindquist, Erik Lindqvist, Jens Ludwig, Daniel Martin, 
Sendhil Mullainathan,
and numerous seminar participants for their valuable comments and suggestions.
The Princeton Center for Health and Wellbeing provided generous funding for this project. This study is part of a wider research program on the use of risk modeling in human services led by Rhema Vaithianathan and Emily Putnam-Hornstein. Evaluation is one of several safeguards in the implementation of these tools. This project was prepared using a Limited Data Set (or De-Identified Data Set) obtained from Colorado Hospital Association (CHA) and does not reflect the opinions or views of CHA.  The experiment was pre-registered on the AEA RCT Registry under ID AEARCTR-0006311. }}

\author{Marie-Pascale Grimon\thanks{Swedish Institute for Social Research, Stockholm University, Stockholm 10691, Sweden} \hspace{1cm}
Christopher Mills\footnote{Department of Economics, University of Notre Dame, Notre Dame, IN 46556, USA}

}
 \date{\fontsize{12pt}{12pt}\today}
 
 \maketitle

\begin{abstract}
\begin{spacing}{1.2}

Despite algorithms' potential to improve public services, adoption has been limited by concerns about effectiveness and equity. We conduct a randomized controlled trial ($N=4,681$) providing real-time algorithm support to Child Protective Services (CPS) workers allocating investigations. Algorithm access reduced maltreatment-related hospitalizations, especially among disadvantaged groups, while reducing CPS surveillance of Black children. Notably, child injury admissions decreased by 21 percent. Workers reallocated investigations toward children at greater likelihood of harm, without mechanically following algorithmic predictions. Discussion notes suggest the algorithm shifted worker attention to complementary information. Counterfactual exercises show that human-algorithm complementarity would outperform algorithmic automation in efficiency and equity.

\end{spacing}
\vspace{1.5em}
\vspace{1cm}
\end{abstract}


\tikzset{   
        every picture/.style={remember picture,baseline},
        every node/.style={anchor=base,align=center,outer sep=1.5pt},
        every path/.style={thick},
        }
\newcommand\marktopleft[1]{%
    \tikz[overlay,remember picture] 
        \node (marker-#1-a) at (-.3em,.3em) {};%
}
\newcommand\markbottomright[2]{%
    \tikz[overlay,remember picture] 
        \node (marker-#1-b) at (0em,0em) {};%
}
\tikzstyle{every picture}+=[remember picture] 
\tikzstyle{mybox} =[draw=black, very thick, rectangle, inner sep=10pt, inner ysep=20pt]
\tikzstyle{fancytitle} =[draw=black,fill=red, text=white]

\thispagestyle{empty}  
\clearpage              
\setcounter{page}{1}    

\section{Introduction} 


Despite the promise of algorithms to improve allocative efficiency \citep{ludwig2024}, their diffusion into social policy settings has been slow. A majority of the public are not optimistic about the benefits of AI, for example, in criminal justice settings (\textit{``How the U.S. Public and AI Experts View Artificial Intelligence''} Pew Survey, 2025). Experts are similarly divided on the benefits of algorithms for social policy, with leading concerns being embedded bias \citep{rambachan2020a} and questionable overrides when humans retain final discretion over decisions \citep{albright2019a,angelova2022algorithmic}. One proposed reason that algorithms have been less helpful than anticipated is that they have not been deployed to help humans leverage their comparative advantage \citep{ludwig2021}. In particular, humans may retain more information than is available to an algorithm and consider payoffs beyond accuracy, such as equity.

This paper provides the first comprehensive field experimental evidence in a social policy setting that human oversight of algorithms can achieve efficiency and equity gains. The trial involved hundreds of social workers using an algorithm tool in real time 
in a challenging, high-stakes policy context: Child Protective Services (CPS). 
Each year, CPS agencies in the United States decide how to respond to four million calls of alleged maltreatment involving a tenth of the U.S. child population (\citealt{hhs2019}), 
a group acutely at risk of negative later-life outcomes (\citealt{doyle2007child}, \citealt{helensdotter2024}).
Social workers must decide in a matter of minutes which reported families to investigate based on limited information from a caller's present concern for the child and the family's prior history with CPS. Agencies do not have the capacity to visit all families, and 
failure to investigate a child suffering from abuse or neglect can result in costly continued victimization (\citealt{currie2010long}) and, in regular instances, death. 
Misallocation of CPS investigations appears substantial: the majority of first-time confirmed maltreatment victims were previously reported but not investigated.\footnote{Authors' calculations, California investigations from 2014-2019.} 
Consequently, child welfare practitioners have taken an interest in tools that can help improve workers' allocation of investigations. 

We conducted a large-scale field experiment ($N=4,681$) where, for a randomized set of reported families, worker teams that allocate CPS investigations could consult an algorithm decision support tool in addition to their regular information set. The algorithm presented an informative prediction score of children's foster care removal risk, ranging from 1 to 20, based on past child welfare records already accessible to workers  (70 percent had prior history). The tool was only available at this decision stage, tool consultation was voluntary, and workers retained full autonomy in deciding which families to investigate. In practice, workers consulted the algorithm in at least 73 percent of decisions.\enlargethispage{0.75em}

First, we find that providing algorithm support improved child wellbeing. Using a unique linkage of statewide hospitalization records to child welfare administrative data, we show that giving workers access to the algorithm reduced children's future hospitalizations for conditions commonly associated with maltreatment while holding the overall rate of investigation constant. Specifically, relative to children in the control group, the intervention group experienced a 21 percent reduction in injury-related hospitalizations and a 0.05 standard deviation reduction in a combined 
``harm'' index constructed from hospital data that included established proxies for abuse and neglect. 
Access to the algorithm helped workers shield children from the worst (top one percent) tail of the child harm distribution. As over half of children with hospital-confirmed maltreatment belong to this top percentile of the child harm distribution, access to the algorithm appears to have decreased child maltreatment.

Second, access to the algorithmic tool did not increase child disparities in health and CPS surveillance, but rather it suggestively reduced them. In the control group, Black, Hispanic, female, and low-income children experienced elevated levels of child harm, consistent with
prior studies on child mortality \citep{currie2016Mortality} and maltreatment prevalence \citep{putnam-hornstein2021Cumulative}.  
In light of these historic disparities, some U.S. states have forestalled the adoption of algorithm decision aids due to concerns over embedded bias, underscoring a need for evaluation.\footnote{\url{https://www.npr.org/2022/06/02/1102661376/oregon-drops-artificial-intelligence-child-abuse-cases}} In contrast to these concerns, we find that decisions made under human-algorithm interaction halved child harm disparities by race, ethnicity, gender, and socioeconomic status. Although these estimates are not statistical significant at conventional thresholds, they are informative in a Bayesian framework for ethnicity, gender and socioeconomic disparities. Our findings for Black children warrant more caution as they only comprise a modest four percent of our sample. We do however find that Black children in the control group were investigated at much higher rates than other children, in line with national estimates \citep{kim2017lifetime}, even after conditioning on underlying risk factors. Consistent with a behavioral framework of automatic and deliberative thinking \citep{kahneman2011thinking,agan2022algorithms}, access to the algorithm appears to have reduced disproportionate surveillance: workers with the tool investigated low-risk-score Black children at significantly lower rates comparable to other low-scoring children, without a distinguishable offsetting increase in harm.

We identify as a key mechanism that algorithm support helped workers to parse through complementary information when deciding which families to investigate. 
Workers allocated investigations more effectively: investigations increased for harmed children, and decreased for children in better health. Triage improved especially for disadvantaged groups. Applying cutting-edge sentence-transformer based methods to proprietary, anonymized text data written during team decision-making discussions, we show that the algorithm changed workers' discussion patterns.
In exploring how discussions changed, we find that the tool appears to have increased the salience of information complementary to algorithm features, such as family structure and changes in circumstances in light of family history. Gains were concentrated among the set of more complex referrals, including those for which workers expressed uncertainty. 
This suggests that algorithm support may have reallocated worker effort away from extraneous cognitive tasks of looking for family history to instead focusing on the current allegation in light of family history. 

Child outcomes improved without changing the overall level of investigations or other services, and without increasing investigations and services for children with high algorithm risk scores. One explanation consistent with our findings is that while the algorithm was trained to predict \textit{levels} of risk based on history and was informative of future child harm, workers also considered the \textit{marginal benefit} (treatment effect) of an additional investigation. This distinction is consistent with the algorithm's emphasis on prior family history clueing workers in to consider changes in child and family circumstances, and to consider whether these changes warranted further intervention. More generally, humans considered a broader set of objectives and had more information at their disposal than was available to the algorithm. 
The multi-pronged evidence for workers' improved allocation of investigations reinforces a core insight: algorithms can facilitate better decision-making that goes beyond strictly following the algorithm's risk score.

These findings raise the question of whether forcing workers to investigate high-algorithm-score children would improve outcomes, and motivate a fourth set of results: We  develop two counterfactual estimation strategies to (1) bound how well the algorithm could have performed without human oversight and (2) test whether users should have relied marginally more on the algorithm score. An algorithm-only treatment arm was legally and practically infeasible in this context, as in many high-stakes environments, motivating the first counterfactual approach. The second approach assessed the policy relevance of imposing a mandatory investigation threshold. On both of these margins, we find that humans with algorithm support likely outperformed counterfactual algorithm-only decisions in terms of both efficiency and equity, despite the algorithm being highly predictive. Worker sensitivity to equity likely played a role in reducing surveillance disparities for disadvantaged groups -- particularly Hispanic and low-income children -- 
compared to an algorithm-only decision rule.

Finally, we consider external validity and identify design features that may have been crucial for human-algorithm complementarity. We use a stylized model of expert decisions under algorithm support to show how our findings could extend to related settings. We then discuss two important features: built-in human expertise through access to exclusive information and implementation as part of a group discussion. The algorithm improved equity more when used in larger teams, complementing recent findings on personnel and fairness \citep{chiang2023}.
A cost-benefit analysis suggests clear policy implications for scaling.

This paper's main contribution is to provide novel experimental evidence that algorithm support can improve efficiency and possibly even equity  without replacing human discretion. Humans and algorithms are complementary in our context; such synergies may not be as ``elusive'' as feared by \citet{zana2022a}. Our experimental findings and counterfactual simulations show that humans paired with algorithms outperformed either party on its own.
Algorithms and human experts compensate for one another's shortcomings with respect to equity: human oversight guarded against potential disproportionate surveillance of Hispanic and low-income children under the algorithm, and the algorithm helped reduce high levels of human surveillance of low-predicted-risk Black children.
The latter is congruent with prior work that algorithms can indeed shift human reference points \citep{mclaughlin2024}.
In short, pairing an algorithm trained on likely-biased historical data together with potentially-biased human decision-makers can still reduce CPS disparities and improve welfare \citep{rambachan2020,gillis2021}.
Our findings align with prior quasi-experimental work showing that experts can partly redress algorithmic biases with additional contextual information and by considering additional objectives \citep{de-arteaga2020,obermeyer2019dissecting,vandonselaar2010Ordering, hoffman2018discretion, gruber2020managing, angelova2022algorithmic, harris2024}.

This study also adds to our collective understanding of when and how human decision-makers benefit from algorithm support.
In light of a pattern of suboptimal human overrides \citep{ye2022,agarwal2023,mullainathan2022Diagnosing,albright2019a}, recent work has focused on how to create more elaborate algorithms that circumvent human biases \citep{mclaughlin2024a}. In contrast, our findings highlight that an alternative approach could be to simplify the algorithm, reminiscent of the literature on information overload. 
 Consistent with \citet{vaccaro2024}, in our setting where workers had a built-in comparative advantage and had clarity on the algorithm's inputs, they were able to integrate the algorithm predictions effectively into their decision-making. They did so by considering other objectives beyond foster care placement risk \citep{kleinberg2024}, such as equity, legal constraints on subsequent action, and seemingly identifying children who could benefit from an additional investigation, distinguishing estimated treatment effects from predicted risk levels (e.g., \citealt{bhatt2024}). An implication of our findings is that algorithms  can improve decisions even when workers do not rotely follow algorithm predictions. The tool was designed for workers' expressed needs in a time- and resource-constrained environment, and its particular efficacy in more complex and ambiguous cases suggests that algorithms may be effective when they reduce extraneous cognitive load.
Beyond its design, the algorithm's deployment in a group-based setting may have also contributed to its effectiveness \citep{galton1907,patel2019Human,chiang2023}.

We are among the first to design an randomized controlled trial (RCT) to evaluate how professionals use algorithmic information in a high-stakes social policy setting, complementing an important emerging set of human-AI studies. 
Discontinuity-based designs have identified effects at a specific cutoff margins \citep{stevenson2024,albright2019,Cowgill2018TheIO}, whereas our RCT design reveals heterogeneous effects of algorithm support across the distribution of child risk. 
Our trial takes place in a real-time, consequential setting involving hundreds of professionals: \citet{imai2020} only evaluate one judge, and medical RCTs that rely on historical case data (e.g., \citealt{agarwal2023}), prospective medical RCTs (see \citealp{rajpurkar2022AI} for a review),  
and online experiments \citep{green2019} 
may not transpose to complex social policy settings. 
Using methodologies akin to pre-post designs, \citet{goldhaber2019impact} and \citet{rittenhouse2024} uncover a decrease in racial disparities in CPS interventions under algorithm support, with less clear findings for welfare. 
 We complement a concurrent evaluation of a similarly-styled algorithmic tool in a different child protection agency by \cite{fitzpatrick2020}, who find a small reduction in team discussion time and some change in investigation rates, but due to COVID-related disruptions and a different randomization strategy are statistically underpowered to detect downstream outcomes. Our paper in contrast  examines the impacts on efficiency using  external data sources on child health, considers consequences for equity, and can speak more generally to human-algorithm complementarity.

Finally, this study offers new evidence that 
appropriately-targeted CPS investigations 
reduce child maltreatment. 
Approximately 37 percent of children in the United States are investigated by CPS before adulthood \citep{kim2017lifetime}. While prior research has evaluated the effects of certain downstream CPS interventions on child and parent economic and social outcomes \citep{doyle2007child,gross2022temporary,bald2022causal, grimon2022parents}, this paper contributes to an emerging literature examining the impacts of CPS on children's health and wellbeing \citep{doyle2013,chorniymills,lacey2024,helensdotter2024,heath2024}.
Despite the high prevalence of CPS contact in high-income countries, many maltreatment incidents are never reported, and children often poorly or selectively recall maltreatment \citep{newbury2018Measuring}. In light of these challenges, our study proposes a novel index of suspected child maltreatment constructed from hospital records. The advantage of this measure is that it relies on objective data not subject to recall bias and can be constructed for the full child population. Appropriately measuring child harm allows us to reliably estimate effects of algorithm support on child wellbeing and disparities. 
Our study provides new, indirect evidence of the 
impacts of CPS investigations allocated by the far-reaching ``front end'' of the child welfare system, complementing ongoing work by \citet{lacey2024}. 

The remainder of the paper is organized as follows. Section 2 introduces the context of Child Protective Services and the algorithm tool. Section 3 describes the study data, experimental design, and empirical specification. Section 4 shows presents the main results on efficiency and equity. Section 5 details the mechanisms underlying observed improvements. Section 6 presents counterfactual human-algorithm interaction estimates. Section 7 discusses the generalizability of core findings. Section 8 concludes.

\section{Empirical Context}

The goal of the child welfare system, also referred to as Child Protective Services (CPS), is to keep children safe from abuse and neglect.\footnote{Technical definitions of abuse and neglect vary slightly by jurisdiction. Abuse refers to direct harm inflicted on a child by a perpetrator, for example physical or sexual abuse, whereas neglect entails failing to protect a child from danger (e.g., drugs, unsafe home environment) or not tending to a child's medical or developmental needs (See Figure \ref{track_assignment} for examples). In this paper, we use the terms ``child welfare'',  ``Child Protective Services'', and ``child protection'' interchangeably.} 
This section provides a brief description of the child welfare system, accompanied by a description of the machine learning tool developed to aid decision makers.

\subsection{Child Protective Services in the Evaluation County} 
\label{sec:cpsbackground}

Appendix Figure \ref{cps_diagram} illustrates how CPS operates in our setting. CPS learns about a child who is at risk of maltreatment through reports (``referrals'') to a child protection hotline. Three-fourths of referrals in our setting originate from mandated reporters, including school employees, law enforcement, and medical professionals, who are required by law to report any instances of suspected abuse or neglect.
A hotline worker records allegations of maltreatment. 
Most referrals in the study county are sent to a decision team that meets the following weekday morning.

A decision team is a group of social workers led by at least one supervisor. In the study county, three to four teams meet each weekday morning to each review referrals that arrived since their last meeting. Each team consists of about five members taken from a designated pool of workers (about 170 workers appear during the trial), with exact worker composition varying non-randomly by day. For each referral, a team collaboratively records notes on key details about the family, discusses a recommended course of action, and writes down additional steps for following up on the referral. When reviewing a referral, typically one team member looks up a certain data source at a time while another team member writes down main points in a shared document in real time. The team makes note of key details from the allegation recorded at hotline, past child welfare history for the family, criminal justice contact and other internal records, and other relevant details known from past investigations.
Ultimately, a team will choose either to send an investigator to the family (``screen in'') or not (``screen out''). Approximately 30 percent of referrals are screened in for investigation in the analysis sample, and the remainder are screened out. For some screen-outs, CPS may send a family visitor, other types of optional light-touch voluntary services, or law enforcement. 
Otherwise, the screened-out family receives no contact: they often remain unaware that they were even reported to CPS. 

Finally, if a family is screened in for investigation, they are assigned to an investigator. An investigator records additional relevant information after visiting the home and speaking with alleged victims, perpetrators, and reporters. As is the case in many child welfare agencies, the investigator is required to fill out risk and safety forms within 60 days, which document family strengths, weaknesses, and a child's current and future risk of harm. The investigator decides whether to substantiate any of the allegations and whether to recommend any voluntary services, mandatory services, or in extreme situations to remove a child from their home and place them in foster care.

\subsection{Algorithm Tool}

The decision aid tool was developed and implemented by Rhema Vaithianathan and colleagues at the Centre for Social Data Analytics (CSDA). It has received favorable and independent support from the scientific community \citep{goldhaber2019impact}. Similar to previous iterations in Allegheny County (PA) and Douglas County (CO), the tool predicts a child's likelihood of being removed from their home (placed in foster care) by Child Protective Services within two years of a referral, an outcome that implies high risk of maltreatment. Foster care was selected as the model's outcome because -- despite some discretion over foster care placement -- it is one of the more relatively firm indicators of severe concern available internally to CPS agencies. Random forests and XGBoost were considered for the algorithm, but a LASSO regularized logistic regression specification was chosen for runtime purposes.\footnote{All data processing and score generation occurred on state servers. This was a deliberate choice by the implementing partner to protect data privacy.} The model was trained on a dataset of past child welfare referrals in the trial county and other statewide counties using historical data from approximately 500,000 child-referrals from 2010-2014. The area under the receiver operating characteristic curve (AUC) of the model is 0.76, which indicates good discrimination.\footnote{Although the set of variables used to predict removals in the Colorado model was more limited than in Allegheny County due to differences in data availability, the training sample size was much larger. The model in our setting was trained using the full state instead of just one county (larger $n$ but smaller vector of features $X$). As a result, the AUC of the model in our county is slightly better than in Allegheny County, where the AUC was 0.72 (Internal Allegheny DHS report, 2019).}

The model uses several hundred features relating to demographics (excluding race, ethnicity, and disability), past child welfare involvement of household members, as well as limited case management data with social services.\footnote{Though the model's full feature set is public, the subset of features selected by the LASSO is not disclosed, in accordance with an agreement between the tool designers and the county.} Of these, CSDA reports that most of the algorithm's predictive power comes from features that encode child and household members' levels of past CPS involvement. 71 percent of our sample indeed had prior family history with CPS, and the remaining 29 percent still had some variation in other predictors. The present referral allegation is not an input to the algorithm, which relies primarily on past referral allegations. Text details for incoming referrals are not uploaded to state servers and cannot be used to build the algorithm tool. The choice of which features to include in the model was made in agreement with local implementing partners. Additionally, nearly all information used by the algorithm could be looked up by social workers prior to the trial and during the trial. The algorithm only summarized the information by giving a numerical prediction of removal. As in the two prior deployments of similar tools, predictions were given to workers in ventile (1-20) risk bins, with a score of 1 assigned to children least likely to be removed, and 20 assigned to children at the highest risk of being removed.

Algorithmic information was designed as a complement to -- and not a replacement for -- human decision making. Social workers were made aware of this goal and were encouraged by their managers to consult the tool during team meetings, but were not required to use it. Workers were specifically trained with examples prior to the trial to ensure that they did not make decisions solely using the tool. For example, a child with no past CPS history but with a clear allegation of sexual abuse must be investigated under state law. On the other hand, a child with substantial CPS history would likely receive a high risk score, but the child should not be investigated if the incoming allegation does not constitute child abuse or neglect by state legal standards (Appendix Figure \ref{track_assignment}). 

The algorithm score was only available during team meetings where workers decided whether or not to screen in a family. The tool was not available to investigators.\footnote{In theory, investigators could have had knowledge of the risk score if they consulted team discussion notes or happened to have taken part in the discussion themselves. In practice, however, we did not hear of any such instances of investigators consulting the score when visiting a family. Additionally, we do not find any marked evidence of tool access changing investigator behavior. We show that improvements in outcomes can be explained by changes in screening decisions and not investigator behavior (Section \ref{sec:targeting_gains}).} 
During worker team meetings discussed in Section \ref{sec:cpsbackground}, after the full allegation call narrative had been read out loud, one of the workers in the team meeting -- typically the supervisor -- consulted the algorithm score for each child on the referral. Each child's score was listed on the tool interface, but the maximum score of all children in the household was made most salient (Figure \ref{tool_display2}). Although workers knew the inputs to the algorithm and what it predicted, they did not receive an explicit explanation for why a score was high or low. The predicted likelihood of a child being removed within two years was also displayed in the bottom-left corner of the score interface: ranging from less than 0.5 percent for a child with a score of 1, to 14 percent for a score of 15, to 48 percent for a child with a score of 20. Workers were aware of the nonlinear relationship between algorithm risk scores and likelihood of removal. In practice, however, workers appear to have exclusively discussed the 1-20 risk score itself (not the probability of removal). After consulting the tool, workers deliberated until they reached a unanimous decision about the most appropriate response (examples in Appendix Table \ref{table_reasons}).

\subsection{Validity of the Algorithm Tool}

We assess the usefulness of the algorithmic tool using proprietary state hospital records that were not available to the algorithm developers (Figure \ref{fig:toolvalidation}). Subfigure (i) plots the relationship between algorithm risk quintile and a standardized aggregate index of potential maltreatment-related hospitalizations that is used as a primary outcome (described in Section \ref{sec:outcomes_detail}). A greater algorithm risk score predicts higher levels of child harm in the study's control group sample. Likewise, other subfigures of Figure \ref{fig:toolvalidation} present a robust, monotonic relationship between the tool's risk score and other indicators of child maltreatment risk such as child removals and investigator ratings of risk conditional on visiting a family. The patterns are consistent across the primary race-ethnic, gender, and socioeconomic status groups used for heterogeneity analyses. These findings confirm that the algorithm provided workers with a signal of future child outcomes.

\section{Data, Sample Characteristics, and Experimental Design}

\subsection{Data Sources}
\label{sec:datasources}

The trial county maintains an extensive administrative data system, which we linked to statewide hospitalization records and a database tracking algorithm randomization. Implementation partners at the Centre for Social Data Analytics (CSDA) shared algorithm risk scores for all incoming referrals, along with each household's randomized treatment assignment. 

The county's child welfare integrated data system includes identifiers and demographics for CPS-involved children and families over time, as well as information about each referral and subsequent CPS actions. 
Demographic characteristics of social workers and text transcriptions of reporter allegation calls were unavailable. However, we did gain access to coded allegation categories and information included in de-identified text data from team discussion notes, recorded when teams were deciding whether to investigate a family.

Statewide hospital inpatient and emergency room (ER) records were linked through a partnership with the state's hospital association,  spanning January 2020 through June 2022.\footnote{ Records were matched probabilistically using person-level identifiers (full name, date of birth, and gender) and include fields for: dates of admission and discharge, point of origin, facility type, admission priority, charged amount, diagnosis codes, and procedure codes. Match probabilities were 95 percent or greater for nearly all children.} 
Notably, we requested information on all hospitalizations in the state -- not just hospitalizations in the trial county -- to address concerns about differential migration. 86 percent of matched hospitalizations occurred within the trial county, and 93 percent occurred within either the trial county or a neighboring county.

\subsection{Sample Characteristics}
\label{sec:Sample}

The trial ran for 17 months from November 1, 2020 to March 29, 2022. During the trial, $4,681$ unique children, appearing on $2,832$ referrals, were successfully randomized into either having the algorithm available for social workers or not. These children were then followed over time, retaining their initial treatment status for any subsequent referrals.
Appendix Table \ref{SampleSelectionTable} describes the sample selection procedure and how the analysis focuses on the initial randomized referral and a child's outcomes thereafter. Section \ref{sec:trial} describes the randomization protocol. Figure \ref{timeline} provides a timeline for the study.

Appendix Table \ref{Desc_sampfull} presents descriptive statistics for the full analysis sample. 
Our sample is evenly split in terms of gender. Two-thirds of children are recorded as white in the data system, 3.6 percent as Black, and race is unknown for most of the remaining children. Race and ethnicity are not mutually exclusive. 
Although ethnicity is unknown for 34 percent of children, at least 17 percent of children in the sample are Hispanic. As is true of many child protection agencies, Black and Hispanic children are overrepresented in the sample relative to their share of county population.

The average referral has 2.24 (median of 2) children listed. Children are referred on average 2.2 times during the trial, and two percent of children are removed.
Overall, three quarters of referrals have at least one allegation of neglect, and 43 percent have at least one allegation of abuse (Appendix Table \ref{Desc_sampfull}, Panel B). 43 percent of children have a hospital visit after their referral, with an average of 1.7 hospital visits (Appendix Table \ref{Desc_sampfull}, Panel C).

For context, we briefly compare the trial county with national population and child welfare characteristics.\footnote{Our comparisons use statistics from the 2023 U.S. Census Bureau and \url{https://www.acf.hhs.gov/sites/default/files/documents/cb/cm2022.pdf}.} The trial county is one of the largest counties in Colorado, spanning urban, suburban, and rural areas. It ranks above the national average in socioeconomic status, with median household income approximately 15 percent greater than the national average, with a predominantly white population (92\%). The county's CPS agency operates similarly to others nationally, though with a screen-in rate (30\%) that is below the national average (50\%) but also not outside the norm for many U.S. agencies. The county uses group decision-making, which is typical in Colorado. Although the trial began during the COVID-19 pandemic (November 2020), schooling had returned to in-person and hybrid formats, and reports to CPS had returned to at least 80 percent of typical rates. Discussions with social workers indicate that COVID had minimal impact on integrating the algorithm tool into workflow. Results are robust to controls for seasonality.

\subsection{Experimental Design}
\label{sec:trial}
We designed a randomized controlled trial (RCT) to estimate the causal effect of providing algorithmic risk predictions on CPS worker decisions and child outcomes. Access to the algorithm was randomized at the household level, as the decision of whether to investigate a household is common to all children in the home. The first time a household was seen during the trial, it was randomized into either the intervention group or control group, with nearly identical workflows (Appendix Figure \ref{tool_workflow}). For each referral, a social worker would look up the referral ID  in the machine learning tool interface, read back the scores to the group, and the group note keeper would typically record scores in the discussion notes. When a family was in the intervention arm of the trial, the scores of all children in the household were shown (usually highly correlated), with the maximum being emphasized. When the family was in the control arm of the trial, the interface did not show any of the scores and workers would continue their standard discussion of the referral. Workers were aware of the trial and that the score would be (un)available for certain families at random.

One novel contribution of our study design is that randomization was instantiated at the household level, such that re-referred children retained their original intervention or control status. We specifically randomized mothers to address the unresolved problem of tracking non-nuclear, unstable households.  This is a key difference from \cite{fitzpatrick2020}, where randomization was conducted at the day-team level such that re-referred families -- which comprise most referrals -- were re-randomized and thus became more likely to be treated. 
Our trial design ensured that the control group would not have a score shown during the trial period, 
allowing us to compare a world where the tool is consistently available to a counterfactual world without any access to algorithmic information.

\subsection{Empirical Strategy}

The empirical strategy in this paper estimates the effect of algorithm support ($\delta_1$). For child $i$ first scored on referral $r(i)$, we estimate:

\begin{dmath}
	\label{eq:eq1}
	Y_i=\delta_0 + \textcolor{blue}{\delta_1 AlgoAccess_{r(i)}} + \gamma_{s(i)} + \epsilon_{i}
\end{dmath}

\noindent where $Y_i$ is the outcome of interest and $\epsilon_{i}$ an error term. Due to an unintentional coding error from the implementing partner, a small subset of observations were randomized only conditional on sibling group size $s(i)$, leading us to include corrective  sibling-group-size fixed effects $\gamma_{s(i)}$, hereafter called ``randomization controls,'' in our  specifications (see Appendix C for more details). The key variable, $AlgoAccess_{r(i)}$, is a binary indicator equal to one if the algorithm tool was available to the team during their discussion of the child's first referral $r(i)$ during the trial, and zero if the algorithm was not available.  Standard errors are clustered conservatively at the household (overlapping family network) level.\footnote{A household cluster is defined conservatively as any extended network of individuals who ever appeared jointly on a referral. 
In practice, there is no difference from clustering at the referral level (Appendix Table \ref{tab:HealthRobustness}).}

 \paragraph{Validation of Algorithm Use and Random Assignment}

The effectiveness of algorithmic decision aids depends on whether workers choose to use them. Since workers were strongly encouraged but not required to consult the tool in our setting, the policy-relevant treatment effect is the intent-to-treat (ITT) effect of access to the algorithm. Using de-identified discussion notes, we ascertain that workers wrote down algorithm scores in at least 73 percent of instances when available in our sample (Table \ref{tab:firststage}). This fraction of observed and recorded scores is relatively constant across the algorithm risk score distribution (Table \ref{tab:firststage}, column 2). The recorded share is likely an underestimate: workers may have consulted the tool but forgotten to write down the score in their meeting notes. Our estimates show that workers regularly chose to consult the algorithm, even though there were no professional repercussions for not using it.

Random assignment is important for estimating unbiased causal effects of algorithm availability on outcomes. We check for evidence of random assignment with a balance test, regressing assignment status on a vector of salient pre-randomization observable characteristics.  Table \ref{tab:balance} confirms that the algorithm was randomized successfully (\textit{p}-value of the F-test is 0.74).

\subsection{Outcomes: Proxies for Child Harm}
\label{sec:outcomes_detail}

We measure effects on child wellbeing using a pre-specified set of hospitalization outcomes across the entire state. These include: a child's number of high-priority hospital admissions (emergency, urgent, or trauma-related), admissions with injury ICD (diagnosis) codes, preventable emergency room visits, visits related to substance abuse and exposure, visits with ICD codes for intentional harm  (assault or self-harm), and visits with ICD codes for maltreatment, which are rare. Preventable ER visits are defined as the number of child admissions for ambulatory care sensitive conditions (ACSC, classified using Appendix 8 of \citealt{carey2017hospital}). For example, some ER visits for asthma may have been avoidable if a child had received proper preventative treatment and supervision.

These categories of hospitalization have been clinically and empirically associated with child maltreatment. Maltreatment ICD codes are a measure of physician-confirmed maltreatment, but notoriously underestimate overall child maltreatment \citep{scott2009, schnitzer2011identification}. Consequently, researchers have primarily used and validated injuries as a leading proxy for maltreatment \citep{schnitzer2011identification, vaithianathan2020}. Prior research has also demonstrated a link between maltreatment and intentional injury, namely self-harm, which can often result from adverse childhood experiences such as parental neglect and sexual abuse (\citealt{liu2018}, \citealt{vaithianathan2020}). High-priority admissions are a related signal of potential acute harm, and avoidable ER visits are particularly associated with possible neglect and insufficient parental supervision. Substance abuse and exposure (i.e., from parents or youths' direct abuse) are one of the primary drivers of child neglect and CPS involvement more broadly. Our outcome categories therefore cover a range of maltreatment types: physical harm, psychological harm, and neglect.

Our preferred outcome is an aggregated ``harm index'' of pre-specified hospitalization outcomes. A high Cronbach's alpha ($\alpha = 0.85$), reflecting internal consistency across the outcomes, confirms that combining these outcomes into a composite index is meaningful. 
The index is the standardized average of the six pre-specified variables described above, each themselves standardized. Variables are standardized (mean 0, variance 1) using the control sample, and a lower value of the harm index is considered better for the child. This index was chosen as the most standard and agnostic way to combine these outcomes, in the absence of a pre-specified composite measure. Its advantage is to equally weigh all of the six proxies for maltreatment, regardless of each outcome's baseline prevalence. This is in contrast to measuring the number of distinct (single-counted) hospital visits that include at least one category from the index (Row 5 in Table \ref{tab:HealthRobustness}), which in practice only picks up on the most prevalent harm index component, the number of high-priority visits (correlation $\rho=0.98$). The harm index weighs equally prevalent and less prevalent outcomes and aligns better with the idea that one of these health incidents is not per se an indication of maltreatment, but rather that having an unusually high number of visits on \textit{several} of these margins at the same time is suggestive of concern. In Appendix Table \ref{tab:HealthRobustness} we consider alternative indices: an optimally weighted index if all components are equally impacted \citep{o1984procedures}, a first principal component, and two ways of counting hospital visits for harm-index-related categories. Per our pre-analysis plan, all outcomes are measured following 30 days after randomization to accommodate a period of potential mechanical increases in child receipt of medical care from CPS interactions, such as post-investigation medical care for previously untreated conditions. Results are robust to excluding the first 60 days or to including the first 30 days (Table \ref{tab:HealthRobustness}).

\section{Results}

Despite Child Protective Service's primary objective of protecting children, few studies have produced causal estimates of the impact of CPS policies on child injury and other health outcomes. We link statewide hospitalization records from the trial state's hospital association (Section \ref{sec:datasources}) to measure the effect of algorithmic information on ground-truth outcomes. The algorithm's score, a summary measure of family history with social services, appears to have helped workers respond more effectively to current family concerns, targeting visits to children in greater need, regardless of their past contact with CPS (see Section \ref{sec:mechanisms} for more detail on mechanisms). This section covers three main findings: a reduction in child harm, a reduction in child harm disparities, and a reduction in racial CPS surveillance disparities.

\subsection{Impact of Algorithm Availability on Child Health and Wellbeing}
\label{sec:healthimpacts}

Table \ref{SERVER_hospitaloutcomes1} presents estimated effects of algorithm support on child wellbeing for the outcomes specified in the previous subsection, using equation (\ref{eq:eq1}). Column 1 of Panel A shows that tool availability reduced a child's future harm by 0.05 standard deviations (SDs) compared to children in the control group ($p=0.046$). Effects seem attributable to both the intensive and extensive margins (see Table \ref{tab:otheroutcomes}, Column 2 for a harm index constructed using just the extensive margin).   
Estimates on the full sample likely underestimate our treatment effects. Children referred late in the trial are  more likely to have limited or no hospital interactions prior to the trial end date,\footnote{\label{fn:attenuation_fn}At the end of the trial, the algorithmic tool became available to all children being (re-)referred, attenuating expected treatment effects.} leading to non-classical measurement error in the outcome which would bias our estimates toward zero.  
As a result, when we restrict to the sample of children referred in the first year of the trial, effect sizes increase to $0.06$ SD and several of the reductions in the harm components (Panel B) become significant at the 95 percent confidence level. As we did not pre-register the sample in column 2, we report estimates using the full sample throughout.

A 0.05 SD reduction in the harm index is sizable in this context. For comparison, children removed to foster care during the trial -- a proxy for severe maltreatment or significant danger -- had an accumulated harm index that was 0.46 standard deviations greater than those children who were not removed. The intent-to-treat effect from the algorithm is approximately 11 percent of this benchmark and -- assuming our first stage does not underestimate tool use -- the effect on compliers (-0.07 SD; Table \ref{tab:HealthRobustness}) is as large as 15 percent of the benchmark.

Column 3 of Panel A shows that the algorithmic tool was particularly effective at reducing extreme values of child harm (-60\%, $p=0.0031$, not pre-registered): Significantly fewer children placed in the top one percent of the harm index (defined over the intervention and control groups together) when the tool was available. Accumulated harm for this top percent ranged from 2.6 to 29 standard deviations above the mean, with the median child in this group having a high priority hospital visit every three months. Relatedly, benefits of the algorithm appear to be greatest for children in the top quintile of algorithm-predicted risk scores, who may have been at greatest risk of maltreatment (Appendix Figure  \ref{SERVER_bar_treatcontrol_indexpost30_std}). 
The majority of children with an explicit maltreatment-related hospital ICD code had a harm index in the top one percent of the harm distribution, which suggests that the algorithm may have reduced child maltreatment.

\begin{table}[H]
    \centering
    \fontsize{10pt}{10pt}\selectfont
	\caption{Access to the Algorithm Tool Reduced Child Harm \vspace{0.25em} \\
	{\small \textit{Hospital Encounters after Randomization}}}
	\label{SERVER_hospitaloutcomes1}
	\vspace{-0.25em}
    \begin{tabular}{l*{6}{c}}
\toprule
\midrule
\addlinespace
\multicolumn{7}{l}{\textit{Panel A: Combined Harm Index}} \\
\addlinespace
                    &\multicolumn{1}{c}{\begin{tabular}{@{}c@{}}(1) \\ Full \\ Sample\end{tabular}}&\multicolumn{1}{c}{\begin{tabular}{@{}c@{}}(2)\\ Longer\\ Follow-Up\end{tabular}}&\multicolumn{1}{c}{\begin{tabular}{@{}c@{}}(3) \\ Top 1\% in \\ Harm Index\end{tabular}}&            &            &            \\
\midrule
Algorithm Available &      -0.050** &      -0.062** &     -0.0088***&            &            &            \\
                    &     (0.025)   &     (0.029)   &    (0.0031)   &            &            &            \\
                    &[\textit{p} = 0.046]   &[\textit{p} = 0.032]   &[\textit{p} = 0.004]   &            &            &            \\
\addlinespace Randomization Controls &         Yes   &         Yes   &         Yes   &            &            &            \\

Control Mean        &       0.000   &       0.000   &       0.015   &            &            &            \\
Effect (\%)         &               &               &         -60   &            &            &            \\
Design Power               &        0.85   &        0.91   &        0.86   &            &            &            \\
Observations        &       4,681   &       3,431   &       4,681   &            &            &            \\
\midrule
\midrule
\multicolumn{7}{l}{\textit{Panel B: Harm Index Components}} \\
\addlinespace

                    &\multicolumn{1}{c}{(1)}&\multicolumn{1}{c}{(2)}&\multicolumn{1}{c}{(3)}&\multicolumn{1}{c}{(4)}&\multicolumn{1}{c}{(5)}&\multicolumn{1}{c}{(6)}\\
                    &\multicolumn{1}{c}{\begin{tabular}{@{}c@{}}High \\ Priority\end{tabular}}&\multicolumn{1}{c}{Injury}&\multicolumn{1}{c}{\begin{tabular}{@{}c@{}}Avoidable \\ ER\end{tabular}}&\multicolumn{1}{c}{\begin{tabular}{@{}c@{}}Substance \\ Exposed\end{tabular}}&\multicolumn{1}{c}{\begin{tabular}{@{}c@{}}Maltreatment \\ ICD\end{tabular}}&\multicolumn{1}{c}{\begin{tabular}{@{}c@{}}Intentional \\ Injury\end{tabular}}\\
\midrule
Algorithm Available &      -0.069   &      -0.034*  &      -0.016   &      -0.034*  &     -0.0019   &     -0.0052   \\
                    &     (0.043)   &     (0.019)   &     (0.017)   &     (0.018)   &    (0.0034)   &    (0.0036)   \\
                    &[\textit{p} = 0.112]   &[\textit{p} = 0.068]   &[\textit{p} = 0.335]   &[\textit{p} = 0.063]   &[\textit{p} = 0.585]   &[\textit{p} = 0.142]   \\
\addlinespace Randomization Controls &         Yes   &         Yes   &         Yes   &         Yes   &         Yes   &         Yes   \\

Control Mean        &       0.543   &       0.166   &       0.137   &       0.079   &       0.012   &       0.015   \\
Effect (\%)         &         -13   &         -21   &         -12   &         -44   &         -16   &         -34   \\
Design Power               &        0.83   &        0.73   &        0.60   &        0.61   &        0.30   &        0.52   \\
Observations        &       4,681   &       4,681   &       4,681   &       4,681   &       4,681   &       4,681   \\
\bottomrule \end{tabular}

\end{table}
\vspace{-1em}
{\footnotesize\setlength{\parindent}{0pt}\begin{spacing}{1.0} Notes: This table presents estimates of the impact of algorithm availability on child injuries and other hospitalization outcomes. In Panel A, we report effects on a combined harm index, whereas in Panel B we report effects on each of the pre-registered subcomponents of the harm index before standardization. The harm index is constructed as a standardized average of the following standardized outcomes: number of high-priority admissions (emergency, urgent, and trauma), number of admissions with listed injury, number of avoidable ER visits, number of substance abuse-related visits, number of admissions with a confirmed maltreatment code, and number of visits with an intentional injury (assault, self-harm) code. A lower value of the harm index is considered better. In Panel A, beyond our main estimate in Column 1, we also report in Column 2 the effects on a sample of  children with a longer follow-up (referred in the first year of the trial) so that we can observe several months of hospital data. Column 3 estimates the effect on the likelihood of being in the top one percent of the harm index.  Randomization procedure controls are included throughout. We report effects as a percent of the control mean and the power to detect significant effects at the bottom of the table. Power is calculated using observed means and standard deviations in treatment and control groups, assuming 1,400 household clusters per group, an average cluster size of two, a coefficient of variation (CV) of 0.56, an intraclass correlation (ICC) of 0.4, and a significance level of 10 percent. Standard errors are clustered at the household level. Significance reported as: * p$<$0.1, ** p$<$0.05, *** p$<$0.01.\end{spacing}}
\vspace{1em}

Panel B of Table \ref{SERVER_hospitaloutcomes1} present effects on each of the six outcomes that make up the harm index. Point estimates are negative across all six outcomes, and significant or nearly significant for the two outcomes where we have highest power ($p=0.112$ for high priority visits and $p=0.068$ for injuries). Furthermore, the point estimates are large relative to the means in the control group, ranging from -12\% to -44\%. In particular, the algorithmic tool reduced the average number of future injury admissions by 21 percent, or 0.03 admissions relative to a mean of 0.17. In the appendix, we report how informative the estimates of these six components would be from a Bayesian perspective (details provided in the next section), assuming as a prior that effects could have ranged between plus or minus one hundred percent (Appendix \ref{fig:bayesharm}). Even for outcomes where our findings are not statistically significant at conventional levels, our findings are informative for the likely range of effects. Our final pre-registered proxy for maltreatment was a previously identified list of ICD codes that are suggestive of maltreatment for a subset of child ages, following \citet{schnitzer2011identification}. As these are age-specific, this outcome is only available for half of our sample, and consequently cannot be included in the full sample index. We are underpowered on this margin, but the percentage effect is consistent with a halving of incidents suggestive of maltreatment (Column 1 of Appendix Table \ref{tab:otheroutcomes}).

Reassuringly, we find no effects of the tool for a natural placebo group: for 11 percent of children assigned to treatment, workers tried but could not see the score due to temporary unavailability, and resulting  outcomes were comparable to the control group (Table \ref{tab:otheroutcomes}, Column 7 placebo specification). We additionally present a series of robustness tests for the main effects of the algorithm on child harm and injury-specific hospital visits in Appendix Table \ref{tab:HealthRobustness}. The findings are robust to: a permutation test ($p = 0.04$); to controlling for all baseline characteristics (including prior hospitalizations); excluding the first 60 days after a referral when prior injuries could be discovered during a CPS-initiated health examination; to including all days after a referral; to alternative clustering of standard errors; and to a bounding exercise for treatment non-compliance.\footnote{As discussed in Appendix C, $n=80$ children in the control group had the algorithmic tool available due to changes in family structure.} The findings are also robust to other ways of combining outcomes into a harm index, such as using a summation of visits from each harm index category, a summation of distinct (non-overlapping) harm index visits, using an optimally weighted index if all components are equally impacted \citep{o1984procedures}, and using a first principal component. 
The least statistically significant effect is for the distinct count of harm visits where the point estimate suggests a marginally significant ($p=0.100$) 0.07 decrease (13\% decrease). Despite its intuitive appeal, this count of harm visits turns out to be a near proxy of the number of high priority visits ($\rho=0.98$). In contrast, our preferred outcome, the harm index, combines all margins of potential harm, including those with lower prevalence, and takes into account hospitalizations with multiple causes for concern (e.g., with both injury and substance exposure codes). 

The reduction in child harm does not appear to be driven by changes in willingness to seek care. Tool access did not significantly reduce the overall incidence or frequency of hospital interactions (Columns 3 and 4 of Appendix Table \ref{tab:otheroutcomes}). Our pre-registered placebo outcome of child cancer-related hospitalizations (pre-specified following \citealt{vaithianathan2020}) is insignificant, though under-powered. We also find null results for better-powered, non-pregistered placebo results for a specification where workers reported not seeing a score even when assigned to treatment, and when using lower-priority hospital visits as an outcome (columns 6 and 7 of Appendix Table \ref{tab:otheroutcomes}). Consistent with our pre-analysis plan, providing algorithm support specifically decreased hospitalizations for reasons more closely associated with maltreatment. We find an imprecise reduction in charges (cannot observe negotiated bill; column 5 of Appendix Table \ref{tab:otheroutcomes}).

We do not detect significant effects of providing the algorithmic tool on the incidence or number of re-referrals and removals (Appendix Table \ref{tab:CPSOutcomesTable}). The absence of detectable impacts on removals should not be over-interpreted, as our study is underpowered to detect even large effects (e.g., power $\approx 0.2$ to detect a 15 percent change). Moreover, the short-term impact of better-targeted investigations on removals is theoretically ambiguous. Our point estimates indicate a small possible reduction in the number of re-referrals (–5 percent), though the lack of statistical significance is more surprising for this outcome than for removals. This pattern suggests potential discrepancies between re-referral–based measures of maltreatment and hospital-based measures, reminiscent of the long-standing literature documenting differences between self-reported and substantiated CPS maltreatment \citep{negriff2017}. Prior research has shown that the informational quality (``signal'') of CPS referrals varies by reporter type \citep{nadon2023}. In line with this evidence, when we restrict to re-referrals from law enforcement -- a reporter type with a high substantiation rate and known to be relatively reliable -- we find suggestive declines consistent with our hospital-based outcomes (–12.5 percent decrease in law enforcement re-referrals, $p=0.20$). In contrast, re-referrals from less reliable reporters, such as education professionals, show no decline (+7.6 percent, $p=0.47$). Given that CPS reporting is selective and intertwined with exposure to reporters, our findings suggest that re-referral counts may be an imperfect and noisy proxy for child welfare. Outcomes external to CPS, such as hospitalization, are therefore essential for corroborating the effectiveness of CPS interventions.

\paragraph{Exploring Spillovers and Learning.}

Importantly, we do not find evidence that tool access crowded out resources provided to children in the control group. Table \ref{spillovers_withinday} tests for the presence of short-run, within-day spillovers (i.e., contamination) by leveraging the share of referrals that were treated during the day, excluding a child's own referral. If benefits had come at the expense of the control group,  then, assuming fixed resources per day, one might expect to see: (1) worse outcomes for children in the control group as the percent treated in the day increases, and possibly also (2) weaker effects for treated children as the percent of children treated in the day increases. Though our confidence intervals are large,  we find no evidence of statistically significant spillover effects in terms of screen-ins, discussion time, or child harm. Overall, we interpret the results as suggestive, reassuring evidence that the algorithm benefited children in the treatment group without imposing costs on the control group.

Our specification may underestimate the benefits of the algorithm if workers learned and changed their investigation tendencies from using the tool. In this scenario, we would expect indirect benefits for children in the control group and an attenuation of treatment effects over time.  While our study is unable to assess learning that might have occurred after the end of the trial, we can observe differences in treatment effects for children referred at different points during the trial. Splitting the sample into four equally-sized groups by date of first referral in the trial, we find that harm is reduced by similar amounts for children referred in the first three quarters of the trial. There appears to be no impacts for children referred in the final quarter (Figure \ref{fig:LRlearning}), possibly owing to mechanical attenuation from a limited window to observe hospital outcomes for treated versus control during the trial (see footnote \ref{fn:attenuation_fn}). 
The stability of estimates for the first three quarters of the trial are consistent with a lack of learning. However, these results should be interpreted with caution given the large confidence intervals by quarter and the fact that chronically-referred children were more likely to appear in earlier quarters of the trial. Although we cannot rule out learning within the trial, anecdotal evidence from interviews with workers suggest that workers did not consciously change their investigation strategy for control referrals. To the extent that there was unobserved learning by workers, our study estimates would likely be a conservative lower bound on the true impacts of providing the algorithm. Our discussions with supervisors at the end of the trial also suggest that workers did not become overly reliant on the tool, as supervisors correctly recalled that the tool only summarized history and not all dimensions of ``future maltreatment risk.''

\subsection{Impact on Group-Level Disparities in Health and Wellbeing}
\label{disparities_health}

In this section, we provide empirical evidence of how algorithm support affects disparities. A significant concern about algorithmic decision aids is their potential to perpetuate and widen inequality. Algorithms trained on past decisions have been repeatedly shown to embed historical bias  \citep{obermeyer2019dissecting} especially when historical data was from automatic decision-making \citep{agan2022algorithms}. They may also be less informative for members of disadvantaged or under-represented groups \citep{larrazabal2020Gender, blattner2022How}. Humans have been documented to override algorithm recommendations in ways that reinforce existing biases \citep[e.g.]{albright2019a}. On the other hand, algorithms may be \textit{less} biased than status-quo human decision-makers (c.f., \citealt{hoffman2018discretion}, \citealt{kleinberg2018}, \citealt{arnold2024building}). 
Absence of rigorous empirical evidence on whether algorithmic tools indeed increase disparities has been a major hindrance to their adoption for social policy.

\begin{figure}[H]
	\caption{Algorithm Tool Reduced Harm the Most for Historically Disadvantaged Groups}
	    \label{fig:HealthDisparity}
	\begin{subfigure}[b]{0.5\textwidth}
		\caption{Race}
		\includegraphics[trim=0 20 10 15, clip,width=\textwidth]{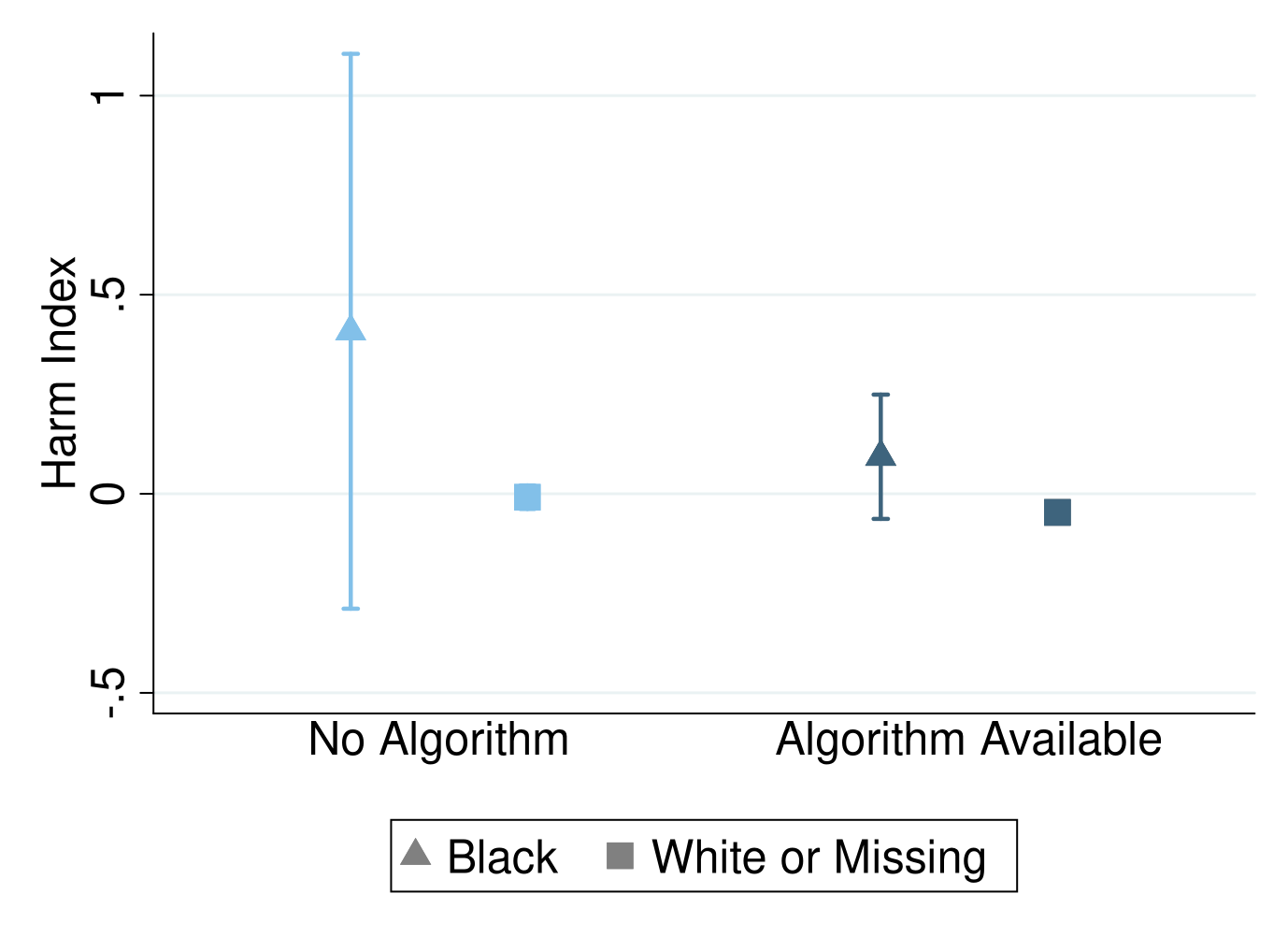}
		\label{fig:blackhealthdisparity}
	\end{subfigure}
	\begin{subfigure}[b]{0.5\textwidth}
		\caption{Ethnicity}
		\includegraphics[trim=0 20 10 15, clip,width=\textwidth]{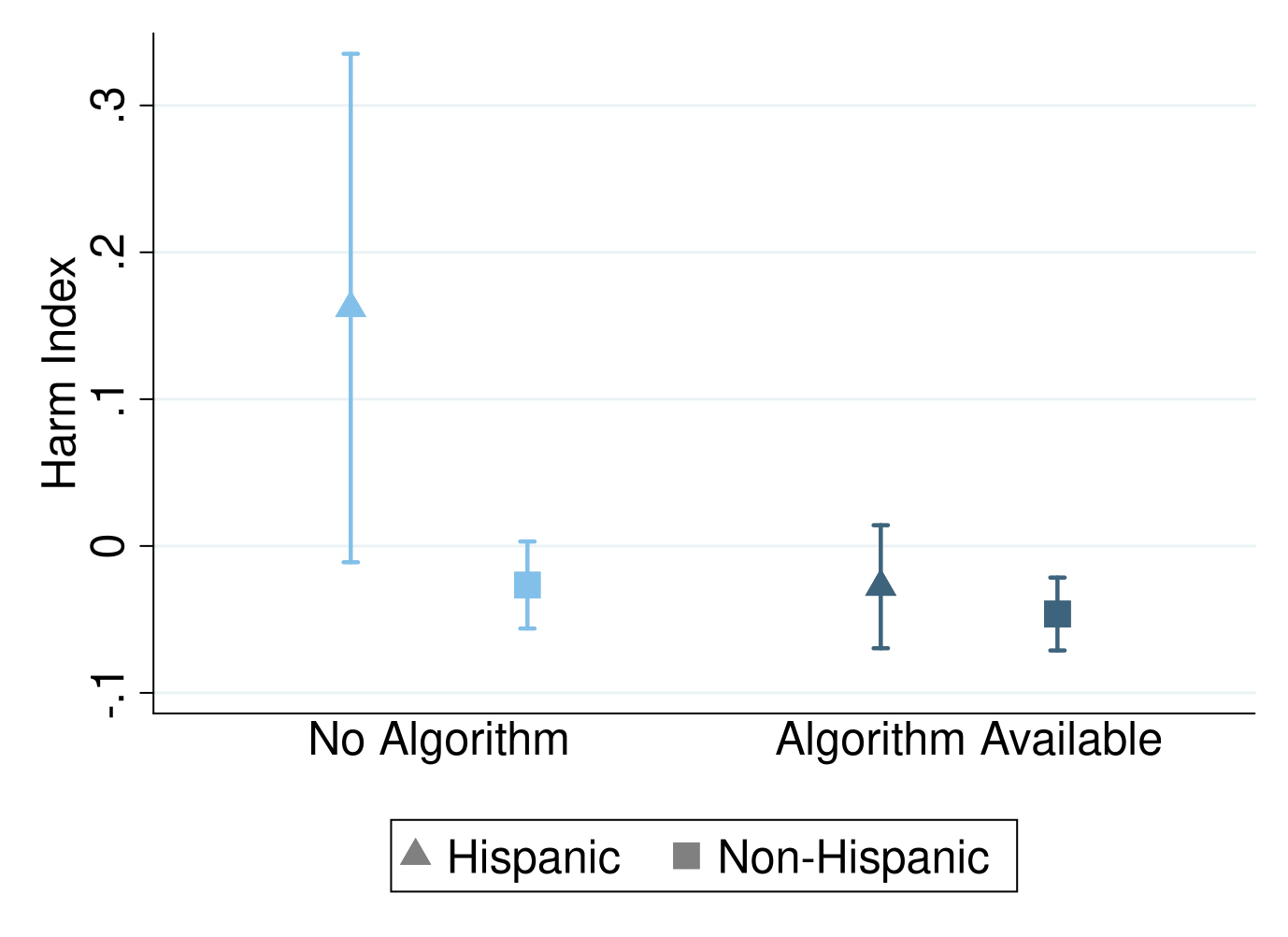}
		\label{fig:hispanicdisparity}
	\end{subfigure}

	\begin{subfigure}[b]{0.5\textwidth}
		\caption{Gender}
		\includegraphics[trim=0 20 10 15, clip,width=\textwidth]{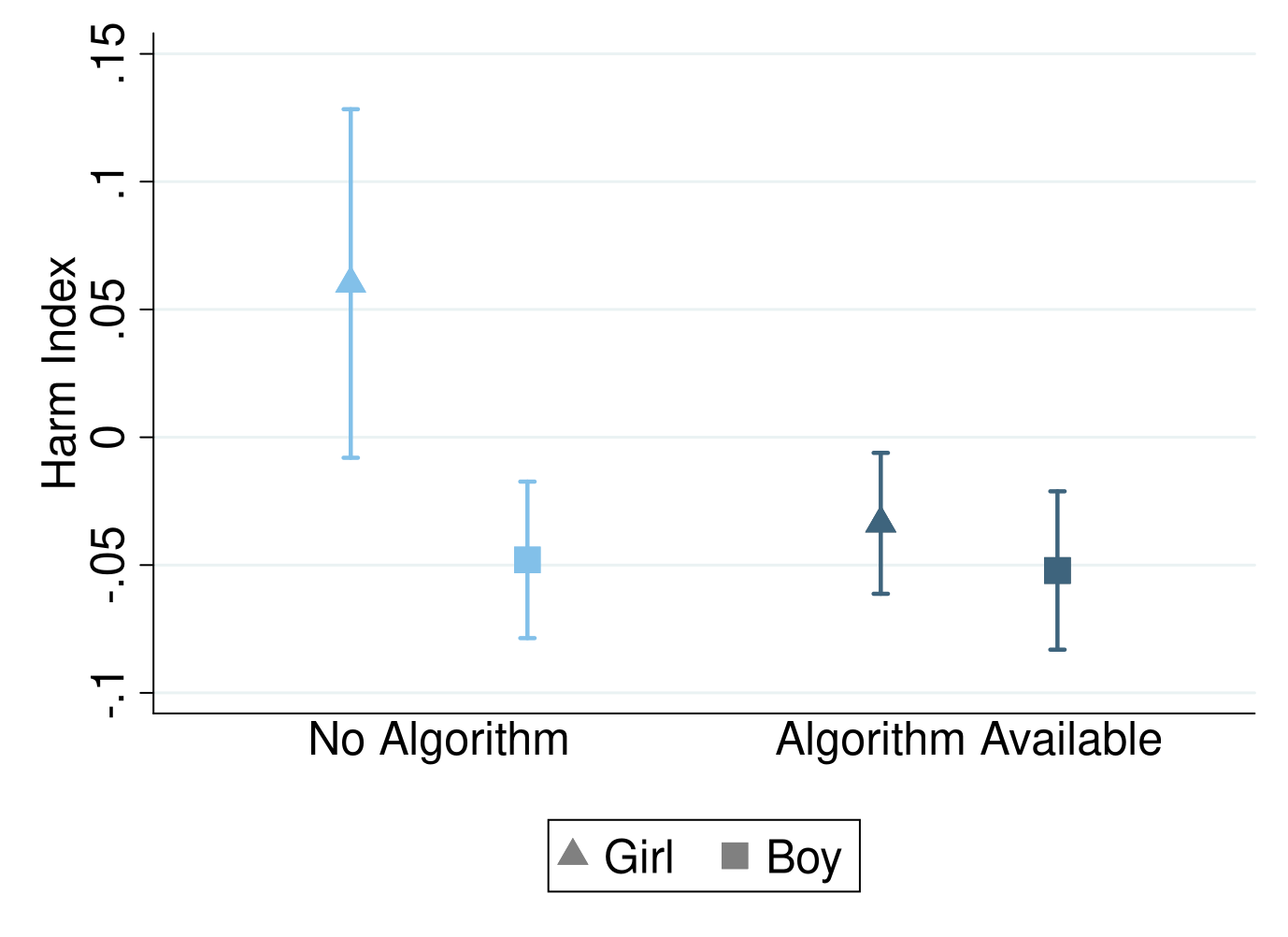}
	\label{fig:genderdisparity}
	\end{subfigure}
	\begin{subfigure}[b]{0.5\textwidth}
		\caption{Socioeconomic Status}
		\includegraphics[trim=0 20 10 15,clip,width=\textwidth]{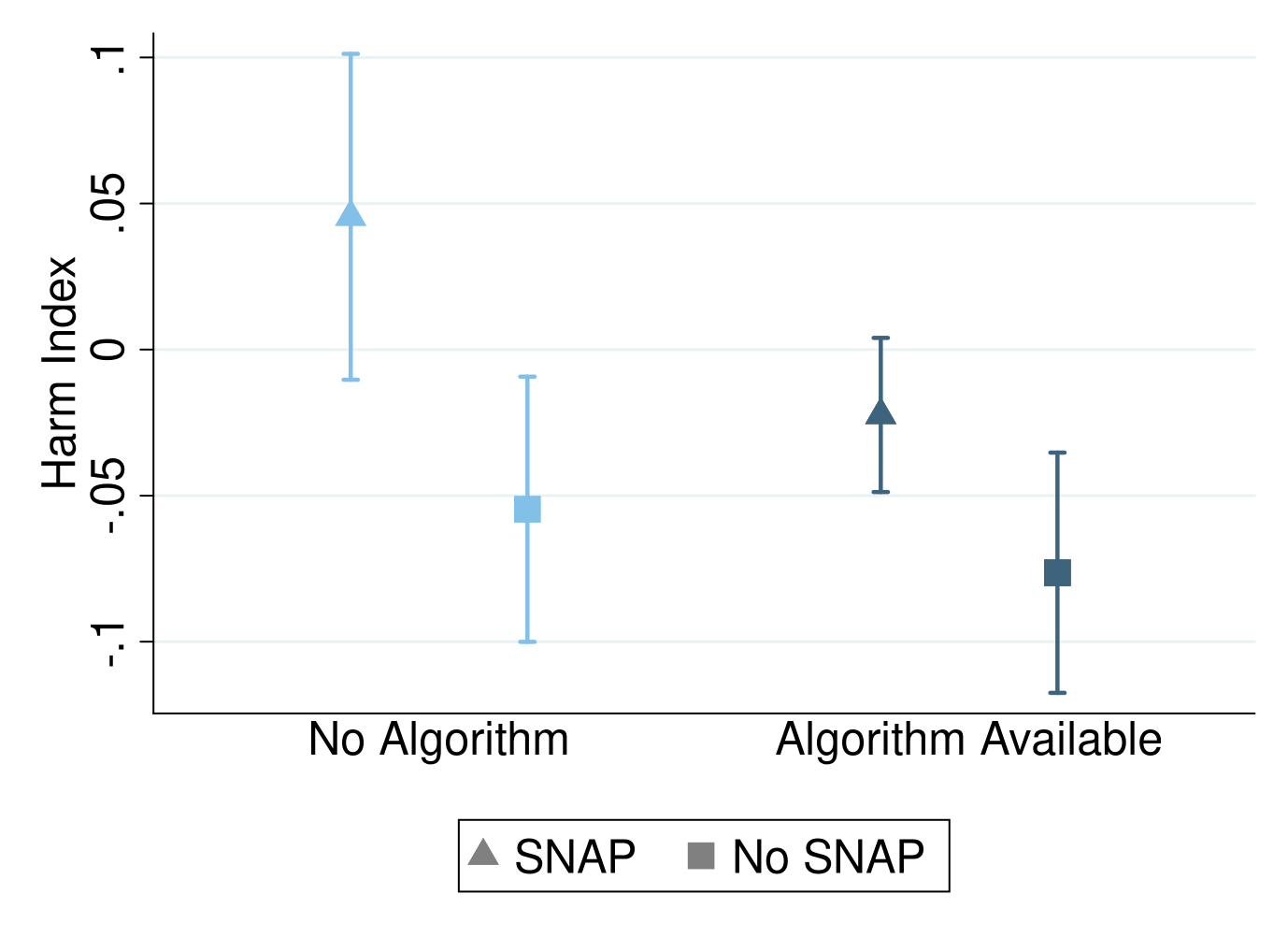}
        \label{fig:sesdisparity}
	\end{subfigure}	
\end{figure}
\vspace{-1em}
{\footnotesize\setlength{\parindent}{0pt}\begin{spacing}{1.0}Notes: These figures show the mean value of children's harm index by demographic category and tool availability, derived from regression estimates. SNAP (food stamp) receipt is a proxy for lower socioeconomic status. White children in Panel (i) have confidence intervals that are too small to be visible. 90\% confidence intervals are reported and randomization controls are included throughout. Regression estimates with \textit{p}-values are provided in Table \ref{tab:healthdisparity}. \end{spacing}}
\vspace{1em}

We provide some of the first field experiment evidence of the impact of algorithm tools on disparities in a social policy setting. Specifically, we report effects of algorithm availability on disparities in health by race, gender, ethnicity, and socioeconomic status (Figure \ref{fig:HealthDisparity} and Table \ref{tab:healthdisparity}). When the tool was unavailable, as shown in light blue on the left side of each panel of Figure \ref{fig:HealthDisparity}, children's harm index was elevated for Black children, Hispanic children, girls, and children from low-income families as measured by SNAP (food stamp) receipt. Table \ref{tab:healthdisparity} (row ``Type'') formalizes this result: in the control group, differences between the specified demographic group and out-group are statistically significant for girls, Hispanic children, and low-income children, and large but imprecise for Black children. These disparities mirror known statistics showing children with these demographic characteristics tend to face heightened risk of child mortality \citep{currie2016Mortality} and/or maltreatment \citep{putnam-hornstein2021Cumulative}.

Point estimates suggest that access to the algorithm reduced harm for children in disadvantaged categories, namely Black, Hispanic, female, and low-income children (compare light blue triangles to dark blue triangles in Figure \ref{fig:HealthDisparity}). The out-groups (square dots) also experienced improvements in outcomes, but they were smaller in magnitude. Precise estimates are reported in Table \ref{tab:healthdisparity}, and imply a reduction in disparities in the harm index on the order of 50 to 80 percent (Table \ref{tab:healthdisparity}, bottom row). Though large, these reductions are not statistically distinguishable from zero.\footnote{Earlier versions of the paper used Stata's \texttt{suest} command, which reported a highly significant joint test $p<0.01$. However, \texttt{suest} does not accommodate different variables temporarily sharing the same name across specifications (i.e., a loop over four disadvantaged groups of interest), leading to an error when compared with hand-calculated stacked regression specifications.} 

Our findings are nevertheless informative in a Bayesian perspective. 
 In the absence of prior evidence, suppose a policymaker would have assumed effects on disparities could be of any value. We model this with a uniform prior distribution where algorithm support could reduce disparities by up to 200 percent or increase them by that same amount. Following \citet{brannlund2024}, we then calculate the posterior distribution given the data we observe, assuming normally distributed point estimates with observed variance.\footnote{The posterior is a truncated normal distribution over [-200,200]: $p(\theta|\hat{\theta})=\frac{1}{\Phi(\frac{200-\hat{\theta}}{\sigma})-\Phi(\frac{-200-\hat{\theta}}{\sigma})}\frac{1}{\sqrt{2\pi\sigma^2}} \exp\left(-\frac{(\hat{\theta}-\theta)^2}{2\sigma^2}\right)$ if $\theta \in [-200,200]$ and zero otherwise. $\Phi$ denotes a normal cumulative distribution function.} We report these posterior distributions with 90 percent credible intervals in Appendix Figure \ref{fig:bayesdisparity}. Our estimates reveal that in most cases we can rule out a worsening of disparities for ethnicity, gender, and socioeconomic status.

\subsection{Impact on Racial Disparities in CPS Screen-Ins}
\label{disparities_surveillance}

Providing algorithmic information may affect disparities in agency surveillance. CPS investigations can sometimes be intrusive and traumatic, and prior work has documented substantial racial disparities -- in some cases unwarranted -- in CPS contact across the U.S. (\citealt{kim2017lifetime}, \citealt{edwards2021contact}, \citealt{baron2023}). Though a small share of the sample, Black children are nearly twice as likely to be investigated as other children in the sample, even after conditioning on algorithm risk score (Table \ref{tab:screenintable}). In contrast, we find no evidence of screen-in disparities by gender, ethnicity, or socioeconomic status even after controlling for algorithm-predicted risk  (Table \ref{tab:screenintable}). We therefore focus our analysis of surveillance disparities on race, specifically Black children relative to children who are not Black (primarily white).

\begin{figure}[H]
	\caption{Algorithm Tool Reduced Racial Disparities in CPS Surveillance for Low-Risk-Score Children}
	    \label{fig:screeninrace}
	\begin{subfigure}[b]{0.5\textwidth}
		\caption{Below Median Algorithm-Predicted Risk}
		\includegraphics[trim=10 20 15 90, clip,width=1\textwidth]{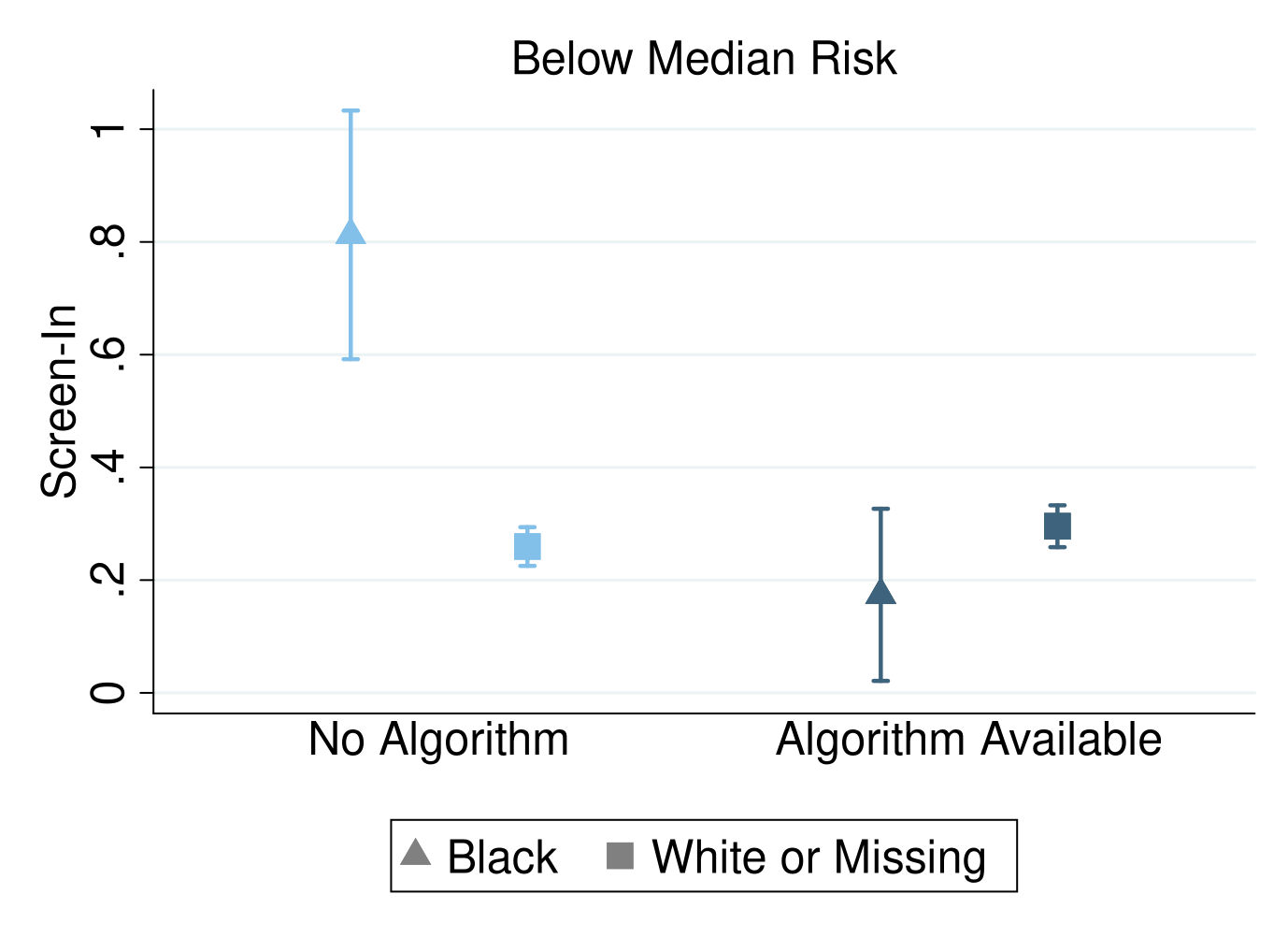}
		\label{fig:lowriskblacks}
	\end{subfigure}
	\begin{subfigure}[b]{0.5\textwidth}
		\caption{Above Median Algorithm-Predicted Risk}
		\includegraphics[trim=10 20 15 90, clip,width=1\textwidth]{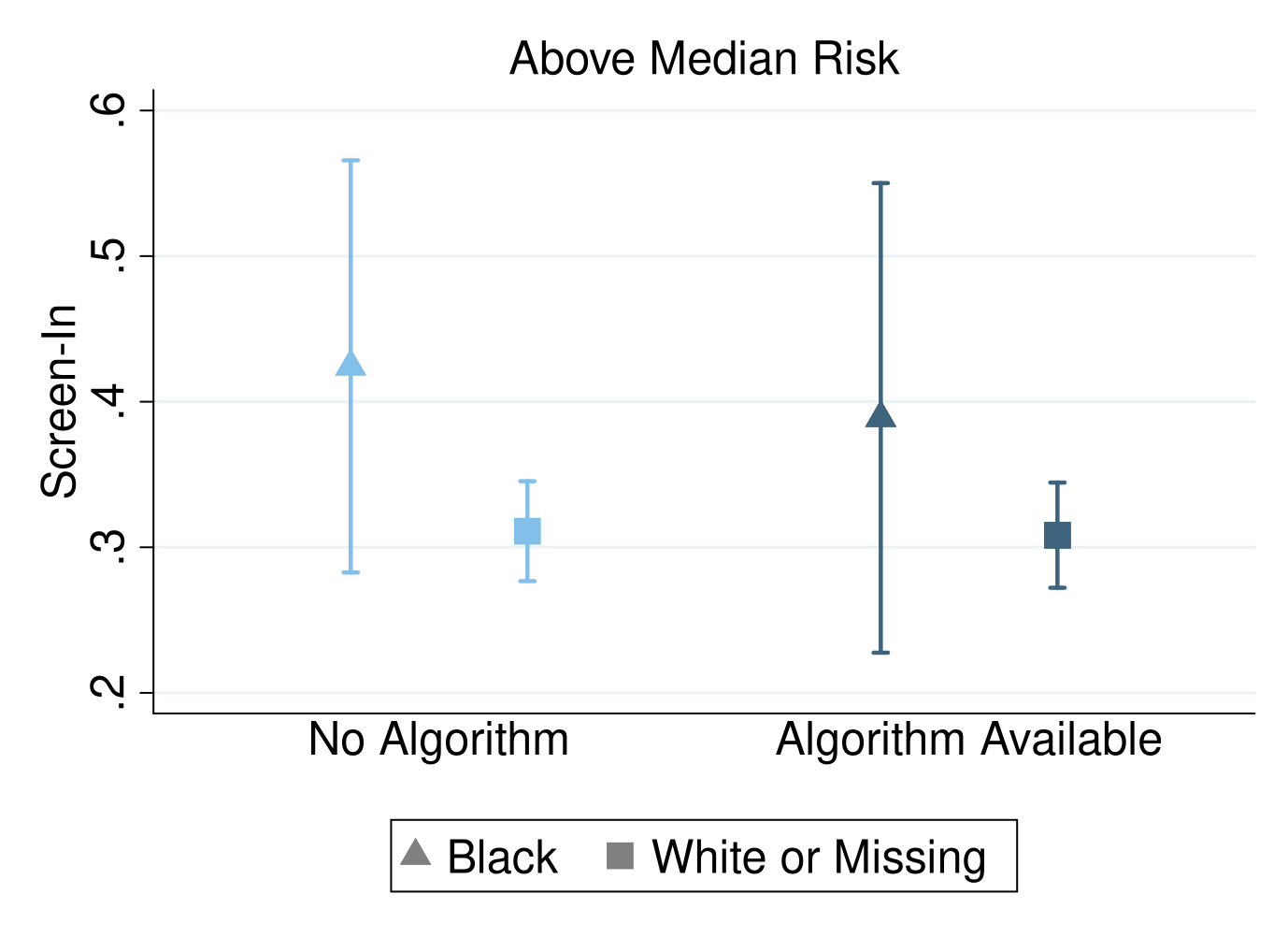}
		\label{fig:highriskblacks}
	\end{subfigure}
\end{figure}

Figure \ref{fig:screeninrace} shows that, in the control group, Black children at below-median predicted risk were \textit{60 percentage points} more likely to be screened in than other (primarily white) children of similar risk level. 
However, when the algorithm was available, below-median-risk Black children were screened in at much lower rates, comparable to rates for other children. Column 2 of Appendix Table \ref{tab:RaceRobustness} shows that the algorithm tool altogether reduced screen-ins by 13 percentage points for Black children across the entire sample, though point estimates are imprecisely estimated. These findings suggest that access to the algorithm reduced unwarranted racial disproportionality in investigations among low-risk children. Although the trial county has a relatively small Black population, there were still $n = 170$ Black children in the full sample when using CPS-derived race alone, with at least 23 children in each race-by-treatment-status by above or below median risk cell. Regression estimates and robustness to including a permutation test and alternate ways of defining race are reported in Appendix Table \ref{tab:RaceRobustness}. 
 Although the exact magnitude of screen-in disparity reductions -- and corresponding power to detect significant effects -- depends on the specification, all estimates suggest the algorithm reduced surveillance disparities conditional on risk and, taking the imprecise point estimates at face value, may have canceled out disparities entirely. Though we are statistically underpowered, we do not detect an offsetting increase in the harm index among low-scoring Black children (Appendix Figure \ref{fig:blackharmbymedian}).

\vspace{-1em}
{\footnotesize\setlength{\parindent}{0pt}\begin{spacing}{1.0}Notes: These figures present rates of CPS screen-in (investigation) by race and tool availability. Above and below-median predicted risk are defined using the algorithm referral risk score. Figures define race as recorded by Child Protective Services. 90\% confidence intervals are reported and randomization controls are included throughout.\end{spacing}}
\vspace{1em}

These results align with an alternative approach for measuring unwarranted disparities (UDs) that makes no assumptions about whether high or low risk scores merit investigations: comparing harm for screened-in versus screened-out Black and white children, respectively. Under the algorithm, screened-out Black children saw disproportionately greater reductions in harm relative to screened-out white children, with no offsetting change for screened-in children. We show in Section \ref{sec:mechanisms} that targeting of investigations (measured by isolating harm for screened-out children) improved more for Black than white children (Figure \ref{fig:harm_target_by_group}), implying a potential reduction in UDs.

Providing an algorithm risk score could alter stereotypes that workers use to make assessments in the absence of an external benchmark. This is consistent with theoretical work by \cite{bordalo2016} on how stereotypes form from a representative agent. For low-scoring Black children, workers may have been prompted by the algorithm to change their reference means to that of other low-risk-score children. 
Discussions with workers underscored their cultural sensitivity and aspiration for greater equality. Future work should examine whether similar reductions in disparities occur in settings with different family demographics or worker beliefs and training.

\section{Mechanisms}
\label{sec:mechanisms}

Providing workers with an algorithm that predicts child risk improved child outcomes and reduced racial disparities.
In this section, we link these benefits to improved worker allocation of investigations and rule out a number of alternative mechanisms. 

\subsection{Improved Worker Allocation of Screen-Ins}
\label{sec:targetingsection}

The intervention did not increase the total number of investigations (``screen-ins'') or any other measurable level of services (Figure \ref{fig:cpslevel}). Workers did not increase overall levels of interventions and did not significantly change discussion length, as is sometimes observed due to asymmetric valuation of false positives and false negatives (\citealt{almog2024}, \citealt{fitzpatrick2020}).
In our setting, levels of intervention may have been relatively inelastic given legal, financial, and time constraints.

Notably, workers also did not increase their rate of investigation for children with high algorithm risk scores. Figure \ref{fig:screeninbyscore} shows that the relationship between screen-in rates and risk scores is essentially flat in the control group and changes little with the introduction of the tool.\footnote{An earlier draft of this paper showed a small-in-magnitude, marginally significant ($p=0.098$) reduction in the screen-in by risk score gradient, when workers had access to the algorithm relative to the control group. This was due to the inclusion of $n=544$ children in our sample (those sent to a residual, overflow meeting group) where workers did not access the tool. Including this set of children introduced a misleading statistical anomaly, and our current sample excludes them. We have confirmed with text data from discussion notes that workers in these overflow groups did not consult the tool.}  Although this may appear counterintuitive -- especially because the algorithm risk score is strongly predictive of downstream CPS involvement and child harm -- there are several reasons why a flat gradient can be consistent with both (i) workers making informed, non-random screening decisions and (ii) the tool providing meaningful predictive information. 
First, screening decisions are strongly influenced by the caller's present concerns, including potential legal requirements for following up on certain concerns, which are often more determinative of an investigation than family history in isolation. Second, many high-scoring children already have ongoing cases and many have already recently been investigated (e.g., 30 percent of children with the highest algorithm risk score were investigated in the year prior to their referral), potentially reducing the marginal value of a new investigation for them. 
As a result, it is possible for the algorithm score to be informative about a child's history and improve the allocation of investigations without necessarily increasing the slope. (In Section \ref{sec:counterfactuals}, we consider a counterfactual scenario where investigations are explicitly allocated to high-scoring children.) Finally, Figure \ref{fig:cpslevel_score} plots estimates from regressing other types of CPS interventions on algorithm access interacted with algorithm risk score. The tool did not increase the likelihood of other interventions for children with high algorithm scores. The only change is that workers appear to open more (fewer) prevention cases to children with low (high) algorithm scores, but these changes cannot explain the reduction in child harm.

With algorithm support, however, workers \textit{did} reallocate their investigations to children at higher risk of harm along other (non-algorithm-score) dimensions.
Randomization allows us to infer improvements in decision-making by comparing realized outcomes for screened-out children in the intervention and control groups. Intuitively, levels of potential harm are equal in the algorithm and control groups prior to intervention, so a reduction in child harm for screened-out children implies a reallocation of screen-ins toward riskier children. This aligns with a CPS goal of determining which children are at low risk of harm and do not need to be visited.
This analysis is our first and primary targeting test, reported in Column 1 of Table \ref{tab:targeting}. 

The test in Column 1 is our best measure of targeting because it uses realized child harm, whether or not such harm could be predicted by the researchers. It captures the full measurable extent (given our main outcome) of the improvement in allocating investigations.
When workers had access to the algorithm, the eventual realized harm index of children who were screened out on their initial referral was 0.06 standard deviations lower than when workers did not have the algorithm. 
Because CPS did not intervene for these children, differences in post-referral hospitalization outcomes reflect a difference in targeting alone.\footnote{Investigations are the primary institutional and measurable margin of intervention, though targeting of lower-level services could also have improved.} Workers therefore must have improved their assessment of which families were low risk and accordingly screened them out. Results are similar, with loss of precision, when restricting to children never screened in during the entire trial. This targeting improvement is robust and marginally more significant after controlling for all the observables in our balance test, including harm prior to the referral and the number of prior hospital admissions. Algorithm support thus reallocated investigations away from lower-risk children, as measured by ex-post hospitalization. Figure \ref{fig:MarginalHumanMLindexpost}, discussed later in Section \ref{sec:counterfactualmarginal}, provides visual detail for this result.

\begin{table}[H]
    \centering
    \fontsize{11pt}{11pt}\selectfont
	\caption{Algorithm Tool Improved Allocation of Screen-Ins to Children \vspace{0.25em} \\
	{\small \textit{Four Targeting Tests}}}
	\label{tab:targeting}
	\vspace{-.25em}
    {
\def\sym#1{\ifmmode^{#1}\else\(^{#1}\)\fi}
\begin{tabular}{l*{4}{c}}
\toprule
                    &\multicolumn{1}{c}{(1)}   &\multicolumn{1}{c}{(2)}   &\multicolumn{1}{c}{(3)}   &\multicolumn{1}{c}{(4)}   \\
                    &\shortstack{Harm Index,\\ \textit{Screened-Out} \\ \textit{Children}}   &\shortstack{CPS Found Injury,\\ \textit{Screened-In} \\ \textit{Children}}   &\shortstack{Prior Harm Index,\\ \textit{Top Quartile} \\ \textit{Algo Risk Score}}   &\shortstack{Predicted\\Harm\\Index}   \\
\midrule
Algorithm Available &      -0.060*  &       0.051** &      -0.123   &      -0.009   \\
                    &     (0.033)   &     (0.021)   &     (0.094)   &     (0.021)   \\
\addlinespace
Screen-In           &               &               &      -0.240** &      -0.041   \\
                    &               &               &     (0.096)   &     (0.026)   \\
\addlinespace
Algorithm Available $\cdot$ Screen-In&               &               &       0.258** &       0.068*  \\
                    &               &               &     (0.131)   &     (0.036)   \\
\addlinespace
Randomization Controls &         Yes   &         Yes   &         Yes   &         Yes   \\
\midrule
Control Mean        &       0.001   &       0.060   &       0.146   &       0.001   \\
Observations        &       3,266   &       1,065   &       1,320   &       4,662   \\
\bottomrule
\end{tabular}
}

\end{table}
\vspace{-1em}
{\footnotesize\setlength{\parindent}{0pt}\begin{spacing}{1.0} Notes: This table reports four tests of whether the algorithm improved the targeting of investigations to children in the trial. Column 1 estimates the impact of tool availability on the realized harm index for children who CPS screened out during the child's first referral in the trial. Column 2 estimates the effects of tool availability on the likelihood of CPS discovering a child injury, conditional on CPS screening the child in for investigation. Column 3 shows changes in targeting based on children's harm index in the 180 days prior to their initial referral in the trial. This analysis is applied only to a subset of children comprising the top quartile of algorithm risk (score of 16-20). In Column 4, we regress a predicted harm index measure on an interacted model of screen-in and tool availability. Predicted harm is constructed using the leave-one-out procedure of \cite{abadie2018} and predicts a child's future harm index using features determined prior to the referral: family history prior to the referral, hospitalizations prior to the referral, demographics at the time of the referral, and basic referral characteristics. Randomization controls are included throughout. Standard errors are clustered at the household level. Significance reported as: * p$<$0.1, ** p$<$0.05, *** p$<$0.01. We report in Figures \ref{fig:cpslevel} and \ref{fig:screeninbyscore} that overall screen-in levels did not change. \end{spacing}}
\vspace{1em}

In columns 2-4 of  Table \ref{tab:targeting}, we provide further evidence that access to an algorithm improved worker targeting of screen-ins to children in need. First, investigators were more likely to find injuries on screened-in children (Column 2). 
CPS-discovered injuries sustained prior to investigation increased by five percentage points (85 percent), a fifth of which were abrasions, black eyes, bone fractures, bruises, burns, cuts, scratches, and skull fractures. This variable was documented by CPS workers, rather than hospitals, providing an independent outcome margin measured for all screened-in children. The changes appear to reflect differences in the types of children screened in rather than downstream investigator behavior, as investigators were not informed of the algorithm score during their standard investigation protocol. 
Second, the algorithm tool helped workers target screen-ins to children who experienced more maltreatment-suggestive hospitalizations \textit{prior} to the screen-in decision, despite teams not directly observing child medical history (Column 3). This result applies to the set of children in the highest quartile of predicted risk (scoring 16-20), which includes a significant majority of children with prior hospitalization. For children in the bottom three algorithm score quartiles, harm levels in the 180 days prior to the referral are too low to detect any effect. 
Finally, column 4 shows that workers increased screen-ins to children with a greater predicted harm index. Predicted harm is estimated using a rich set of characteristics described in the footnote of Table \ref{tab:targeting} and constructed using the leave-one-out sample procedure described in  \cite{abadie2018} to prevent overfitting predictions in the control group. 
Children screened in using the algorithmic tool had a tenth of a standard deviation (0.07 SD) greater predicted harm index than children screened in without algorithm support.

To summarize, columns 2-4 of Table \ref{tab:targeting} establish that, with algorithm support, workers became more likely to investigate families where children had recent injuries, experienced more prior harm, and had higher predicted risk of future harm. Although targeting of screen-ins could theoretically improve without significant effects on any of these three margins (targeting to children at higher risk that is unobservable on those dimensions, for instance children not yet bearing injuries, children not recently hospitalized, and margins of future harm that our simple prediction model does not capture), these findings further bolster the evidence that algorithmic information improved CPS allocation of screen-ins. 

It may seem surprising that workers better allocated investigations based on potential child harm, despite not increasing investigations for children with high algorithm risk scores. These facts permit a number of hypotheses, and rule out others (Table \ref{table_theories}). 
The hypothesis most consistent with our findings is that workers leveraged expertise that was complementary to the algorithm, which we explore in greater detail in Section \ref{sec:text}. Reminiscent of \cite{bhatt2024}, social workers in our setting may have allocated screen-ins partly based on expected treatment effects (proverbially $\hat{\beta}$) rather than expected outcome levels ($\hat{Y}(0)$). In other words, not all children with high algorithm scores necessarily benefit from an additional CPS investigation, and vice versa.

Finally, the magnitudes for the improvement in screen-in targeting reported in columns 1 and 4 of Table \ref{tab:targeting} (0.06 SD) are large enough to potentially explain the entire reduction in child harm with algorithm support (-0.05 SD). Enhanced targeting of screen-ins could have also led to better targeting of subsequent CPS actions, though we are underpowered to detect this. Targeting of other concurrent interventions (e.g., law enforcement, community services) could also have improved, but screen-ins are the primary margin of CPS attention. 
In either event, child outcomes improved with algorithm support because workers made better assessments of child risk, as evidenced by better screening decisions. We provide additional evidence of this in Section \ref{sec:targeting_gains}.

\subsection{Improved Targeting Reduced Child Harm and  Disparities} 
\label{sec:targeting_gains}

\begin{figure}[H]
      \centering
	  \caption{Groups with Most Improved Targeting had Greatest Reductions in Child Harm \vspace{0.25em} \\
	{\small \textit{Improvements in Outcomes are Tied to Workers' Screen-In Allocation Decisions}}}
	  \label{fig:harm_target_by_group}
	  \vspace{-1em}
      \makebox[\textwidth][c]{\includegraphics[width=.90\textwidth]{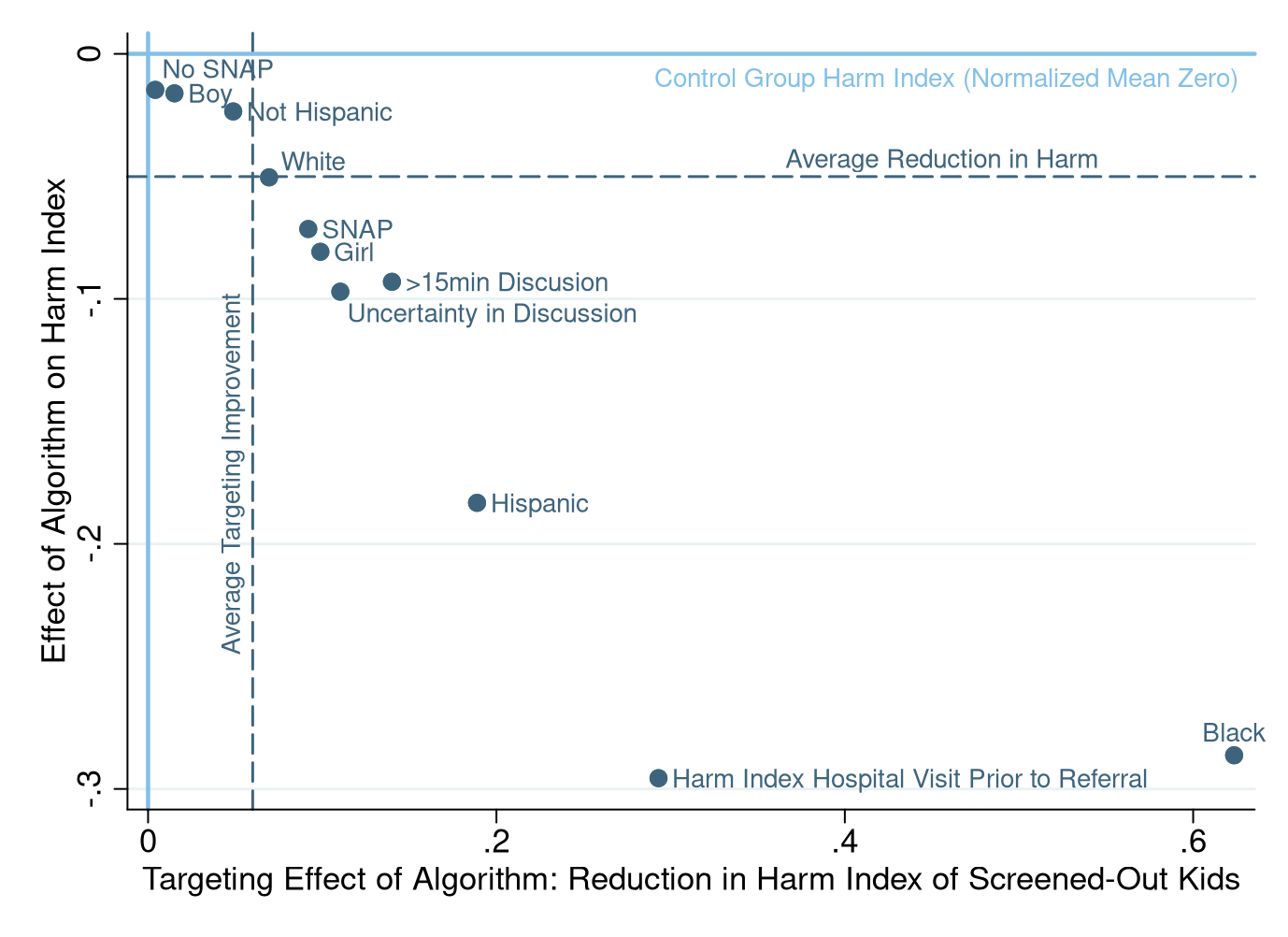}}
	\end{figure}

We find that children in demographic groups that experienced the greatest targeting improvements with the algorithm also experienced the greatest reductions in measured harm (Figure \ref{fig:harm_target_by_group}).  Figure \ref{fig:harm_target_by_group} reports the effect of providing algorithm support on both the harm index (vertical axis) and targeting of screen-ins (horizontal axis; measured as harm conditional on screen-out) for a number of salient groups. 
Providing the tool led to greater improvements in targeting for girls relative to boys, for Hispanic children relative to non-Hispanic children, for Black children relative to white children, and for SNAP recipients relative to non-recipients. The larger targeting improvement for these groups corresponds to greater reductions in child harm, and ultimately a reduction in child harm disparities. These findings more broadly suggest that reductions in harm are attributable at least in part to improvements in targeting of screen-ins.

These findings are not mechanical. Screening out children with lower potential for harm (horizontal axis) implies an offsetting increase in potential harm among screened-in children. 
Since families were randomly assigned to treatment, the average potential for harm among control and treated children was the same by design. 
An overall reduction in child harm (vertical axis) therefore implies that CPS intervention improved child wellbeing more when workers screened in children in greater danger. Better targeting of CPS interventions appears to have reduced child harm and harm disparities. 

To further probe that improved screen-in targeting is a central mechanism for improvements in child outcomes, we conduct an additional test inspired by Figure \ref{fig:harm_target_by_group}. We randomly draw a thousand subsamples of varying size, mimicking sample size heterogeneity in the prior figure, and imposing a minimum sample size of $n=100$. For each random subsample, we then calculate the effect of the algorithm on the overall harm index and on targeting of screen-ins (the respective vertical and horizontal axes of Figure \ref{fig:harm_target_by_group}). Results are presented in Appendix Figure \ref{fig:harm_target_by_random_group}. 
On average, a 1 SD improvement in targeting for screen-outs (corresponding to roughly a 2 SD improvement in targeting for screen-ins, given the 2:1 screen-out to screen-in ratio) is associated with a reduction in overall child harm by 0.72 SD ($p<0.001$). 
The groups that benefit most from the algorithmic tool are the ones where targeting of screen-ins improves the most. This result is generalizable and not specific to the groups we report in Figure \ref{fig:harm_target_by_group}. 
Consistent with novel ongoing work by \cite{lacey2024}, we find that screening in a family for investigation and possible services can improve child health and wellbeing.  

\vspace{-1em}
{\footnotesize\setlength{\parindent}{0pt}\begin{spacing}{1.0}Notes: This figure shows that groups that experienced the greatest improvements in the targeting of screen-ins also experienced the greatest reductions in overall child harm. The thick, light blue lines are normalized control group outcomes in the absence of any improvement from algorithm support. In contrast, the dashed dark blue lines represent average realized effects of providing the algorithmic tool on the harm index (vertical axis) and targeting (horizontal axis).  Each dot represents the average effects of providing the algorithmic tool on harm and targeting for each group, respectively. The horizontal axis measures targeting by leveraging the difference in harm after the referral among screened-out children with the tool compared to screened-out children without the tool (relying on random assignment to treatment and control, which implies that harm would have been the same without algorithm-induced improvements in allocation). This corresponds to the targeting test reported in Column 1 of Table \ref{tab:targeting}.  \end{spacing}}
\vspace{1em}

\subsection{Worker Discussion Text: Changes in Worker Attention}
\label{sec:text}

How were workers able to better target screen-ins? 
In this section we provide suggestive evidence of how access to algorithmic information changed workers' decision-making process. To do so, we leverage rich, unstructured administrative text data from the partnered child welfare agency, taken from discussion notes written during workers' decision-making process (see examples in Table \ref{table_reasons}). Discussion notes vary considerably in detail and quality, but may still provide valuable insight into whether discussions changed. 

Combining these data with recent advances using word embeddings from sentence transformers, we assessed in a principled way whether workers' discussions changed when they had access to the tool. Specifically, we trained a BERTopic model \citep{grootendorst2022} on discussion notes from approximately 11,000 referrals. We then used the generated topics ($t=131$) to predict whether workers had access to the algorithm tool in our trial sample, using three-fold cross-validation. Even with a drastically reduced dimensionality of the text data to 131 topics, the model provides clear evidence that discussions changed when workers had access to the algorithm, as shown in Figure \ref{fig:BERT} ($AUC\approx0.55$).  A linear regression of treatment status on the selected topics gives an F-statistic of $F=77, p<0.0001$ (not reported).

Topics types found by the BERTopic model are not designed to predict what is changing in the discussions and still suffer from lack of explainability. For those reasons, we grouped together categories of words in light of the ways in which the algorithmic tool could modify attention. 
The algorithmic tool was based on data from household members' past history and listed a score for each child on the report. With algorithm support, workers might make fewer mistakes in aggregating child or family history, better understand family history, better contextualize the incoming allegation report, or have a broader household focus that considers the circumstances of all siblings and alleged perpetrators. Each of these interrelated potential mechanisms could improve worker assessment of risk.

 We find that access to the algorithm increased the salience of information that was complementary to algorithm features. Workers paid more attention to all children on the referral and which caregiver had custody, across the entire distribution of algorithm scores (Figure \ref{fig:discussioncontent} Panel I). As reported in Appendix Table \ref{tab:text}, providing the algorithmic tool significantly increased the likelihood of mentioning such proxies of family structure by 4.1 percentage points or 7 percent ($p=0.059$).

We also hypothesized that the algorithm's use of child and family history inputs -- and workers' access to new information from the incoming report not available to the algorithm -- could steer workers to further concentrate on changes in a child's circumstances over time. We find that workers became more likely to use time-related language such as ``new'', ``prior'', and ``immediate'' for children in the top three quintiles of the algorithm score, who have accumulated CPS history (Figure \ref{fig:discussioncontent} Panel II). Proxies for attention to time include words related to the urgency of the call, to family history, and distinguishing between present versus historical information. Such attention to time is about understanding current events in light of families' history. 
 Overall, discussion of time-related proxies increased by 4.3 percentage points or 6.5 percent ($p=0.039$, Appendix Table \ref{tab:text}).
Discussions seemed to have refocused toward referral elements that algorithm support made more salient.

\begin{figure}[H]
	\caption{Access to the Algorithm Changed Worker Discussion Content}
	    \label{fig:discussioncontent}
	\begin{subfigure}[b]{0.5\textwidth}
		\caption{Attention to Family Structure}
		\includegraphics[trim=0 25 0 0, clip,width=\textwidth]{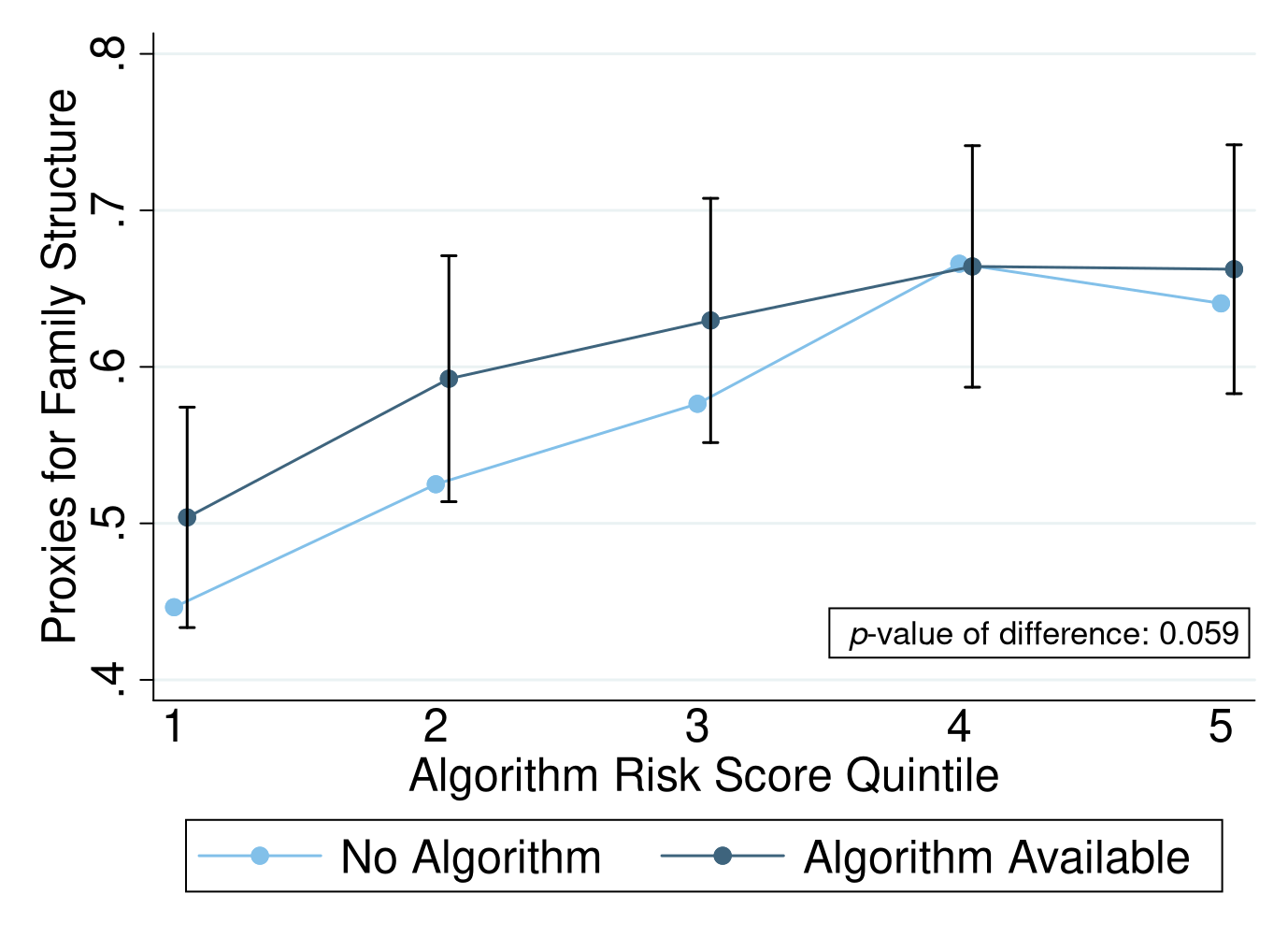}
		\label{fig:mentionfamily}
	\end{subfigure}
	\begin{subfigure}[b]{0.5\textwidth}
		\caption{Attention to Time}
		\includegraphics[trim=0 25 0 0, clip,scale=.2,width=\textwidth]{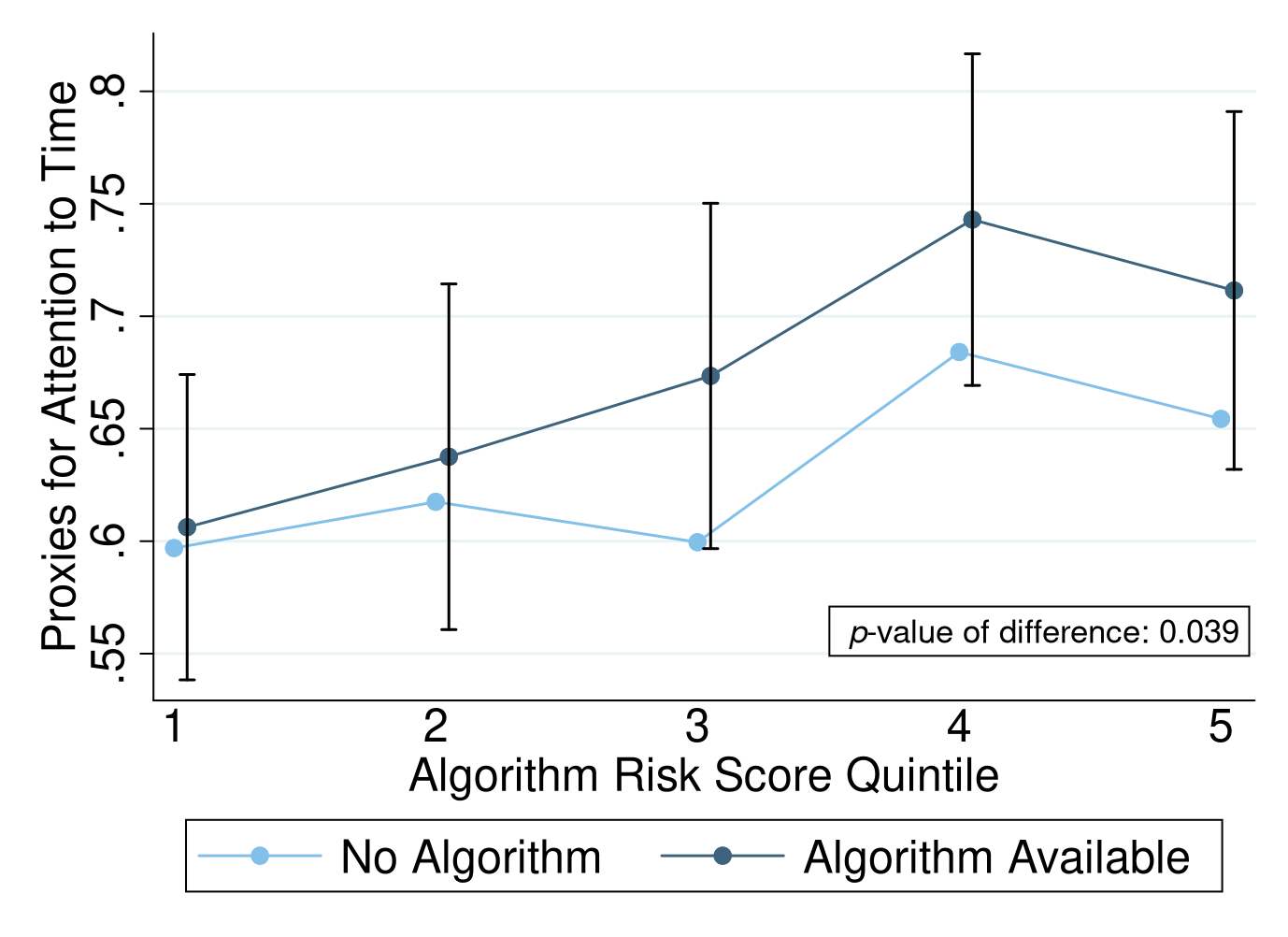}
		\label{fig:mentiontime}
	\end{subfigure}
\end{figure}
\vspace{-1.6em}
{\footnotesize\setlength{\parindent}{0pt}\begin{spacing}{1.0}Notes: This figure illustrates changes in worker discussion content toward topics related to the design of the algorithm. Children are grouped by algorithm score quintiles on the horizontal axis. The vertical axis reports probabilities of whether workers mentioned a proxy for family structure (Panel I) or a proxy for time (Panel II). Proxies for family structure include mentioning multiple children (terms: sibling, other child, children, kids, sib, brother, sister, older child, younger child) and discussing custody (terms:  custody, co parenting, co-parenting). Proxies for attention to time include referring to the urgency of the call (terms: immediate, emergency, imminent, urgen), mentioning the past (terms: last, hx [history], [name of state record-keeping software], past, prior, previous, histor, cw [past child welfare], dhs [past department of human services], same concern), and referring to a new situation (terms: new, current, now, sober). 90\% confidence intervals are reported and randomization controls are included throughout.  \end{spacing}}
\vspace{1em}

\subsection{Human-plus-Algorithm Gains under Complexity}
\label{sec:complexity}

\cite{ehrmann2022} view cognitive load as a critical determinant of whether algorithmic tools are complementary to human expertise. 
Complicated interfaces or unnecessary alerts from algorithmic decision supports could contribute to cognitive overload. In contrast, a machine learning tool designed in collaboration with users could reduce what cognitive theory calls extraneous (modifiable) cognitive load \citep{sweller2011}. In our setting, having an algorithm aggregate vast, disorganized historical data could allow workers to reallocate effort to other inherently difficult parts of their decision task (intrinsic cognitive load). 

According to cognitive load theory, 
one would expect greater improvements in welfare when cases are complex and where workers are more likely to benefit from the algorithm making family history more digestible. 
In Table \ref{tab:het_complex}, we consider heterogeneity in algorithm benefits under three distinct indicators of complexity: (1) discussion notes that mention uncertainty about a child's situation or how to address it, containing expressions such as ``?'', ``unknown'', ``unclear'', and ``uncertain'' (column 1), (2) decisions that take longer than 15 minutes, corresponding to the top quartile of decision length (column 2), and (3) children with a recent prior potential-maltreatment-related hospitalization (column 3). 
Each of the three measures captures a distinct dimension of complexity: the largest correlation coefficient between any two complexity measures is $\rho = 0.16$. Algorithm support did not affect the frequency of uncertainty-related text markers nor the probability of a long discussion, either in levels or when interacted by score, and the third proxy for complexity was exogenous to tool availability. 

\begin{table}[H]
    \centering
    \fontsize{10pt}{10pt}\selectfont
	\caption{Algorithm Benefits were Largest for Complex Referrals
 \vspace{0.25em} \\
	{\footnotesize \textit{Child Harm after Randomization, by Type of Complexity}}}
	\label{tab:het_complex}
	\vspace{-0.25em}
    {
\def\sym#1{\ifmmode^{#1}\else\(^{#1}\)\fi}
\begin{tabular}{l*{3}{c}}
\toprule
                    &\multicolumn{3}{c}{\shortstack{Outcome: Harm Index\\ Type of Complexity: Column Titles}}\\\cmidrule(lr){2-4}
                    &\multicolumn{1}{c}{(1)}   &\multicolumn{1}{c}{(2)}   &\multicolumn{1}{c}{(3)}   \\
                    &\shortstack{Text Marker\\ Uncertainty}   &\shortstack{Long Discussion\\ $\geq 15$ min}   &\shortstack{Any Index Hospital Visit,\\ Previous 180 Days}   \\
\midrule
Algorithm Available &       0.001   &      -0.008   &      -0.001   \\
                    &     (0.025)   &     (0.024)   &     (0.016)   \\
\addlinespace
Type                &       0.088** &       0.127***&       0.552***\\
                    &     (0.045)   &     (0.045)   &     (0.119)   \\
\addlinespace
Algorithm Available $\cdot$ Type&      -0.097** &      -0.098*  &      -0.314** \\
                    &     (0.049)   &     (0.053)   &     (0.125)   \\
\addlinespace
Randomization Controls &         Yes   &         Yes   &         Yes   \\
\midrule
Frequency of Type   &       0.519   &       0.266   &       0.157   \\
Observations        &       4,681   &       3,754   &       4,681   \\
\bottomrule
\end{tabular}
}

\end{table}
\vspace{-1em}
{\footnotesize\setlength{\parindent}{0pt}\begin{spacing}{1.0} Notes: This table presents impacts of the algorithmic tool on child harm across three measures of decision complexity. The outcome in each regression is the standardized harm index of potential maltreatment-related hospitalizations occurring at least 30 days after randomization. Column 1 reports effects of the algorithm by any markers of worker uncertainty in the decision discussion notes (any instance of ``?'', ``unknown'', ``unclear'', ``uncertain'', or a related word). Column 2 reports effects of the algorithm by whether workers discussed the referral for at least 15 minutes (top quartile of discussion time), estimated on the subset of referrals with well-measured timestamps. Column 3 reports effects of the algorithm by whether a child had a recent potential-maltreatment-related hospital visit (high priority visit, injury, avoidable ER, maltreatment ICD) in the 180 days prior to the referral. The means at the bottom of the table show the frequency of each of these three binary measures of complexity. Randomization controls are included throughout. Standard errors are clustered at the household level. Significance reported as: * p$<$0.1, ** p$<$0.05, *** p$<$0.01.\end{spacing}}
\vspace{1em}

Consistent with theory, gains from algorithm support were greatest among complex referrals.
Across all three complexity measures, reductions in harm from the algorithmic tool were almost exclusively concentrated among complex referrals (Table \ref{tab:het_complex}, see rows ``Algorithm Available $\cdot$ Type'' and ``Algorithm Available''). The findings also hold when controlling for algorithm risk score.
Although each of the three complexity measures are correlated with above-average levels of harm in the control group (Table \ref{tab:het_complex}, row ``Type''), these findings are consistent with recent work showing that algorithm support is particularly beneficial to individuals for whom a task is complex \citep{brynjolfsson2023}.

Did the algorithmic tool's informational content (risk score) matter, or did its presence merely serve as a behavioral reminder or nudge to workers? Although difficult to answer definitively, the fact that workers regularly checked and recorded the algorithm score -- or lack thereof -- in both the intervention and control groups (Appendix Figure \ref{tool_workflow}), together with score-based heterogeneity in discussion content and decisions (Figures \ref{fig:screeninrace} and \ref{fig:discussioncontent}) suggest that benefits of the algorithm are at least partially attributable to the process of reviewing the predicted risk scores. Although the specific  1-to-20 value of the algorithm's risk score likely helped workers allocate their attention more effectively, we cannot definitively rule out benefits from a general increase in the salience of family history -- or other characteristics -- upon seeing a listed score, regardless of its magnitude. An important topic for future research is understanding the impacts of how information is presented to users in similar settings.

\section{Human-plus-Algorithm vs. Algorithm-Only Counterfactuals}
\label{sec:counterfactuals}

To what extent do human decision-makers contribute their independent expertise to improve upon algorithms? 
As machine learning decision supports become more widespread, so do questions about human experts' contributions to decisions, and when to automate. 
In this study, outcomes improved when workers were given access to an algorithm. Workers, however, did not target investigations toward children with higher algorithm risk scores, but instead used complementary information to target more effectively on other dimensions.
Given how well the algorithm risk score predicts relevant outcomes (Appendix Figure \ref{fig:toolvalidation}), it is reasonable to ask: Could outcomes have improved \textit{even more} if workers had based decisions more closely on the algorithm risk score? And would there have been implications for equity?

In this section, we conduct two innovative counterfactual exercises to estimate the impact of deferring to an algorithm to make decisions. 
We assess whether child harm and equity would have improved under (i) investigation decisions determined solely by algorithm predictions, and (ii) marginal increased reliance on algorithm predictions. This methodology can be applied in other experimental settings to assess whether users underutilize algorithmic information. Parts of the methodology can also be applied to study hypothetical, alternative decision rules beyond the implemented algorithm.

\subsection{Counterfactual 1: Deferring Entirely to an Algorithm}
\label{sec:algoonly}

The first counterfactual exercise estimates the impact of fully deferring screening decisions to the algorithm using a bounding approach.
The experiment did not include an algorithm-only treatment arm due to legal constraints: the statutory requirements for a screen-in do not always correspond to high algorithm-predicted risk. 
By design, the algorithm provided a prediction of child risk, rather than an explicit recommendation of whether to screen in. For the counterfactual, we assume the algorithm would prioritize screen-ins for children with the highest predicted risk. The counterfactual exercise operates as follows: 
\begin{enumerate}[itemsep=0pt, parsep=0pt]
    \item Sort children by the algorithm's predicted risk score.
    \item  Holding screen-in rates constant with the other treatment arms at 30 percent, screen in the riskiest 30 percent of children as predicted by the algorithm.
    \item Estimate child outcomes conditional on the new screening decisions.
\end{enumerate}
\vspace{-0.5em}

\vspace{2mm}

To estimate counterfactual outcomes under this algorithm-based decision rule, we classify children from the control group in a 2-by-2 grid defined by screen-in decisions made by humans (observed) and by hypothetical algorithm recommendations (inferred above with minimal assumptions). The four states are as follows: humans and algorithm each decide to screen in, humans and algorithm each screen out, humans screen in but the algorithm would screen out, and humans screened out but the algorithm would screen in. The grid is presented in Table \ref{tab:counterfactualAIonly}. We directly observe child health outcomes when both parties, humans and algorithm, agree on the screen-in decision (both screen in, or both screen out). Second, for children screened in by human decision-makers but for whom the algorithm would have screened out, we make an assumption in favor of the algorithm. We assume that child health would not have been any worse if these children had been screened out, allowing us to impute the counterfactual outcome using observed child harm. This gives a best-case bound of outcomes if screening decisions were deferred to the algorithm. 

Finally, we must estimate outcomes for children observed to be screened out by humans, but in the counterfactual screened in by the algorithm. We allow counterfactual outcomes for this cell to vary over a range of values for how much a screen-in reduces child harm, assuming homogeneous impacts of screen-in and a floor value for child harm. 
The counterfactual outcome for children screened out by humans and screened in by the algorithm is calculated as: $max(y_i-R, \underline{H})$, where $y_i$ is the realized outcome for child $i$, $R$ is the hypothetical effect of a screen-in, and $\underline{H}$ is the minimum possible value of standardized child harm (zero visits for all hospitalization types included in the harm index). Though developed independently, this methodology shares similarities with \cite{ben-michael2024} who also propose a bounding approach.

\begin{figure}[H]
      \centering
	  \caption{Child Harm under Counterfactual Algorithm-Only Decisions}
	  \label{counterfactual_boundAI}
	  \vspace{0em}
      \makebox[\textwidth][c]{\includegraphics[width=0.80\textwidth, trim=10 20 0 10, clip]{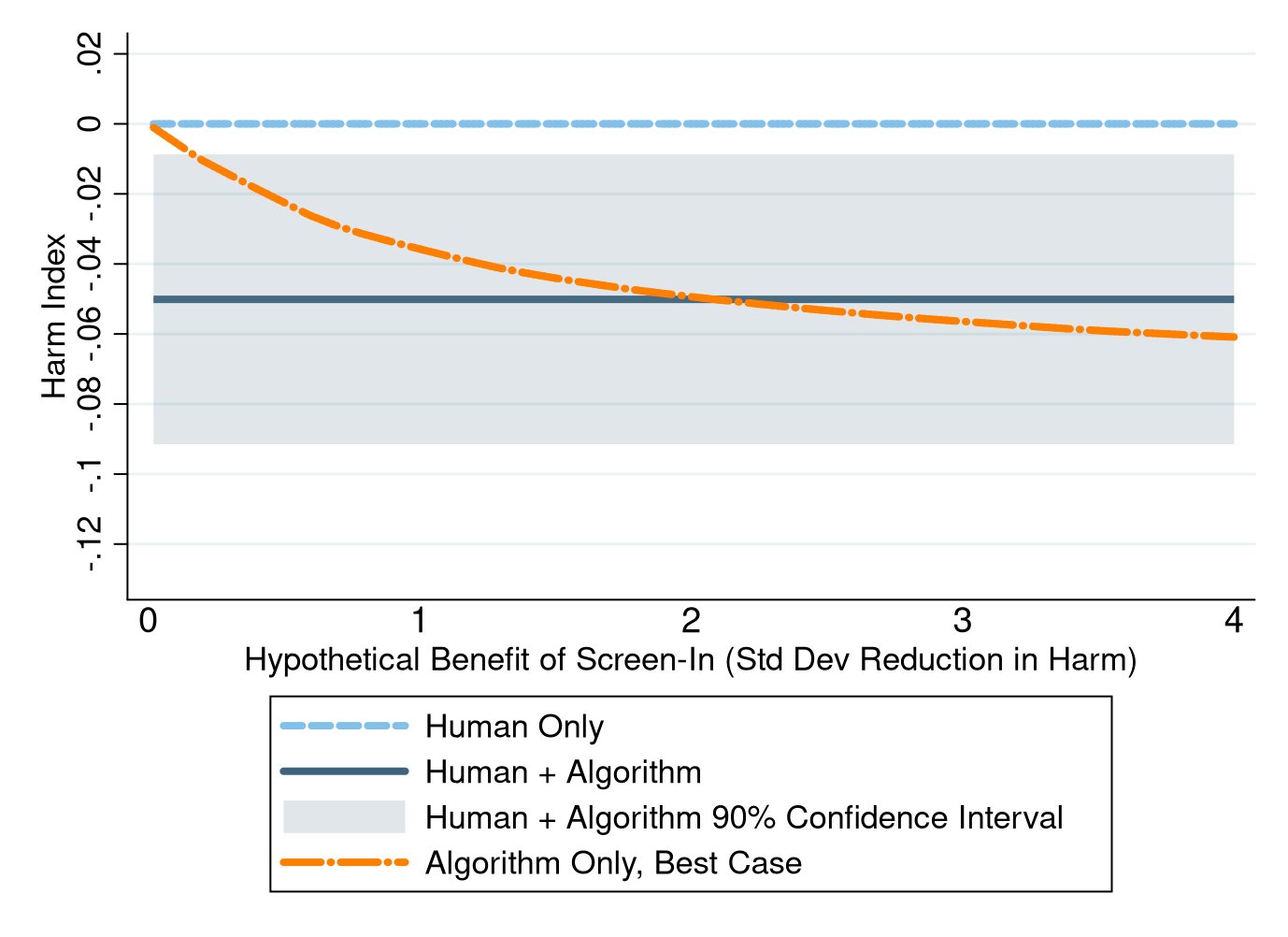}}
	\end{figure}
\vspace{-1em}
{\footnotesize\setlength{\parindent}{0pt}\begin{spacing}{1.0}Notes: This figure shows average child harm under three scenarios. The first scenario is a benchmark control group where humans do not have access to the algorithm (light blue short-dashed line).
The second scenario is when humans have access to the algorithm, which is the experiment's intervention group, shown with a 90\% confidence interval (solid dark blue line with shaded confidence bands; main estimates presented in Table \ref{SERVER_hospitaloutcomes1}). Finally, the figure shows a best-case scenario for the algorithm under a range of assumptions about the benefits of screen-in (orange dash-dot line). The best-case scenario for the algorithm assumes that the children who were screened in by humans, but for whom the algorithm would have screened out, would have been just as well off under the algorithm decision.
Conversely, estimated benefits of algorithm screen-ins for children who were screened out by humans depend on the effects of a screen-in on child harm (range of hypothetical screen-in benefits, in terms of standard deviation reductions in harm, shown on horizontal axis). \end{spacing}}
\vspace{1em} 

Results of this counterfactual exercise are plotted in Figure \ref{counterfactual_boundAI}. We compare child harm under a best case for our algorithm-only scenario (dashed orange line) with outcomes observed under human-only (dashed light blue line) and human-plus-algorithm decision-making (solid blue line with confidence band). The horizontal axis varies the hypothetical effect of a screen-in for children that humans screened out but the algorithm would have screened in. Given the favorable bounding assumptions, the algorithm-only counterfactual mechanically outperforms the human-only scenario. We find that child outcomes under a human-plus-algorithm scenario are better than, though statistically indistinguishable from, this favorable algorithm-only rule under realistic screen-in treatment effects ($<1$ SD), and are also comparable under larger hypothetical treatment effects.\footnote{To benchmark these magnitudes, the difference in harm between children placed in foster care and those who are not is 0.5 standard deviations of the harm index.} This result highlights the limits to fully automating complex social decisions with an algorithm, and provides evidence in favor of human-algorithm complementarity in our setting.

This methodology allows the researcher to estimate bounded effects of alternative screen-in decision rules.
For example, appendix Figure \ref{bestcase_counterfactual_boundAI} presents outcomes under an ``oracle'' screen-in decision rule where children are ranked by ex-post realized harm. According to this decision rule, the 30 percent of children with the highest ex-post realized harm index are selected to be screened in. This rule is impractical, as potential future harm is inherently uncertain at the time of decision, so the estimate represents a best-case scenario for potential gains in child wellbeing. Appendix Figure \ref{bestcase_counterfactual_boundAI} illustrates that this unrealistic ``oracle'' bound outperforms both humans and the human-plus-algorithm condition if one assumes large benefits from screening in (greater than a 0.5 SD reduction in harm). This finding, however, relies on the strong assumption that the two-thirds of families screened in by workers -- but who would have been screened out by the oracle -- are not adversely affected. Nevertheless, this oracle exercise highlights that algorithms could in principle outperform the intervention arm of our trial, perhaps by leveraging richer data sources or being better optimized for human use \citep{ludwig2024a}.

\subsection{Counterfactual 2: Marginal Increased Reliance on Algorithm}
\label{sec:counterfactualmarginal}

In contrast to the first counterfactual exercise, which deferred decisions entirely to the algorithm, the second counterfactual exercise examines whether social workers should have relied \textit{marginally} more on the algorithm's risk score to make screen-in decisions. 
Increased marginal reliance corresponds to screening in additional children that the algorithm predicted as high risk. A real-life analog of this would be a mandatory screen-in protocol for families scoring above a high-risk threshold as was implemented in Pittsburgh, PA \citep{lacey2024}. 

\begin{figure}[H]
      \centering
	  \caption{Realized Harm for Screened-Out Children, by Algorithm Risk Score \vspace{0.50em} 
    \\ \footnotesize \textit{Would a Default Screen-In Requirement Catch Vulnerable Children that Workers Screened Out?    } \vspace{0.50em}}
	  \label{fig:MarginalHumanMLindexpost}
      \makebox[\textwidth][c]{\includegraphics[width=0.71\textwidth, trim=10 20 0 40, clip]{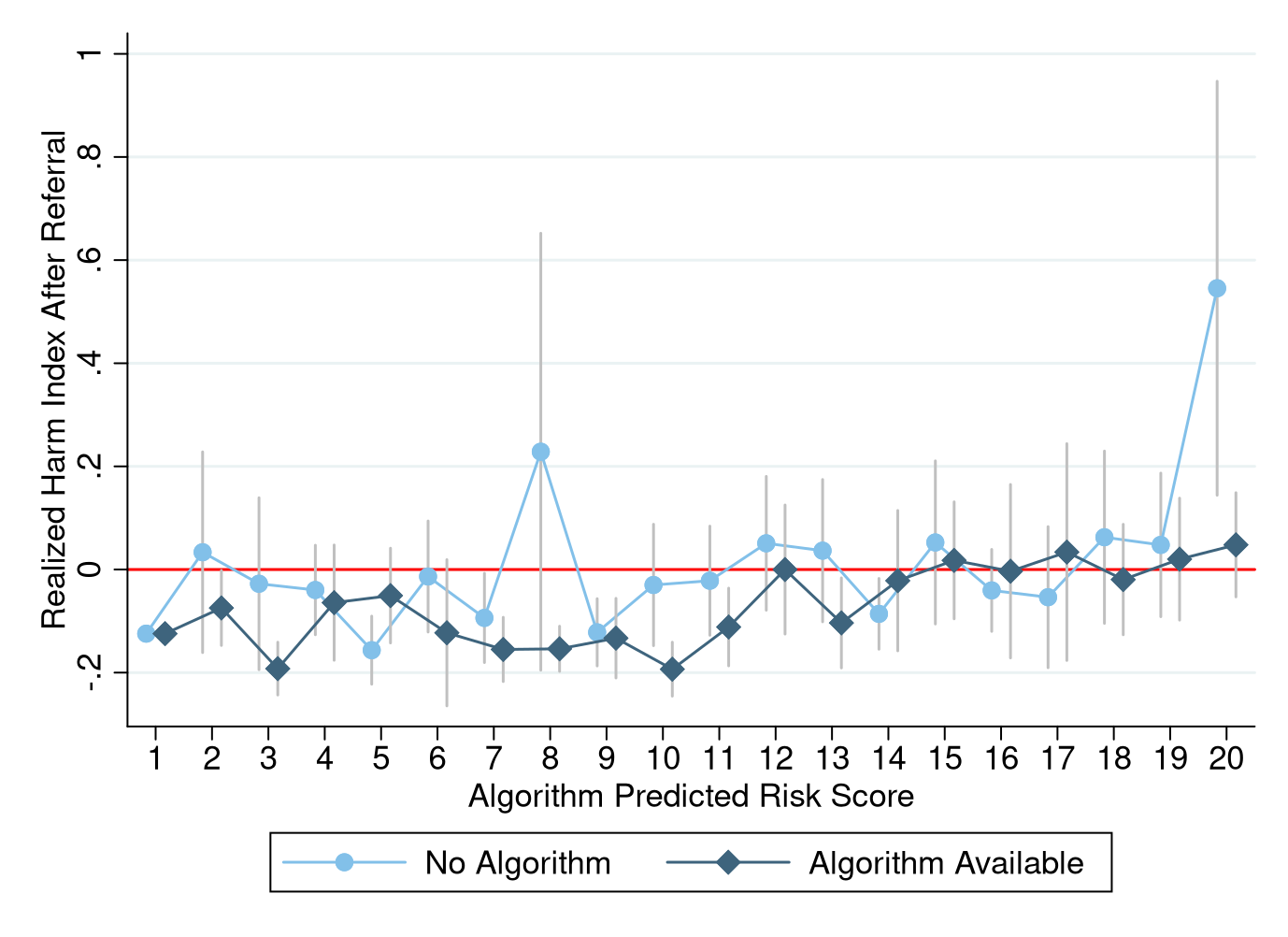}}
	\end{figure}

Figure \ref{fig:MarginalHumanMLindexpost} shows no significant anticipated benefits from deferring marginally more to the algorithm score, relative to observed human-plus-algorithm decisions. 
Among children not investigated by CPS, an algorithm-only decision-maker would first prioritize screening in those with the highest risk scores, on the right side of the figure. In absence of algorithmic decision support for children with a score of 20 who were not investigated, mean harm levels were almost 0.6 SD above the mean (light blue line with circles). Social workers were missing out on a subset of high-scoring children in the control group  who accrued significant harm, such that a mandatory screen-in protocol for these children might have readily improved outcomes. In contrast, when workers had access to algorithmic information (dark blue line with diamonds), screened-out children scoring a 20 only had 0.1 SD greater harm than the control group mean. 
With algorithm access, the reduction in harm for screened-out children on referrals scoring a 20 is statistically significant at the 90 percent confidence level ($\beta=-0.46,p=0.070$). 
Hence, much of the capacity for harm reduction is already realized by social workers using the algorithm freely, such that further gains from a screen-in mandate would be comparatively modest, especially relative to the costs of additional investigations that would result from such a policy.

Figure \ref{fig:MarginalHumanMLindexpost} confirms that although workers did not change their screen-in rates for children with the highest algorithm risk scores when using the tool (Figure \ref{fig:screeninbyscore}), workers did screen in a set of ex-post more vulnerable children. 
Furthermore, in situations where workers ``disagreed'' with the algorithm (high algorithm score, worker screen-out), harm for high-score, screened out children was comparable with the control group harm mean (horizontal red line). 
Therefore, ex-post partial automation of investigation decisions using algorithm predictions alone is likely to be of limited benefit in our setting. Future work with larger study samples could consider partial automation with optimal triage, as proposed by \citet{raghu2019algorithmic}. 

\vspace{-1em}
{\footnotesize\setlength{\parindent}{0pt}\begin{spacing}{1.0}Notes: This figure shows the average realized harm index of children who were screened out, by treatment status and algorithm risk score. The dark blue plot shows the means when the tool is available. The light blue plot shows the means for the control group, where the tool is not available. 90\% confidence intervals reported throughout. The horizontal red line is the average harm in the control group (for all children, both screened in and screened out). 
\end{spacing}}
\vspace{0.5em}

\subsection{Counterfactual 3: Algorithms and Disparities without Human Oversight}

\label{sec:disparitiescounterfactual}
\begin{figure}[H]
		\centering
            \caption{Counterfactual: Deferring to Algorithm would Increase CPS Surveillance Disparities}
		\includegraphics[width =.80\linewidth]{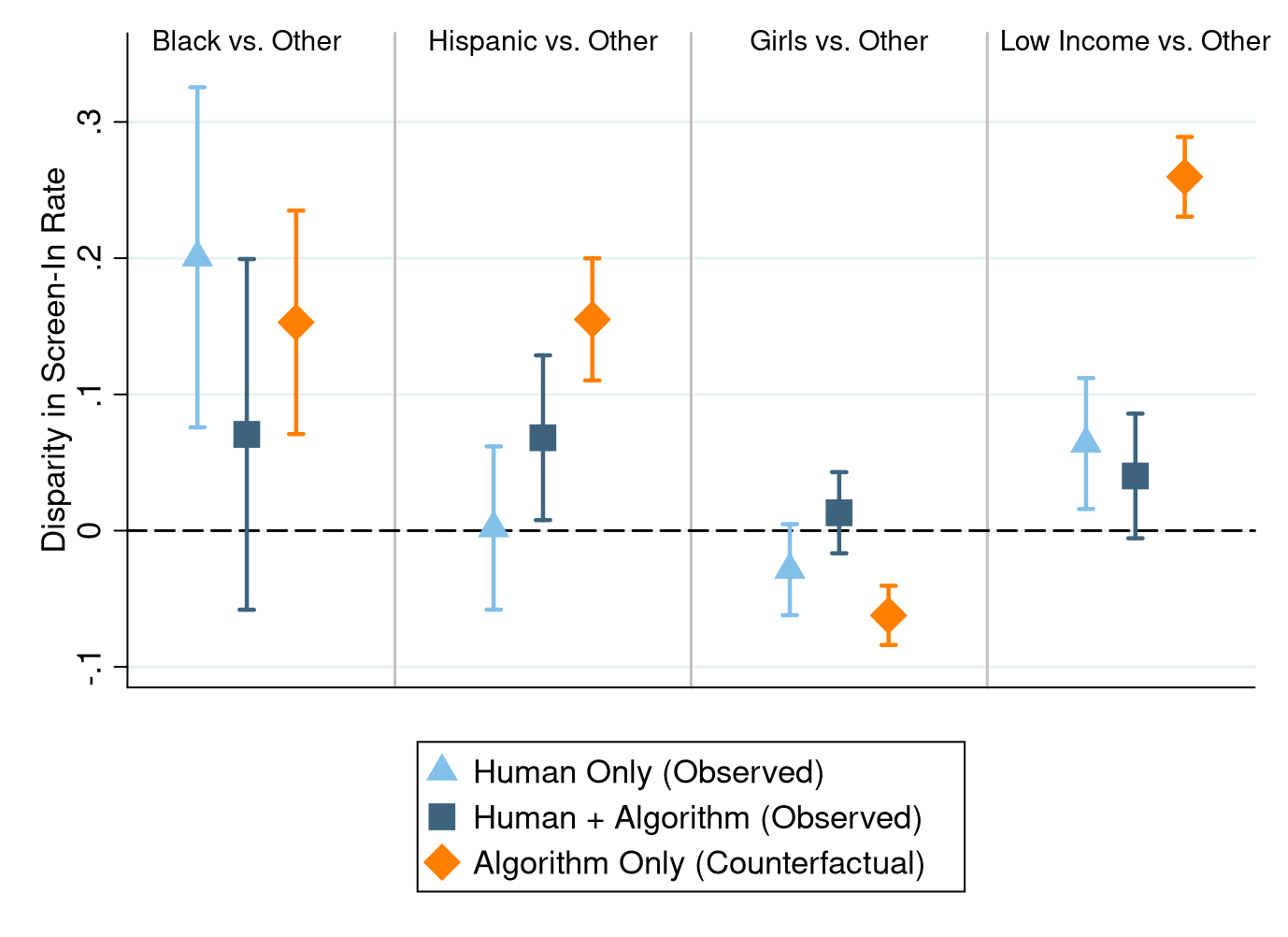}
    \label{counterfactual_disparities}
\end{figure}

Finally, we estimate with minimal assumptions how an algorithm-only decision rule would have impacted screen-in equity. This is possible through the counterfactual algorithm screen-in decision rule described in Section \ref{sec:algoonly}, which only assumes no change in the overall screen-in rate and that the algorithm would have prioritized screen-ins for children with the highest predicted risk. 
Figure \ref{counterfactual_disparities} shows CPS surveillance (screen-in) disparities for relevant demographic subgroups under three decision regimes: human-only, human-plus-algorithm, and algorithm-only.

\vspace{-1em}
{\footnotesize\setlength{\parindent}{0pt}\begin{spacing}{1.0} Notes: This figure shows point estimates and 90 percent confidence intervals for disparities in screen-in rates by demographic category. Human-only and human-plus-algorithm disparities are directly observed in the data. Algorithm-only disparities are inferred with minimal assumptions: control group observations are sorted by algorithm risk score, and mandatory screen-ins are imposed above a score threshold that keeps overall screen-in rates constant. The vertical axis shows the disparity in screen-in rates. A zero point estimate represents zero disparity in screen-ins and is illustrated using a horizontal dashed black line.   For example, in the first panel (``Black vs. Other''), screen-in rates were 20 percentage points greater for Black children compared to other children for decisions where humans did not have access to the algorithm (``Human Only''; light blue triangle on the left). Low income refers to children whose families are receiving SNAP benefits. \end{spacing}}
\vspace{1em}

We find that human-plus-algorithm decisions improved equity not only compared to the status quo of human experts alone, but also compared to counterfactual screen-ins from the algorithm without human oversight. 
Relative to an algorithm-only decision rule, human oversight reduced disparities in CPS investigations by race, ethnicity, gender, and socioeconomic status: the dark blue squares in Figure \ref{counterfactual_disparities} (human with algorithm) are always closer to zero than the orange diamonds (algorithm only), though point estimates are in some cases imprecise.  
Second, the algorithm on its own would have investigated significantly more Hispanic and low-income children relative to the status quo of human experts on their own (light blue triangles). Increased surveillance under the algorithm-only counterfactual is not associated with a greater reduction in harm for these subgroups relative to observed human-plus-algorithm decisions (Appendix Figure \ref{fig:healthoutcomesalgoonly}), suggesting that the algorithm-only disparities may not be warranted. 
In short, the improvements in equity observed during the trial would not have been fully realized -- and for some groups may have worsened -- without human oversight of the algorithm. Providing algorithm support to humans leads to more equitable levels of surveillance by CPS, relative to the algorithm on its own.

Estimating child health disparities under a counterfactual algorithm-only scenario requires additional assumptions about similar effects of screen-ins across subgroups. We present estimates in Appendix Figure \ref{fig:healthdisparityalgoonly}. Human-plus-algorithm decisions appear to be more effective at reducing health disparities than the algorithm on its own, though the differences are not statistically significant. 

We conclude that human oversight of the algorithm for investigation decisions not only both improved efficiency and equity relative to a human-only scenario, but also improved equity relative to deferring all decisions to algorithms. Combining algorithm support with ``in-the-loop'' human expert oversight could be an effective strategy to overcome public skepticism about algorithms and the perpetuation of disparities. 

\section{Generalizability and External Validity}

This section identifies decision environments where our findings are most likely to be informative. We develop a theoretical model that helps rationalize core results, and highlight two contextual features that may have been especially important given prior research. We conclude with a cost-benefit analysis. 

\subsection{Group Decision-Making}

Group-based decision-making is common in high-stakes settings with large amounts of information and professional discretion: for example juries and parole boards, criminal justice and health care expert boards, and hiring committees. Similarly, as required by state legislation, child protection investigation decisions in the partner agency were made in small teams.  
Team decision-making may have contributed to the success of the algorithm tool in our setting. A longstanding literature on the ``wisdom of the crowds'' highlights how groups can often aggregate information more effectively than individuals \citep{galton1907}. The literature on jury size and composition indeed suggests that larger groups reach better and fairer decisions \citep{waters2004,anwar2012}. 
In the specific context of AI support, an emerging literature has found that groups tend to make better \citep{patel2019Human} and fairer decisions than individuals \citep{chiang2023}.

Although we are unable to directly compare individual versus group decision-making, we can use day-to-day variation in group size to provide early empirical evidence on the impact of team size on algorithm benefits. Most teams have between four and six workers, with some outliers. Team size appears relatively balanced on observables, though we caution that team size may not be as good as random in our setting. 
Point estimates for benefits of the algorithm appear to be potentially greater for larger teams, but we cannot statistically distinguish between these estimates (Figure \ref{fig:teamsize_indexpost30_std}). However, consistent with the literature on team size and equity, we do find greater equity gains on some margins when the algorithm was used by larger teams (Figure \ref{fig:teamsize_equity_indexpost30_std}).
Though we cannot detect efficiency gains of increasing team size, our observations of team discussions suggest group social reinforcement (regardless of size) contributed to the algorithm's continued use and benefits, and we find greater equity gains in larger teams.

\subsection{Model of Decisions with Algorithm Support}

\label{sec:model}

We develop a theoretical model to clarify how an algorithm support tool could affect decision-making (Appendix D). In line with state mandates, we model workers as conducting their best possible assessment of whether the noisy signal of child risk they perceive warrants an investigation. Children are harmed when workers underestimate risk, especially when underestimation is sizable. Under simple functional form assumptions (e.g., normally distributed noise), we obtain four key results when the algorithm reduces uncertainty in the perceived signal of risk. Improved signal could occur, for example, by allowing workers to reallocate their time and effort toward higher-order tasks, such as thinking rather than searching for data. The first key result of the model is that algorithm support reduces child harm, consistent with our findings in Table \ref{SERVER_hospitaloutcomes1}. Second, harm is especially reduced for groups with greater variance in risk (Figure \ref{fig:HealthDisparity}). Third, the tool reduces incongruent overestimation of risk (Figure \ref{fig:screeninrace}). Finally, these effects occur because workers make better assessments of child risk (Table \ref{tab:targeting} and Figure \ref{fig:harm_target_by_group}). Details and proofs are found in Appendix D.

Despite its simplicity, this model captures all of our key findings. The model highlights how our results may generalize to settings where workers make informed assessments based on a noisy signal, but where large mistakes in risk assessment are particularly costly. 

\subsection{Design Choices and Human Expertise}

An intentional choice made by the partner agency was that the algorithm's information set was comparatively narrow in our setting, including only history and no explicit information about current concerns. 
Keeping some data features available only to workers was not only more tractable but also helped re-emphasize psychologically the importance of worker diligence and expertise.

These design choices may have contributed to the human-algorithm complementarity we observe. When full automation is legally infeasible as in most social policy settings, designing algorithms with the user in mind can lead to human-algorithm synergies. \citet{vaccaro2024} suggest that settings where humans have more expertise in a task than the algorithm -- and are aware of their area of expertise relative to the algorithm -- are more likely to engender human-algorithm synergy: humans exercise better judgment on when to rely on the algorithm. Our study relatedly suggests that when humans retain final decision-making authority, ensuring that they have some dimension of absolute advantage on the algorithm tool, such as restricting algorithm inputs while retaining predictive accuracy, could be essential for complementarity. Making more data available to the algorithm is not necessarily better when humans are involved, echoing the literature on information overload \citep{edmunds2000}.

\subsection{Cost-Benefit Analysis: MVPF Estimate}

Recent scholarship has shown that algorithmic decision aids exhibit high social returns across a variety of domains \citep{ludwig2024}. 
We evaluate the trial algorithm's estimated benefits against its implementation and development costs with a marginal value of public funds (MVPF) calculation. 

For the MVPF numerator, we note that beneficiary willingness to pay for preventing child maltreatment is unambiguously positive. For the denominator, net government cost, we assume that the algorithm reduced the number of maltreated children by about 20, following our main estimates of 20 fewer children in the extreme right tail of the harm distribution (Table 2, Column 7). A lower-bound estimate of the public costs of maltreatment is \$62,500 per child (2020 USD) in terms of special education, criminal justice, and additional child public health expenses alone \citep{fang2012}.\footnote{We cannot assess whether our findings are from a reduction in the intensive or extensive margin of maltreatment incidence. We are unaware of any study estimating costs of maltreatment for these two margins separately and so are unable to benchmark against the cost of one additional maltreatment incident.}  The resulting estimated public savings from reduced maltreatment ($20 \cdot \$62,500=\$1.25M$) are much larger than the algorithm's implementation costs -- about \$280k over two years with \$15k annual maintenance thereafter -- implying an infinite MVPF of the algorithmic tool. A limitation of our approach is that many of the children involved in the trial may have already been victims of maltreatment, confounding our estimates of government savings from curtailed abuse and neglect. However -- to the extent that the costs of maltreatment are cumulative\footnote{Prior studies have found an association between negative outcomes and number of maltreatment incidents or number of types of experienced maltreatment (e.g., \citealt{currie2012understanding}), consistent with a clinically hypothesized dosage-response relationship, though causality is difficult to establish. } and/or some fraction of children were not prior victims -- even if only a fourth of the lower-bound estimate of government costs for child maltreatment were saved, the resulting MVPF would remain infinite, suggesting some robustness of our findings to alternative cost assessment approaches. A large share of the costs of the tool were up-front fixed costs, implying a potential increase in cost effectiveness from scaling the tool to other agencies, which has occurred since the trial.

\section{Conclusion}

Machine learning tools have drawn increased interest and scrutiny in social service and criminal justice settings, including Child Protective Services (CPS). This paper offers one of the first randomized evaluations of human-algorithm interaction in such a context.  
CPS represents an important and generalizable use case for evaluating algorithmic decision supports. Like many public agencies, Child Protective Services (i) have long-accumulated histories of administrative data that are useful for predictive risk modeling, and (ii) require critical, time-sensitive decisions from workers with limited experience and high turnover. Our trial's high-stakes field setting, reasonably representative sample of children referred to CPS during the study period, 
and the algorithm's scaling potential given its implementation in other counties (in Colorado, Pennsylvania, and California) provides cautious optimism for external validity \citep{list2020non}. 

We find that humans and algorithms are complementary in our setting: both decision quality and equity improve compared to either party's decisions (observed or hypothetical) on their own. When humans had access to algorithm support, child injury-related hospitalizations fell by 21 percent and two thirds of extreme child harm instances (top percentile of child harm index) were prevented. 
The algorithm was designed as an aid rather than a replacement for social workers, who were acutely aware of what information was included in the algorithm versus available exclusively to them. In this context, algorithm support helped workers parse through new, complementary information about families.
The algorithm helped workers reduce their tendency to investigate Black children, particularly those scored as low risk, and conversely human oversight helped reduce CPS surveillance that the algorithm would have otherwise induced for Hispanic and low-income households. Human-algorithm complementarity is thus salient not just for efficiency but also for equity. Our model and documented mechanisms rationalize these findings and suggest they could extend to other decision settings when algorithms reduce the noise in the decision-making process, such as through prompting richer discussion of critical information or reducing extraneous cognitive load. 
Existing literature suggests that design features -- in particular which information was available to the algorithm versus humans, and that the algorithm was deployed as part of a group discussion \citep{galton1907,simoiu2019,chiang2023} -- may have contributed to our findings.

Although the evaluation in this paper took place over multiple years, it is important that future work estimate the effects of algorithmic decision support over even longer time horizons. Algorithmic tools may affect workers' competence, the pipeline of future decisions, and the types of workers that a firm can recruit or retain through algorithms' effects on turnover (\citealt{brynjolfsson2023}). Recent studies have articulated concerns about worker deskilling or turnover when algorithms replace expertise \citep{rinta-kahila2018Consequences,grennan2020Artificial}. However, in a complex social setting such as child protection where complementary worker expertise seems critical, long-run effects could be more consistent with emerging studies showing that humans can learn to make novel and better decisions over time when interacting with AI \citep{shin2023Superhuman}. Longer-run evaluations could yield further insights on how algorithms can be adapted, statistically or dynamically, to human cognitive biases (\citealt{kleinberg2024}).

Finally, although the use of predictive risk modeling is valuable for refining worker decisions, our discussions with workers highlight limits to professional discretion.
In our context, agencies are statutorily required to have supporting (non-algorithm) evidence to investigate a referral for alleged maltreatment, whereas other types of allegations require an investigation regardless of human perception of risk. A statutorily-optimal policy must therefore investigate at least a fraction of predicted low-risk reports, and cannot investigate all high-risk reports. Abstracting away from human oversight and accountability would infringe on legal requirements, highlighting the need for algorithm designers to consider the incentives and behaviors of human decision makers. Consequently, identifying which context-specific features of human-algorithm interaction lead to the greatest efficiency gains remains a valuable area for future research. Promising research directions include identifying optimal decision points for deploying new algorithms, optimal triage of decisions to algorithm-only versus human-only or human-plus-algorithm tracks, improving the explainability and presentation of predictive information, the tradeoffs of predicting treatment effects versus outcome levels, 
and understanding the impact of algorithms on individual versus group decision-making.


\newpage
\begingroup
\setstretch{0.97}
\spaceskip=1\fontdimen2\font 
\xspaceskip=1\fontdimen7\font 
\bibliography{bib}
\bibliographystyle{aernobold}
\endgroup


\setlength\parindent{0pt}

\newpage
\appendix
\section*{Online Appendix}
\section*{Appendix A}
\setcounter{figure}{0} \renewcommand{\thefigure}{A.\arabic{figure}} 
\setcounter{table}{0} \renewcommand{\thetable}{A.\arabic{table}}  

\begin{figure}[H]
    \centering
    \caption{Description of Child Protective Services in the Evaluation County}
    \label{cps_diagram}
    \vspace{-0.25em}
    \makebox[\textwidth][c]{\includegraphics[width=1.0\textwidth]{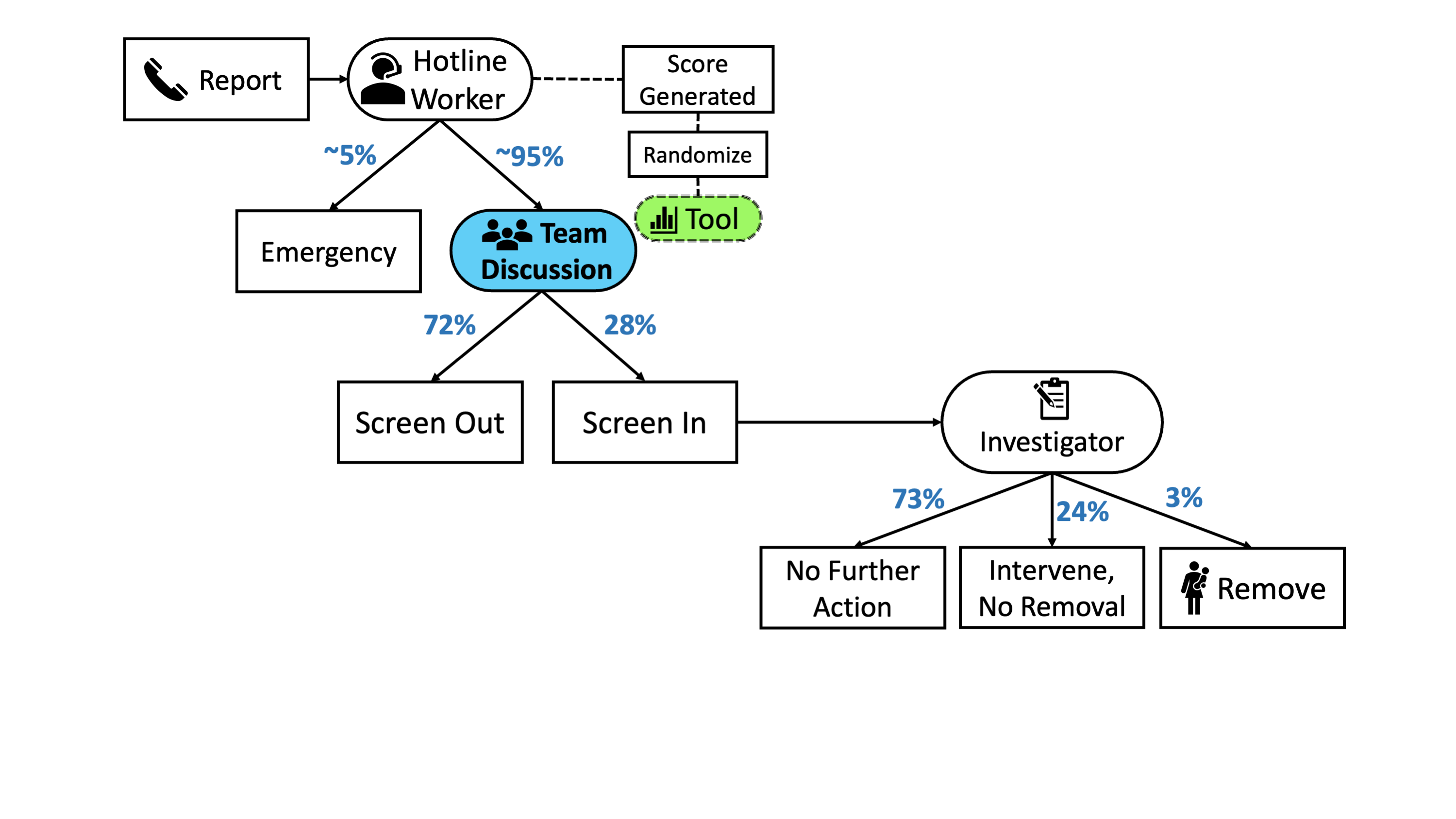}}   
\end{figure}
\vspace{-6em}
{\footnotesize\setlength{\parindent}{0pt}\begin{spacing}{1.0}Notes: Listed percentages are the share of children in the county assigned to the respective decision, conditional on reaching the given node. For example, three percent of investigated children are placed in foster care (removed from home) within 90 days.\end{spacing}}
\vspace{1em}

\begin{figure}[H]
    \centering
    \caption{Algorithm Tool Display for Workers}
    \label{tool_display2}
    \makebox[\textwidth][c]{\includegraphics[width=0.9\textwidth]{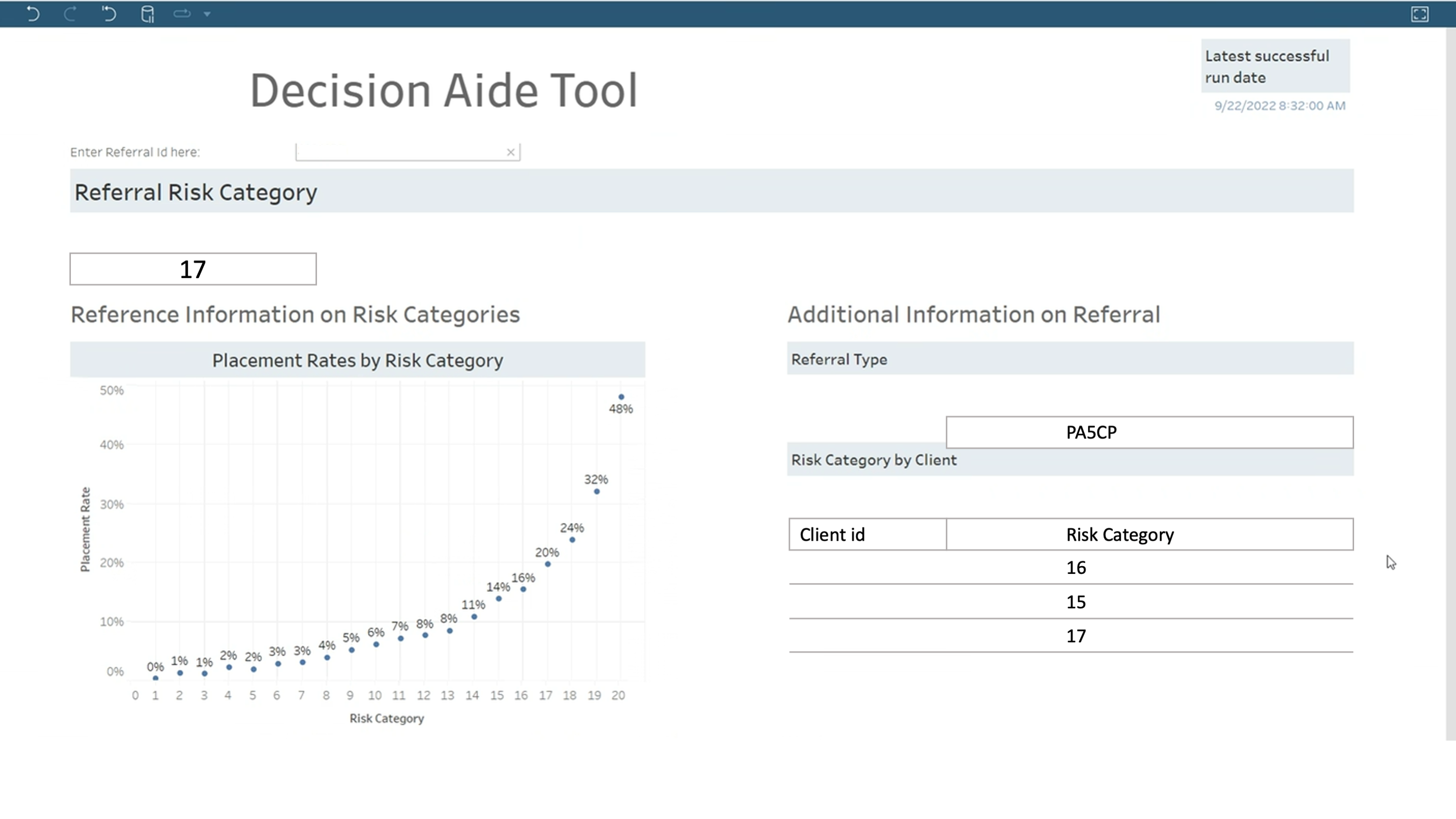}}
\end{figure}
 \vspace{-1em}
 {\footnotesize\setlength{\parindent}{0pt}\begin{spacing}{1.0}Notes: This figure shows an example of the algorithmic tool interface. The right panel shows predicted risk scores for all children listed on the referral (in this case 16, 15, and 17). In the top left of the display, the highest score across all of the children is emphasized to decision makers (in this case 17). The plot on the bottom left reminds workers of the chance that a child is removed from their home (foster care placement) within two years at each score level. \end{spacing}}
\vspace{1em}

\begin{figure}[H]
	\caption{Algorithm Risk Score Predicts Relevant Outcomes \vspace{0.25em} \\ \footnotesize\textit{Control Group}}
	\label{fig:toolvalidation}
	\begin{subfigure}[b]{0.50\textwidth}
		\caption{Harm Index}
		\includegraphics[trim=0 100 20 20, clip,width=\textwidth]{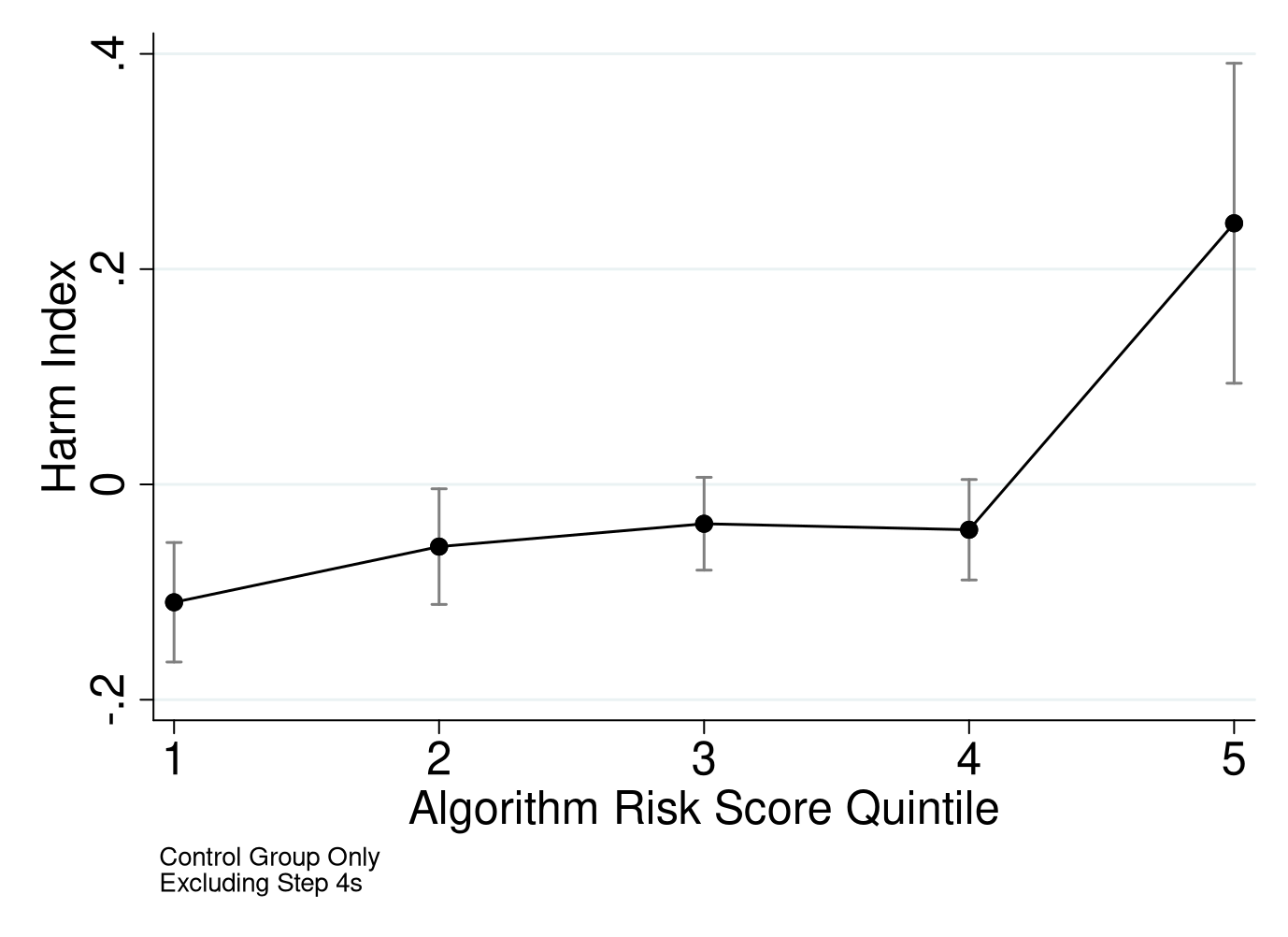}
		\end{subfigure}
	\begin{subfigure}[b]{0.50\textwidth}
		\caption{Removed from Home}
		\includegraphics[trim=0 100 20 20, clip,scale=.2,width=\textwidth]{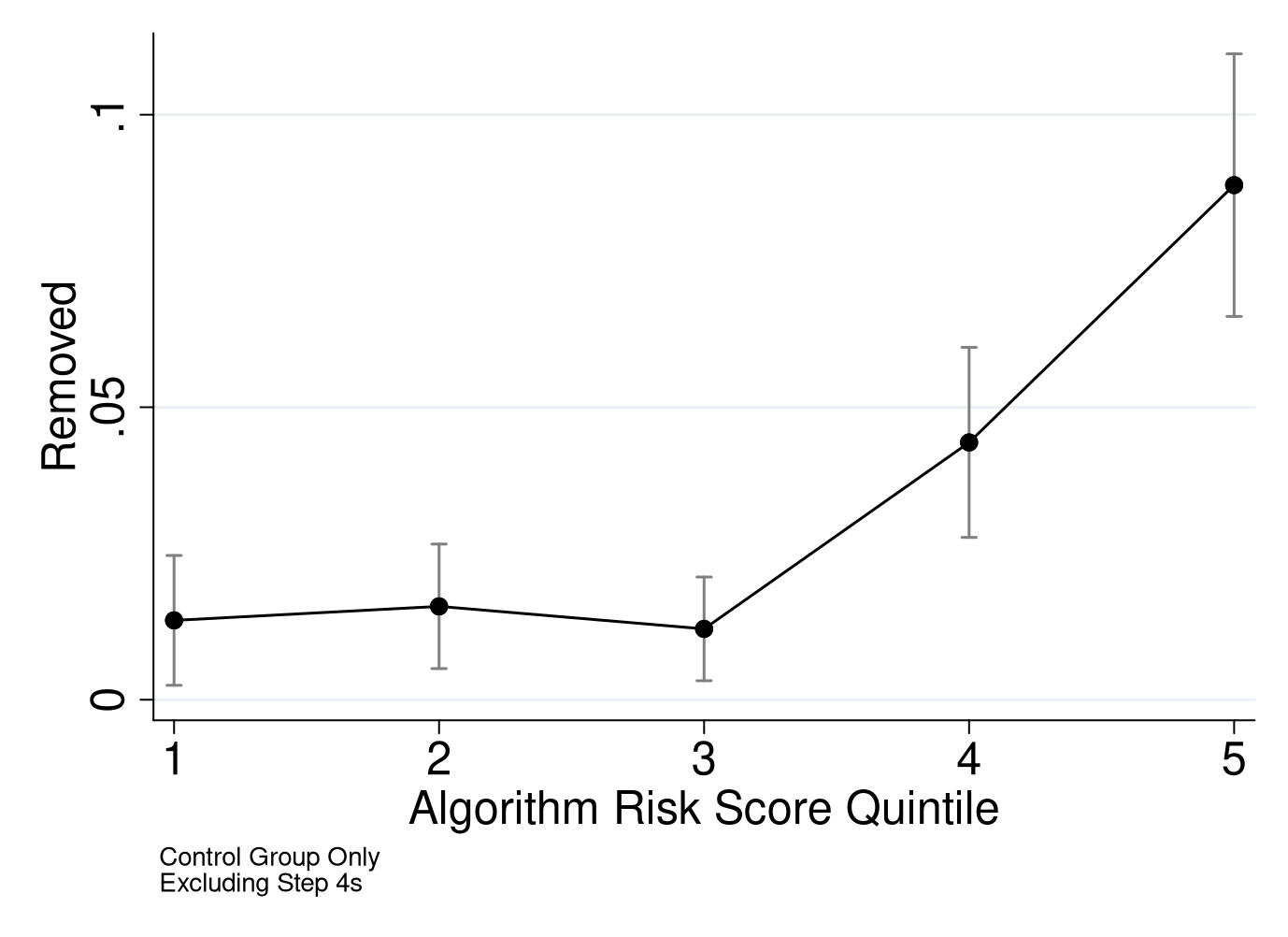}
		\end{subfigure}
  \hfill

  \centering
        \begin{subfigure}[b]{0.50\textwidth}
		\caption{Investigator-Assigned Risk Rating}
		\includegraphics[trim=0 100 20 20, clip,scale=.2,width=\textwidth]{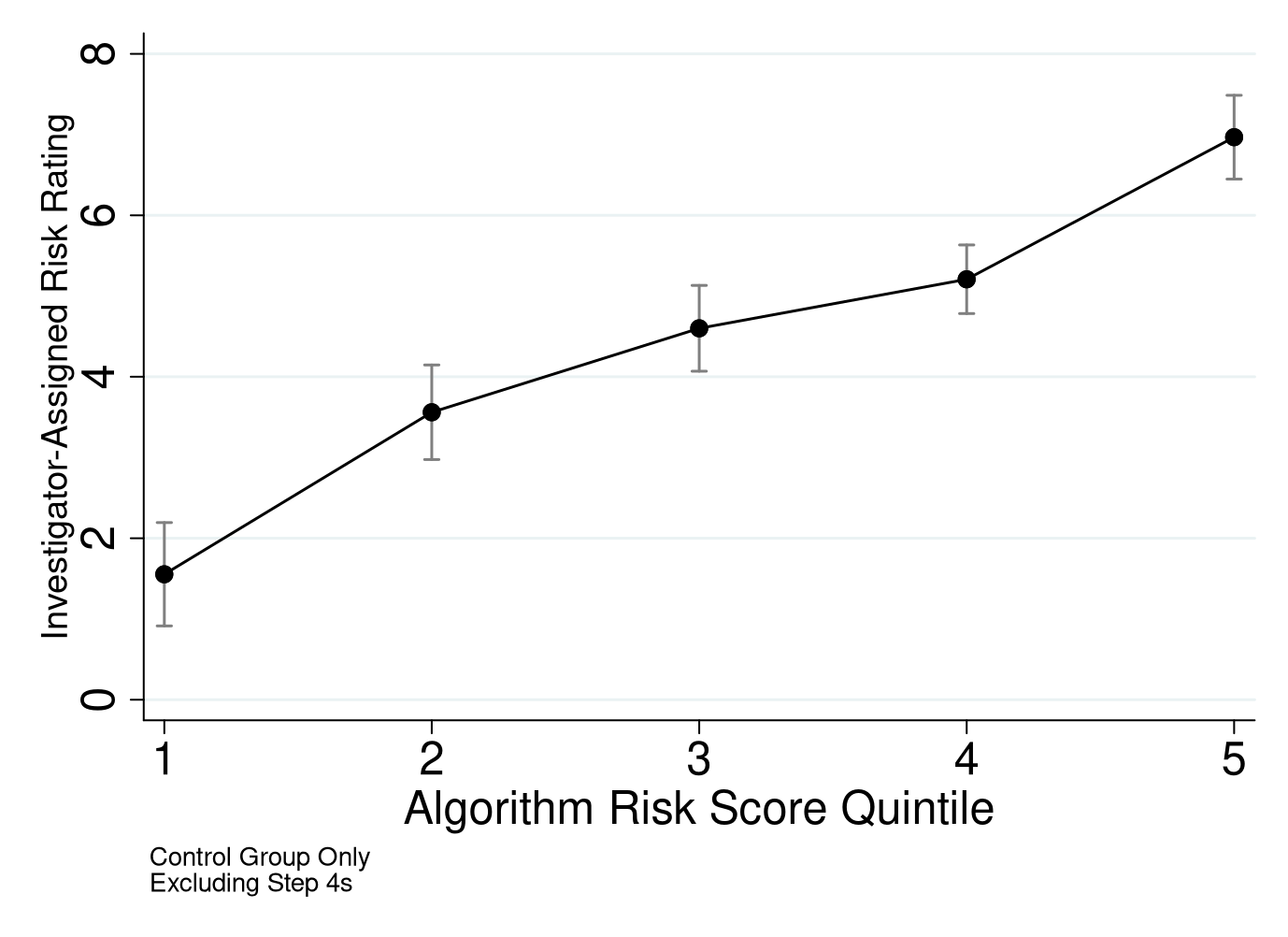}
	\end{subfigure}
\end{figure}
\vspace{-1em}
{\footnotesize\setlength{\parindent}{0pt}\begin{spacing}{1.0}Notes: These subfigures show the relationship between relevant outcomes and the algorithm risk score, binned by quintile for visual clarity. 
Each point represents the mean of the outcome variable in the control group at a given quintile of predicted risk, with a 90\% confidence interval reported. The algorithm was trained on prior data to predict a child's removal from their home to foster care within two years of a referral. Investigator-assigned risk ratings (Panel iii) come from visits to the home and are only available for the set of screened-in children. \end{spacing}}
\vspace{1em}

\begin{figure}[H]
    \centering
    \caption{Trial Workflow}
    \label{tool_workflow}
    \makebox[\textwidth][c]{\includegraphics[width=0.8\textwidth]{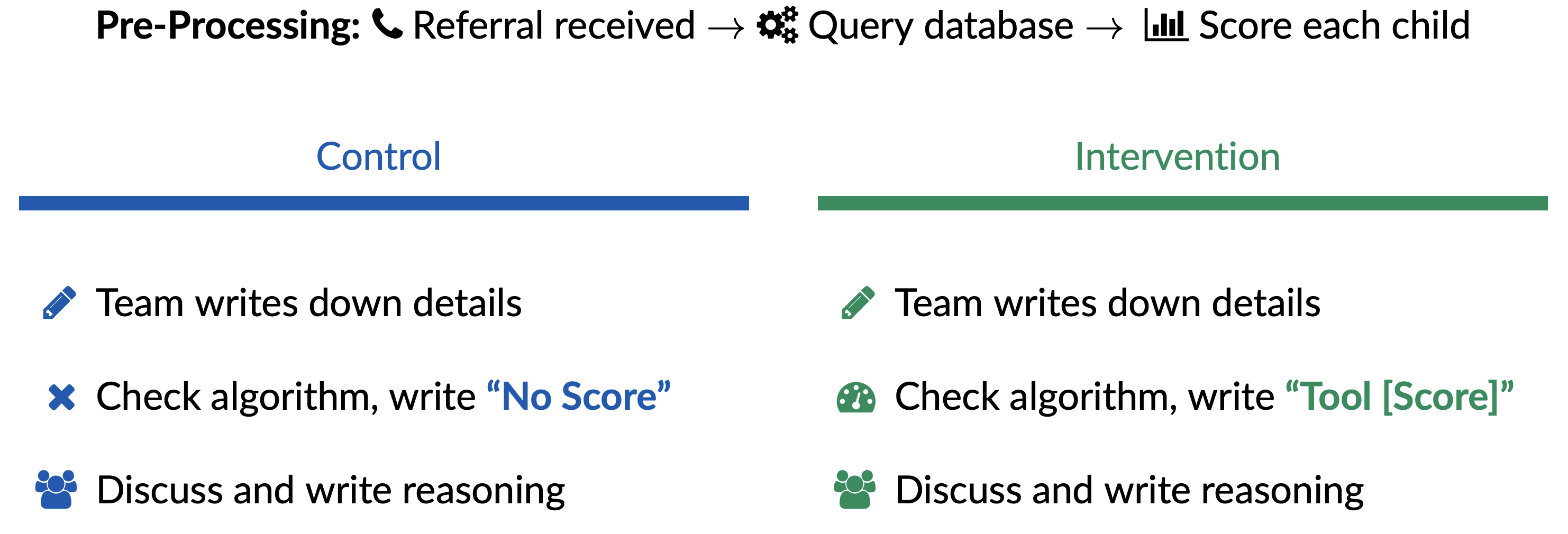}}
\end{figure}
 {\footnotesize\setlength{\parindent}{0pt}\begin{spacing}{1.0}Notes: This figure presents a stylized summary of teams' decision workflow during the trial. Algorithm scores were available for the intervention group, and unavailable for the control group. Workers were encouraged, but not required, to check the algorithm and write down ``No Score'' if the score was unavailable or ``Tool [Score]'' with the specific listed score(s). 
\end{spacing}}
\vspace{1em}

\begin{figure}[H]
        \centering
	    \caption{Impact of Algorithm on Child Harm, by Predicted Risk Level}
	    \label{SERVER_bar_treatcontrol_indexpost30_std} 
		\includegraphics[width=0.70\textwidth,trim=0 80 0 0, clip,]{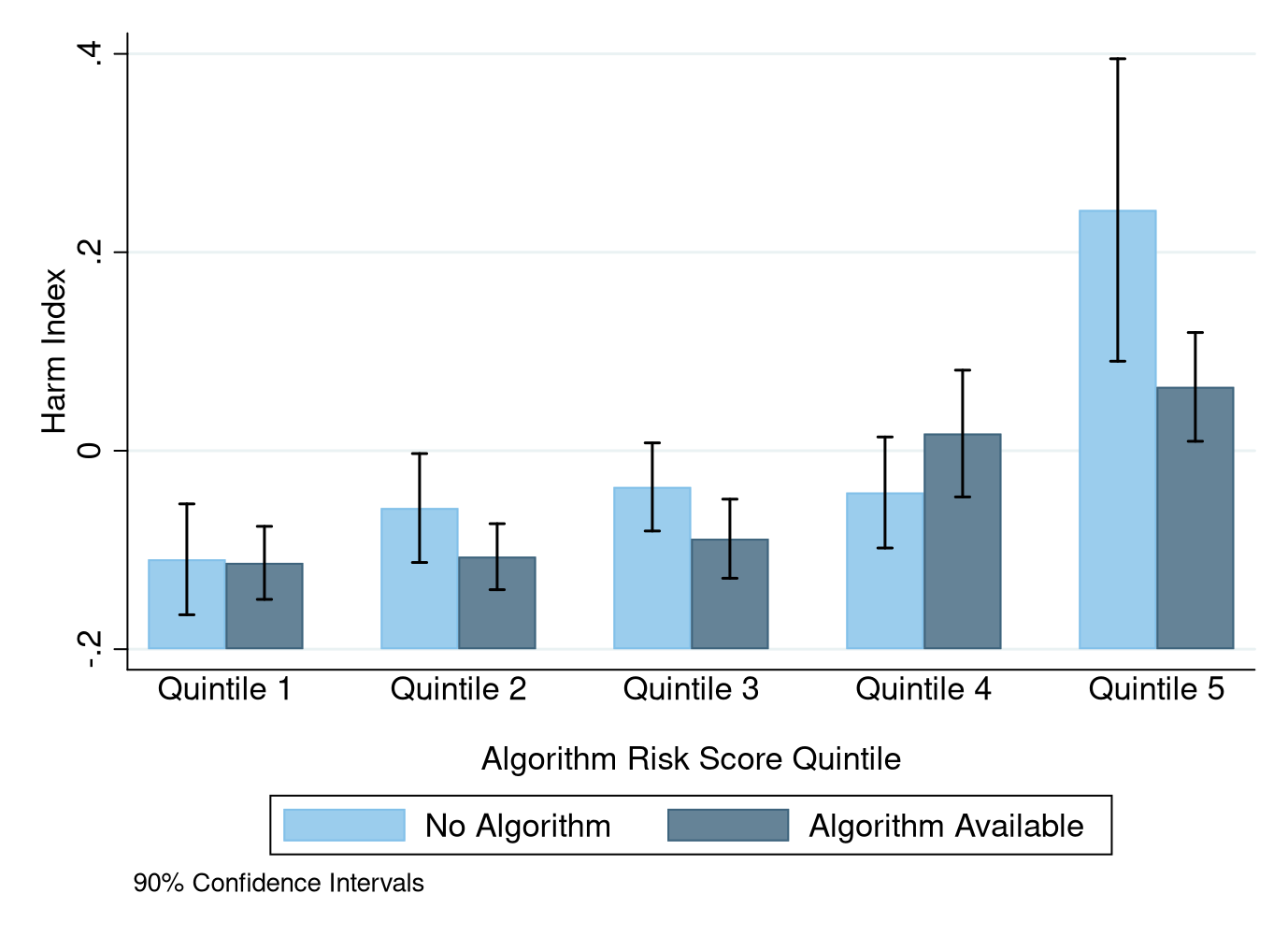}
\end{figure}
\vspace{-1em}
{\footnotesize\setlength{\parindent}{0pt}\begin{spacing}{1.0} Notes: This figure presents mean child harm by algorithm score quintile, with and without the tool available, with 90\% confidence intervals. The sample is restricted to children randomized during the first year of the trial to allow for several months of potential hospitalization following randomization. The harm index is constructed as a standardized sum of standardized outcomes including a child's number of high-priority hospital admissions, number of admissions with a listed injury, number of avoidable ER visits, substance exposure incidents, number of admissions with a confirmed maltreatment code, and number of visits with an intentional injury (assault, self-harm) code. The index is standardized (mean 0, variance 1) on the control group and a lower value is considered better.\end{spacing}}
\vspace{1em}

\begin{figure}[H]
        \centering
	    \caption{Bayesian Posterior Distribution of Effects on Harm Index Components}
	    \label{fig:bayesharm} 
		\includegraphics[width=1\textwidth,trim=0 200 0 0, clip,]{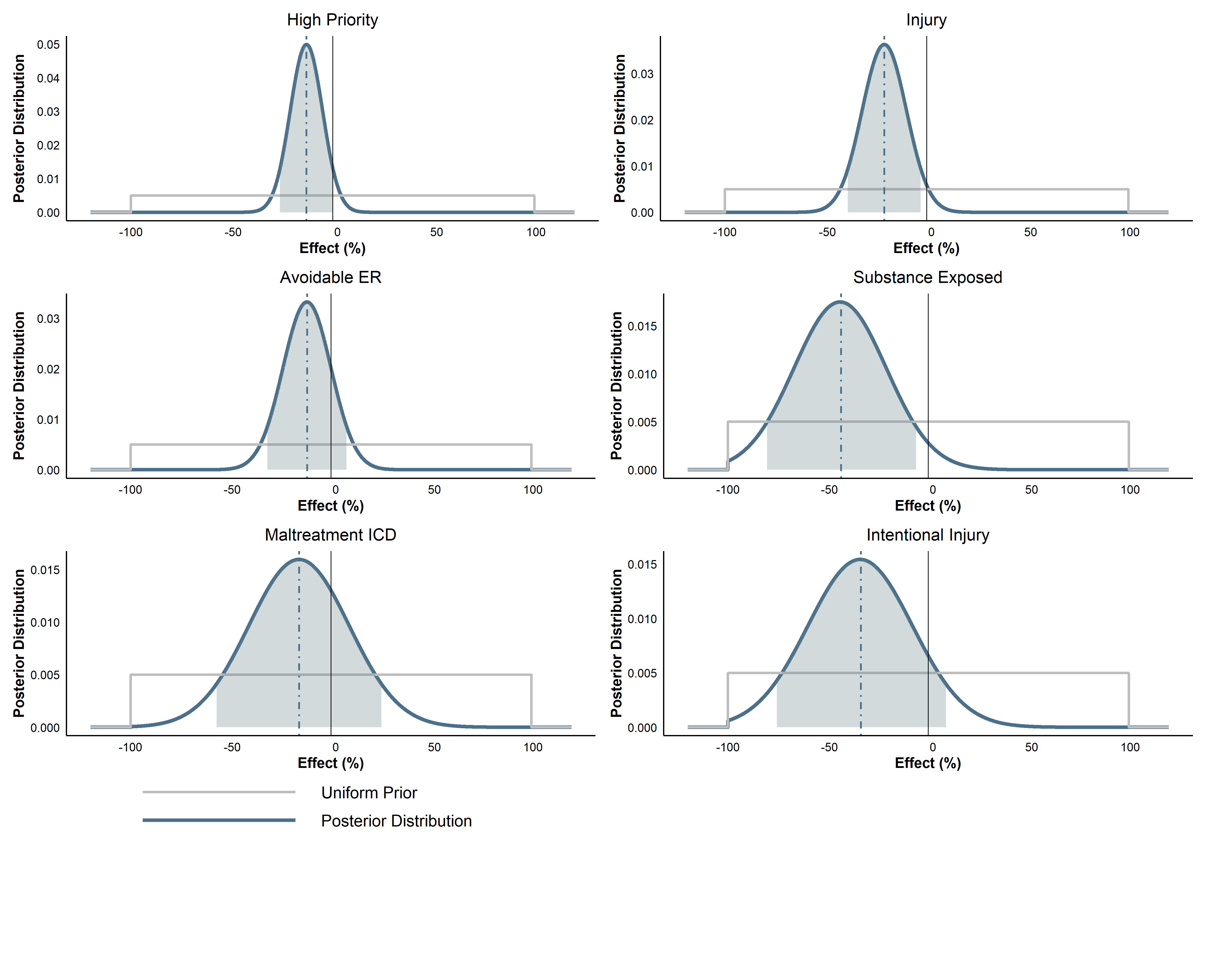}
        
        \vspace{2mm}
        \includegraphics[width=0.25\textwidth,trim=200 190 680 660, clip,]{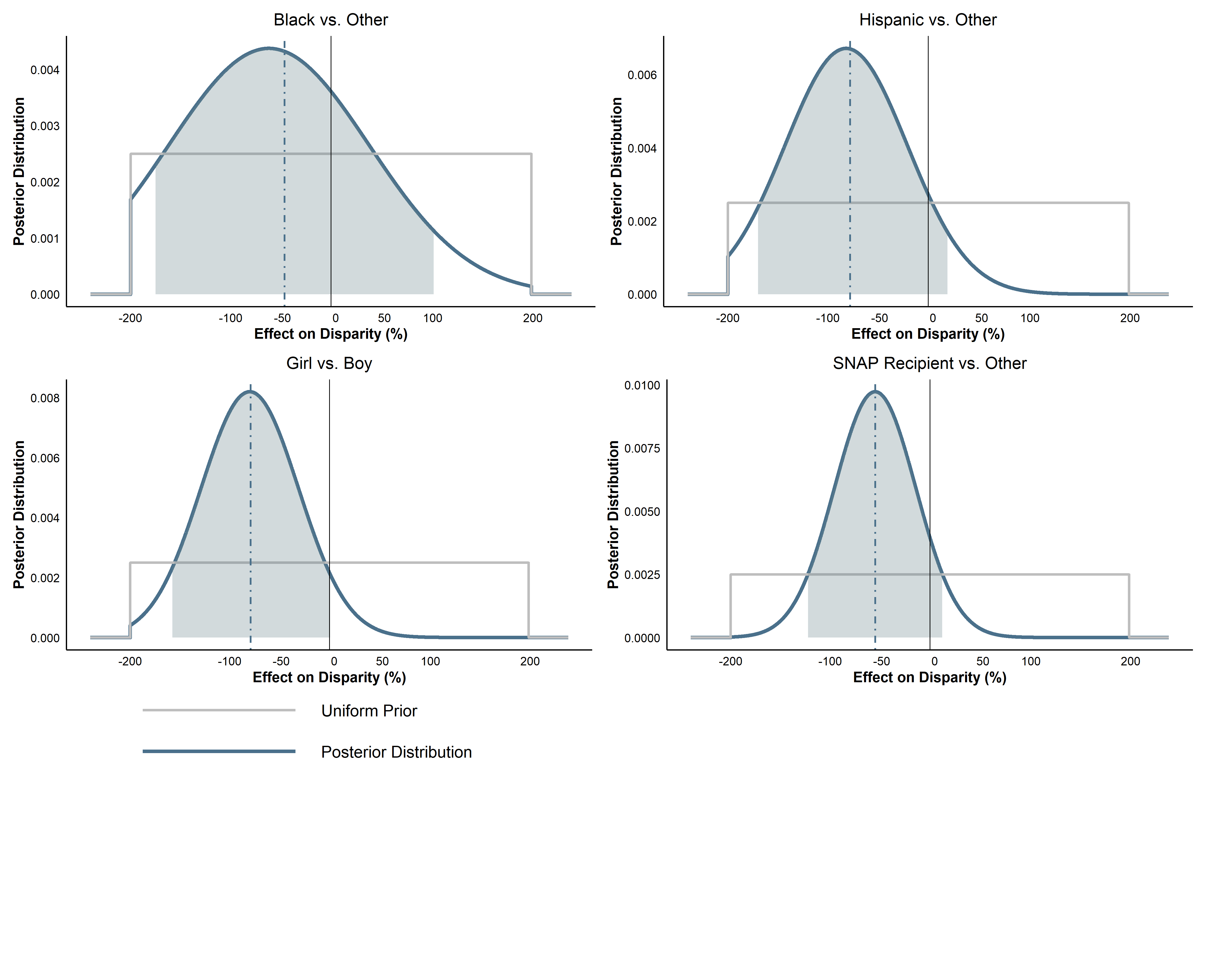}
\end{figure}
\vspace{-1em}
{\footnotesize\setlength{\parindent}{0pt}\begin{spacing}{1.0} Notes: These figures report the posterior distribution of the effect of algorithm support on six components of the harm index: counts of high priority visits, injury visits, avoidable ER visits, substance exposure incidents, visits with explicit maltreatment codes (rare), and visits with intentional injuries. We assume as an uninformative uniform prior that effects could have been anywhere in the range of -100 percent to +100 percent (gray line centered around vertical line at 0). Following \citet{brannlund2024}, we then calculate the posterior distribution given the data we observe (posterior distribution outlined in blue; observed mean marked by dash-dot vertical blue line). 90\% credible intervals are reported as the shaded areas. 
\end{spacing}}
\vspace{1em}

\begin{figure}[H]
        \centering
	    \caption{Estimated Effects by Time of First Appearance in the Trial}
	    \label{fig:LRlearning} 
		\includegraphics[width=0.90\textwidth,trim=0 20 0 0, clip,]{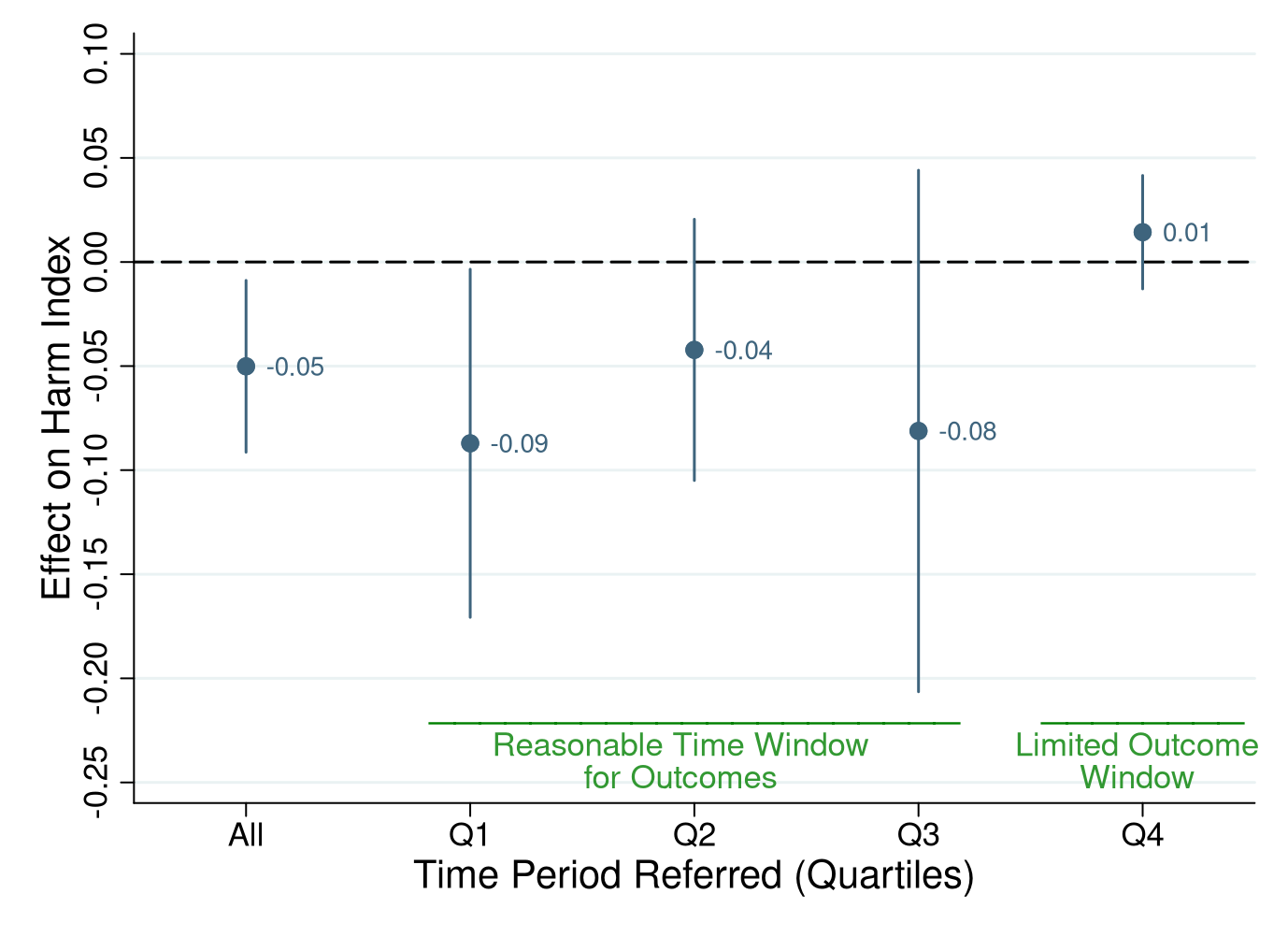}
\end{figure}
\vspace{-1em}
{\footnotesize\setlength{\parindent}{0pt}\begin{spacing}{1.0} Notes: 
This figure presents estimates of the tool on harm by the time a child was first seen during the trial. On the left, we present the main (full sample) point estimate of providing the algorithm tool on child harm (``All''). The full sample is then split into four equally sized groups by date of referral and estimate the effects of providing the tool on the harm index in each of these sample quarters. Q1 is the first quarter of children referred during the trial, Q2 the second, and so on. 90\% confidence intervals are reported throughout. 
Absence of impacts for children referred in the final quarter (Figure \ref{fig:LRlearning}) could be due to a limited window of time to observe hospital outcomes for this subsample during the trial. After the end of the trial, the algorithm tool became available for all children being referred, attenuating effects. 
\end{spacing}}
\vspace{1em}

\begin{figure}[H]
        \centering
	    \caption{Bayesian Posterior Distribution of Effects on Disparities with an Uninformative Prior}
	    \label{fig:bayesdisparity} 
		\includegraphics[width=1\textwidth,trim=0 280 0 0, clip,]{Posterior_Disparity.png}
        
        \vspace{2mm}
        \includegraphics[width=0.25\textwidth,trim=200 190 680 660, clip,]{Posterior_Disparity.png}
\end{figure}
\vspace{-1em}
{\footnotesize\setlength{\parindent}{0pt}\begin{spacing}{1.0} Notes: These figures report the posterior distribution of the effect of algorithm support on four harm index disparity margins: race, ethnicity, gender and socioeconomic status. We assume as an uninformative uniform prior that effects could have been anywhere in the range of -200\% to +200\% (gray line centered around vertical line at 0). Given the data and following \citet{brannlund2024}, we then calculate the posterior distribution given the data we observe (posterior distribution outlined in blue; observed mean marked by dash-dot vertical blue line). 90\% credible intervals are reported as the shaded areas. 
\end{spacing}}
\vspace{1em}

\begin{figure}[H]
	\caption{Tool Reduced Racial Disparities for Low-Risk-Score Children \vspace{0.25em} \\ \footnotesize\textit{Using CPS-Known Race and Hospital-Known Race}}
	  \begin{subfigure}[b]{0.5\textwidth}
		\caption{Below Median Predicted Algorithm Risk Score}
		\includegraphics[trim=0 20 15 85, clip,width=\textwidth]{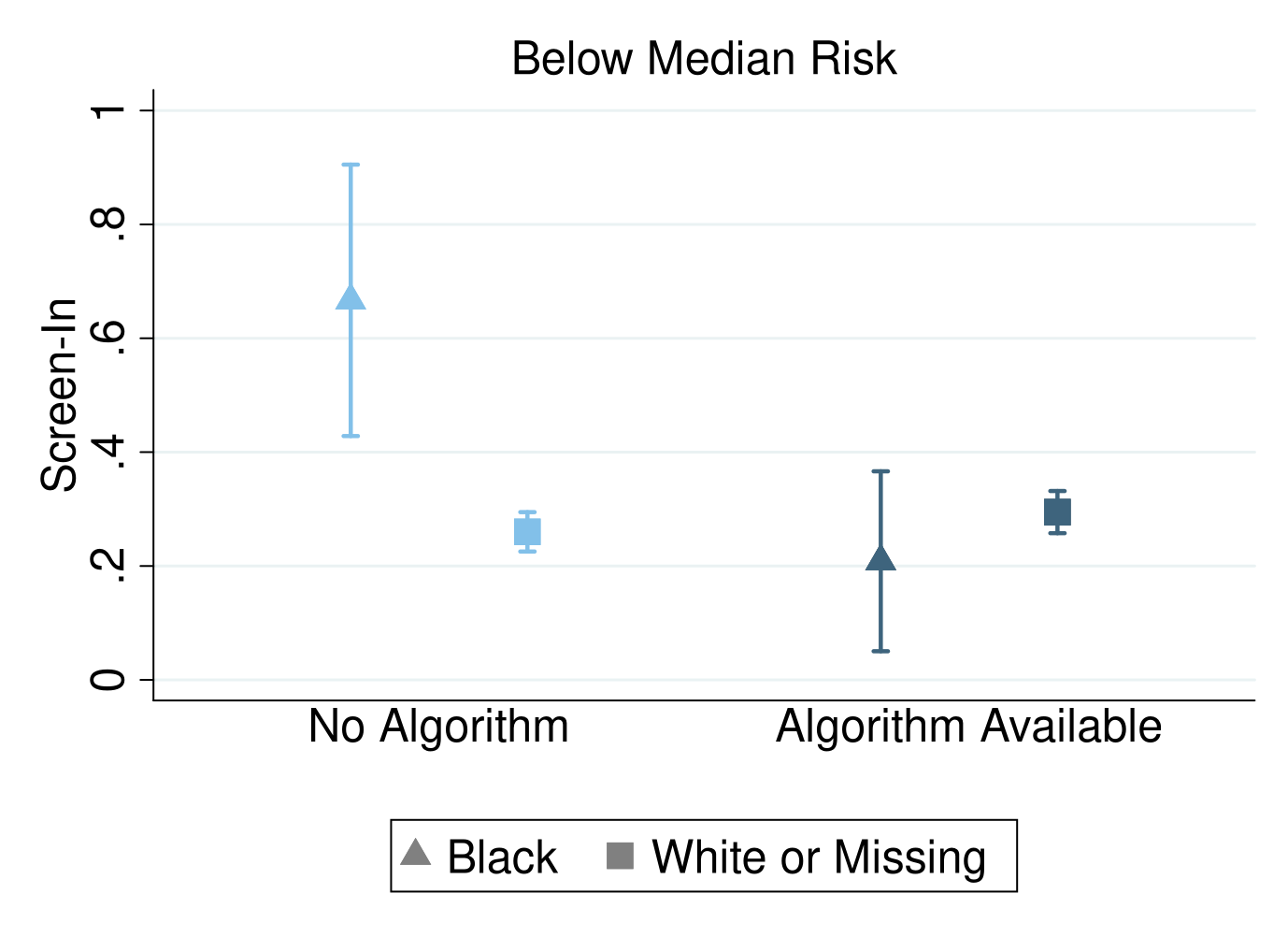}
		\label{fig:hospracescreenin0}
	\end{subfigure}
	\begin{subfigure}[b]{0.5\textwidth}
		\caption{Above Median Predicted Algorithm Risk Score}
		\includegraphics[trim=0 20 15 85, clip,scale=.2,width=\textwidth]{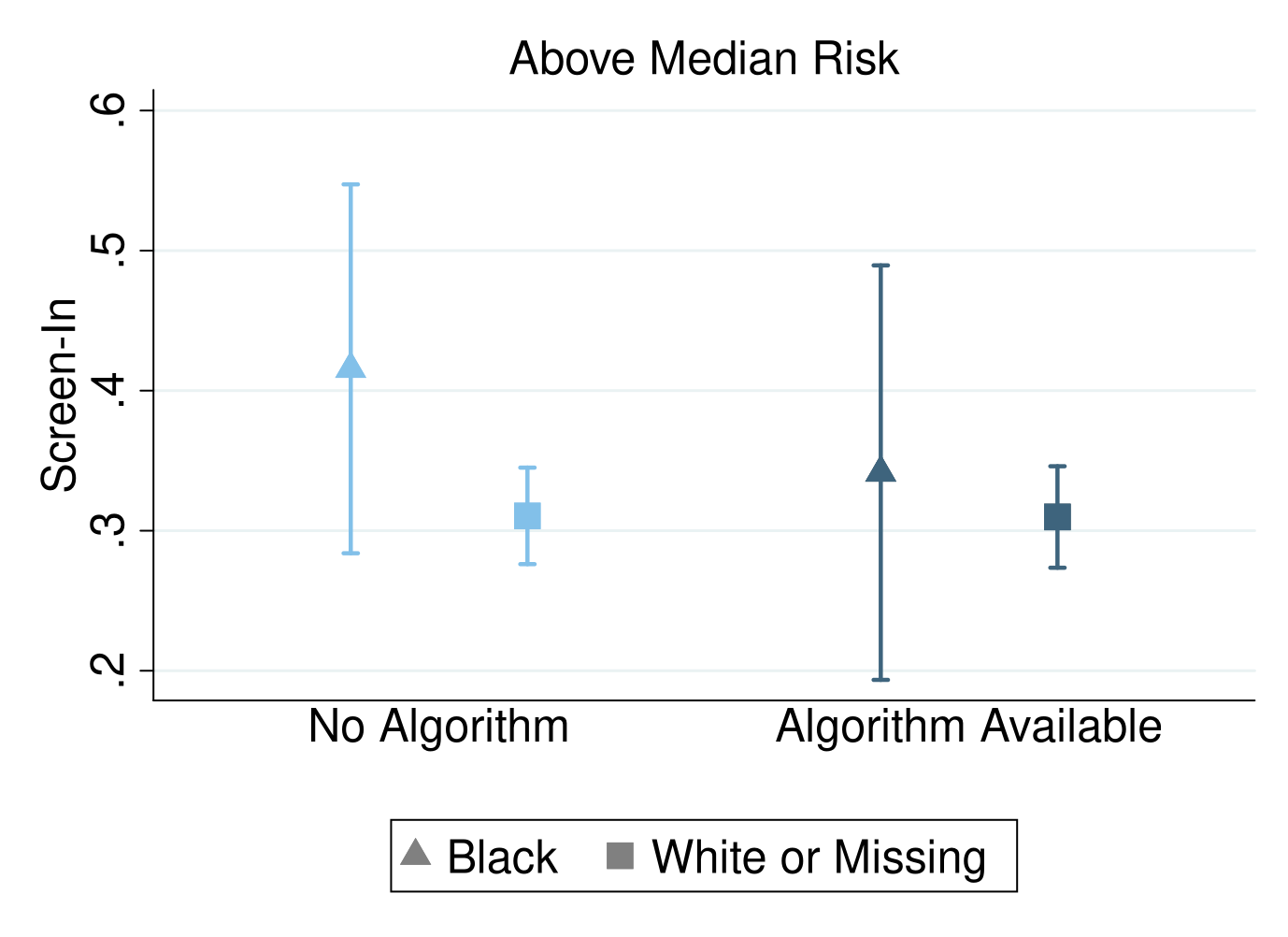}
		\label{fig:hospracescreenin1}
	\end{subfigure}
\end{figure}
\vspace{-2em}
{\footnotesize\setlength{\parindent}{0pt}\begin{spacing}{1.0}Notes: These figures show the estimates of mean harm index by demographics and tool availability. Figures define race as ever being recorded as Black by Child Protective Services or hospitals. 90\% confidence intervals are reported and randomization controls are included throughout.  \end{spacing}}
\vspace{1em}

\begin{figure}[H]
	\caption{No Significant Increase in Harm from Reducing Screen-In Disparities \vspace{0.25em} \\ \footnotesize\textit{Using CPS-Known Race}}
	    \label{fig:screeninhosprace}
	\begin{subfigure}[b]{0.5\textwidth}
		\caption{Below Median Predicted Risk Score}
		\includegraphics[trim=0 20 15 50, clip,width=\textwidth]{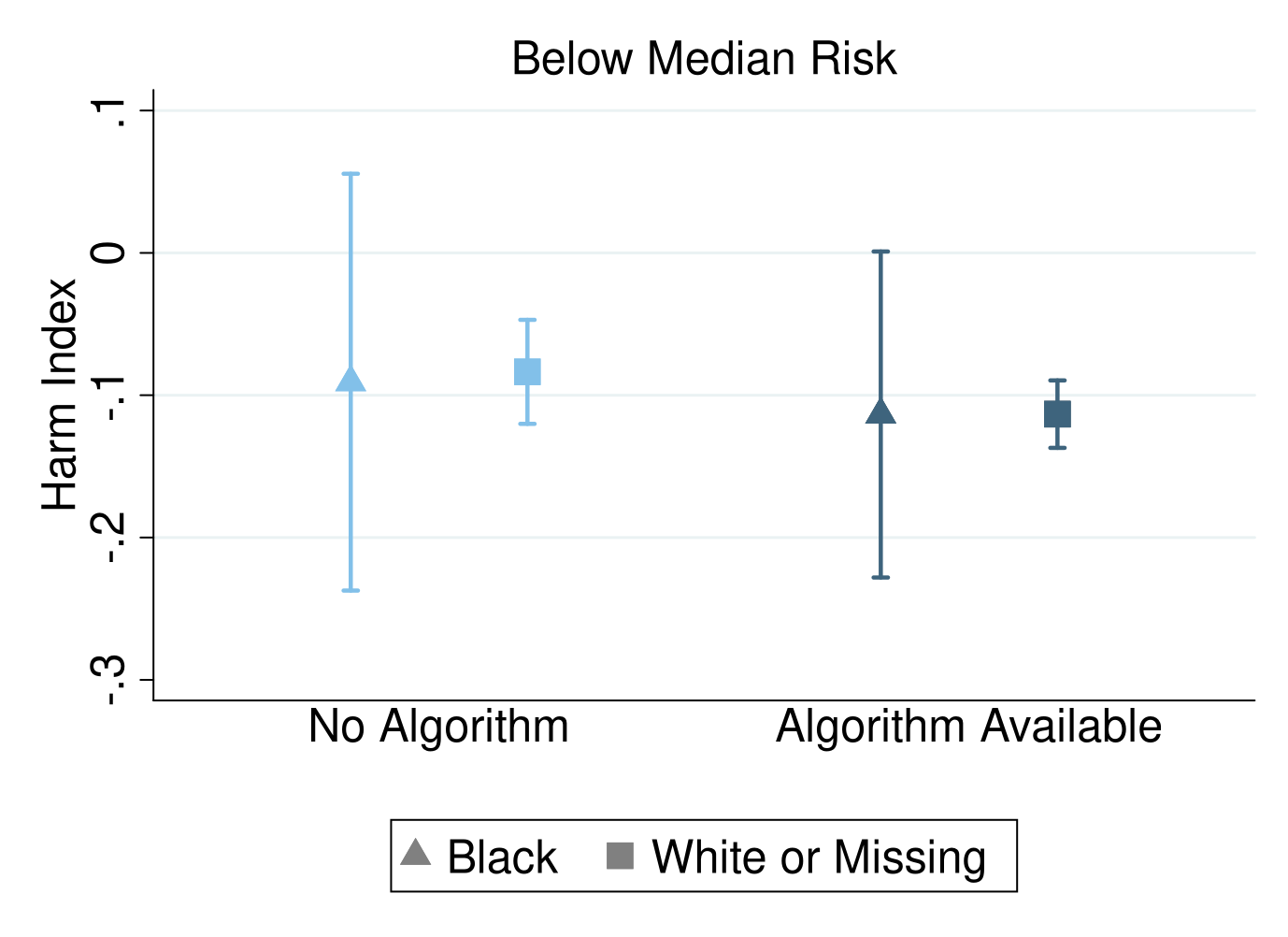}
		\label{fig:blackharmbymedian}
	\end{subfigure}
	\begin{subfigure}[b]{0.5\textwidth}
		\caption{Above Median Predicted Risk Score}
		\includegraphics[trim=0 20 15 50, clip,scale=.2,width=\textwidth]{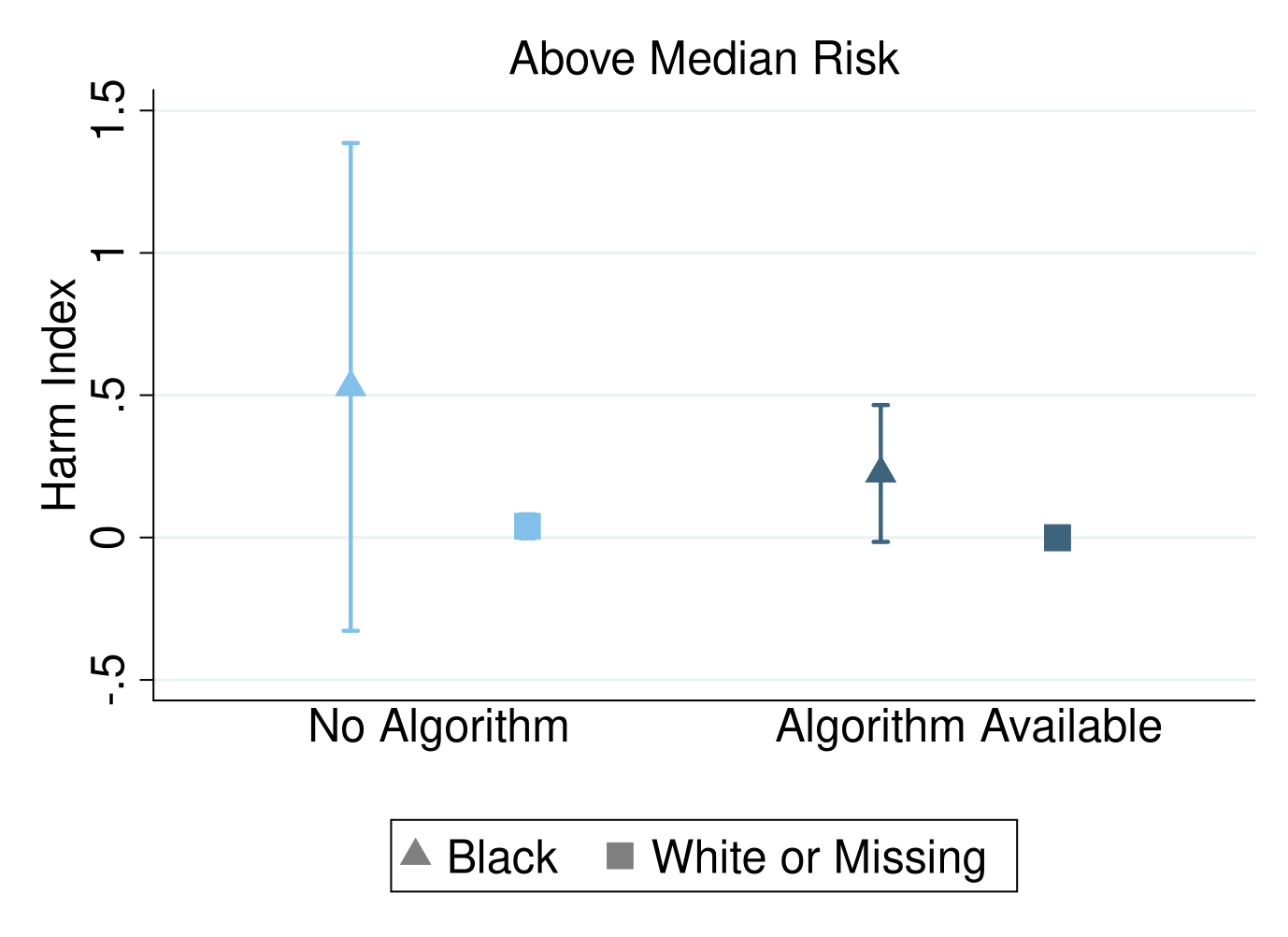}
		\label{fig:blackharmbymedian1}
	\end{subfigure}
\end{figure}
\vspace{-1em}
{\footnotesize\setlength{\parindent}{0pt}\begin{spacing}{1.0}Notes: These figures show the estimates of mean harm index by demographics, tool availability, and whether algorithm-predicted risk was above or below median. Figures define race as recorded by Child Protective Services. 90\% confidence intervals are reported and randomization controls are included throughout.  \end{spacing}}
\vspace{1em}

\begin{figure}[H] 
    \centering
    \caption{Level Effect of Tool on Margins of CPS Intervention}
    \label{fig:cpslevel}
   
    \makebox[\textwidth][c]{\includegraphics[width=0.80\textwidth,trim= 10 30 10 30, clip]{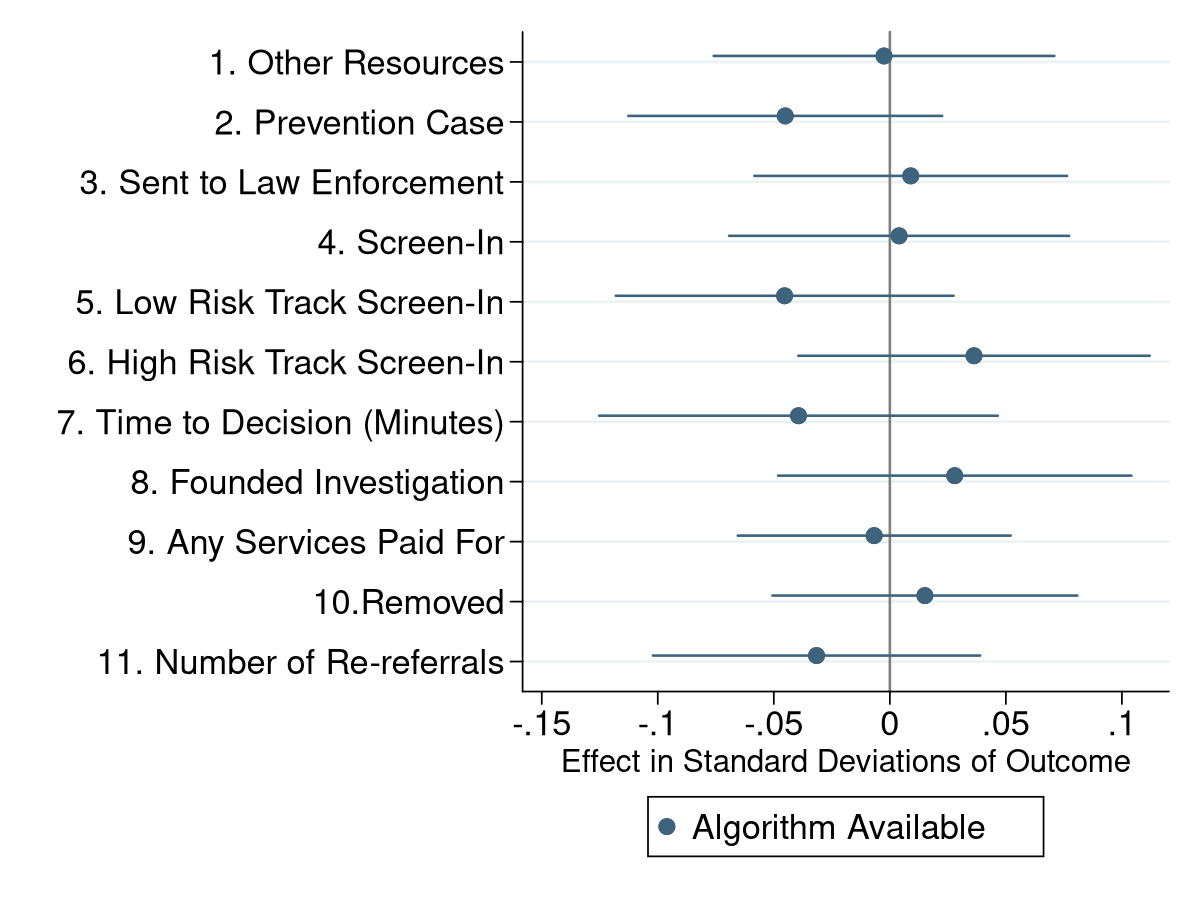}}
\end{figure}
\vspace{-1em}
{\footnotesize\setlength{\parindent}{0pt}\begin{spacing}{1.0}Notes: This figure presents point estimates and confidence intervals of regressing the listed outcomes on tool availability, as in equation (\ref{eq:eq1}), for the full sample of children (unless otherwise specified). Estimates are presented with standardized outcome variables (mean 0, variance 1) for comparability of coefficient magnitudes across different outcomes. We measure margins of CPS intervention, such as high-risk-track screen-ins, using child welfare administrative data. Additional margins of intended intervention, such as voluntary resources sent to screened-out families, are inferred from team discussion notes. Some interventions are unobserved: services paid for by Medicaid, for example, are not included in administrative data. Workers can recommend opening a prevention case (row 2), can send the referral to law enforcement (row 3), and can connect the family to other resources (row 1) instead or in addition to screening in (row 4). If workers decide to screen in a family, they also choose whether the screen-in is on a low-risk or high-risk track (rows 5 and 6). Low-risk track investigations are a type of differential response that is less intrusive for the family. Formal investigations are only run for the high-risk track. As a result, having a founded (i.e., substantiated; row 8) investigation requires having a high-risk track investigation. Row 9 shows the change in any other services paid for directly by CPS. The number of minutes it took workers to make a decision (row 7) is measured using time stamp data and is only available for observations where edits to the notes were not made after normal discussion meeting times. The final two rows (rows 10-11) report estimates on whether a child was ever removed to foster care, and the number of re-referrals. We report more details on these last two outcomes in a separate table (Appendix Table \ref{tab:CPSOutcomesTable}). Randomization procedure controls are included throughout. Standard errors are clustered at the household level and are shown at the 90 percent confidence level.\end{spacing}}
\vspace{1em}

\begin{figure}[H]
		\centering
		\caption{Workers Did Not Increase Investigations for High-Scoring Children}
        \label{fig:screeninbyscore}
		\includegraphics[width=.75\linewidth]{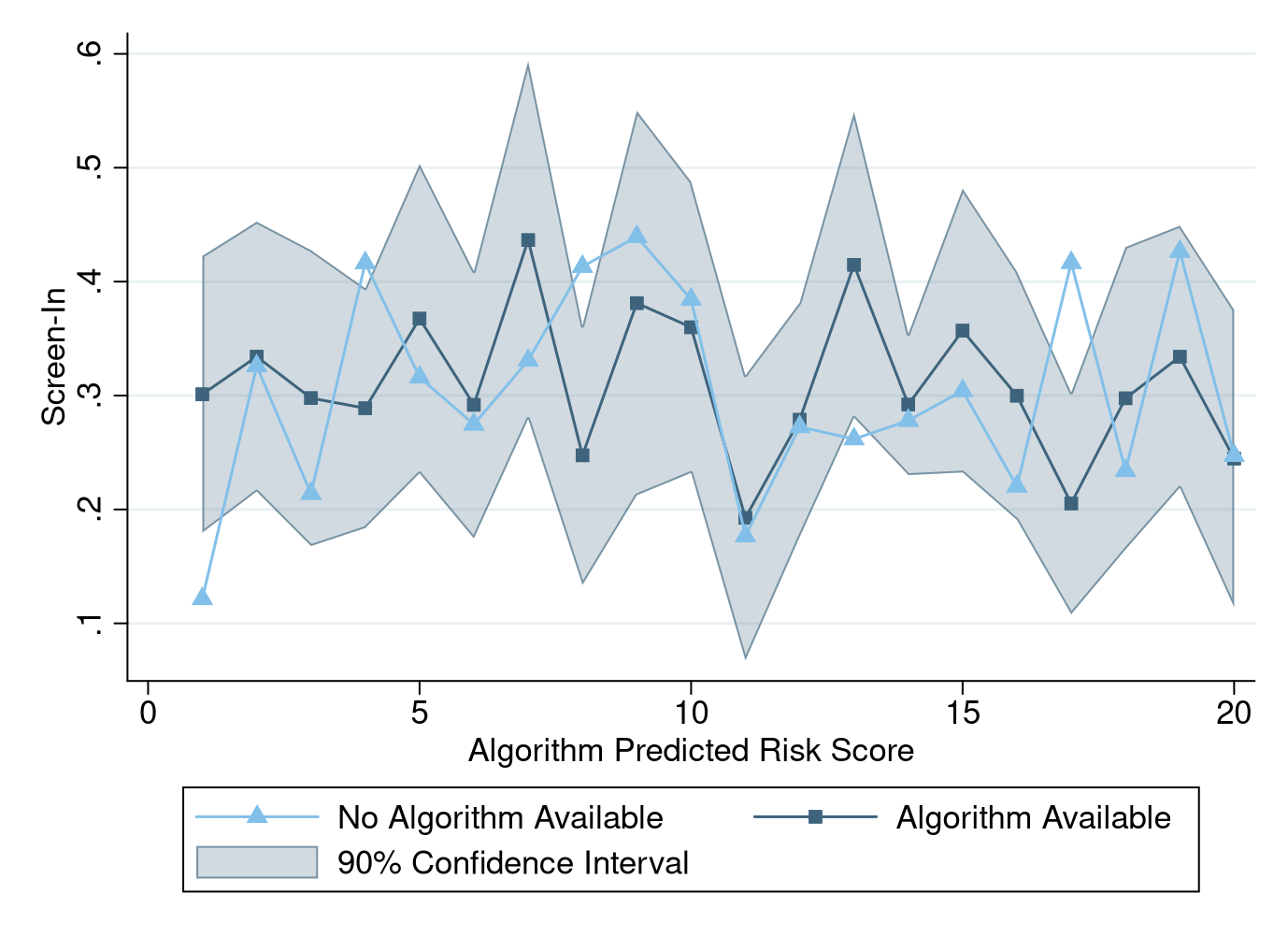}
\end{figure}
\vspace{-1em}
{\footnotesize\setlength{\parindent}{0pt}\begin{spacing}{1.0}Notes: This figure presents screen-in rates with and without the algorithm tool, by algorithm predicted risk score. It includes the full sample of child referrals. For visibility purposes, 90\% confidence intervals are displayed for the intervention group only. Randomization controls are included throughout. \end{spacing}}
\vspace{1em}

\begin{figure}[H] 
    \centering
    \caption{Effect of Tool on Margins of CPS Intervention, Interacted with Algorithm Score}
    \label{fig:cpslevel_score}
   
    \makebox[\textwidth][c]{\includegraphics[width=0.80\textwidth,trim= 10 30 10 40, clip]{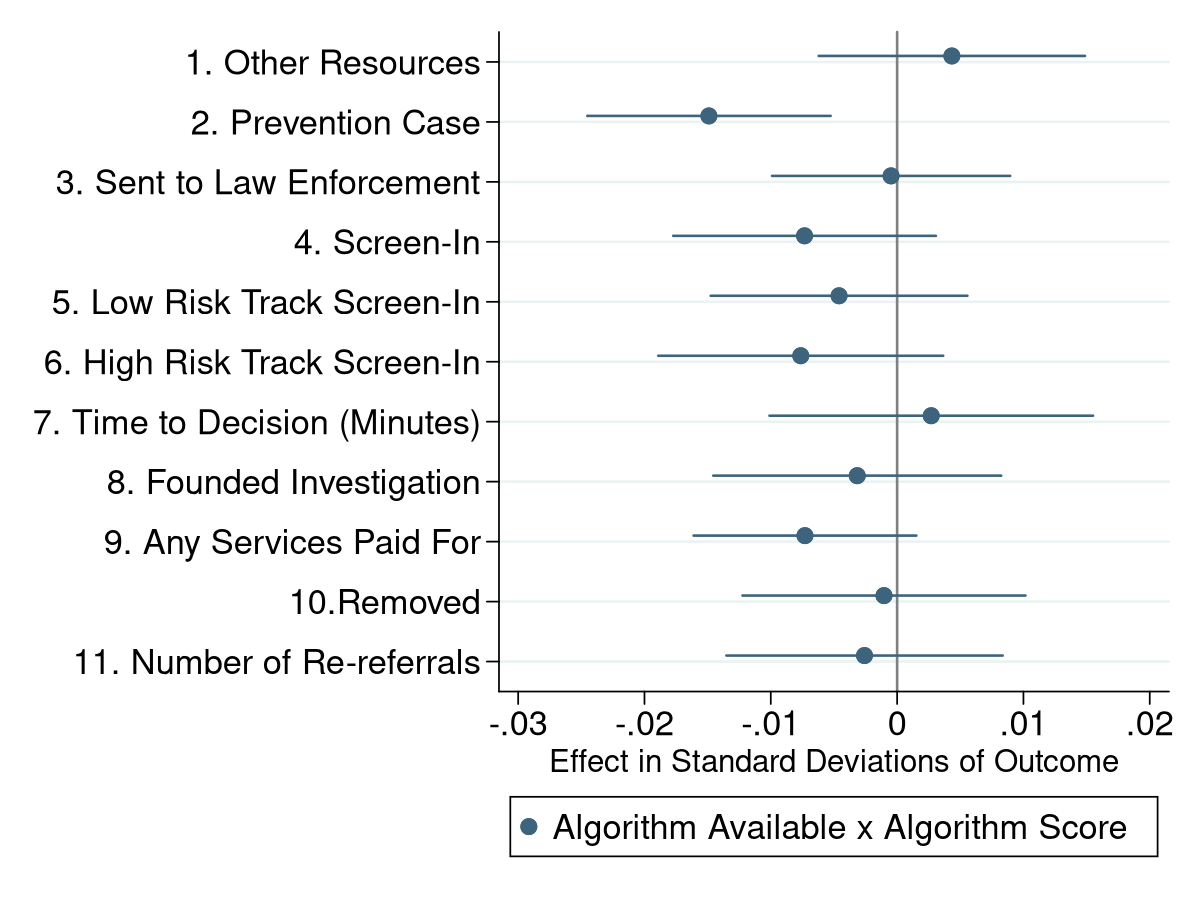}}
\end{figure}
\vspace{-1em}
{\footnotesize\setlength{\parindent}{0pt}\begin{spacing}{1.0}Notes: This figure presents point estimates and confidence intervals of regressing the listed outcomes on tool availability interacted with algorithm score for the full sample of children (unless otherwise specified). We report point estimates and standard errors for the coefficient on the interaction of tool availability with algorithm score. Following our pre-registration plan, we use the maximum score of all children on the referral, as this maximum score was most salient to workers and most of the decisions listed typically apply to all children in a household.  
 Estimates are presented with standardized outcome variables (mean 0, variance 1) for comparability of coefficient magnitudes across different outcomes. We measure margins of CPS intervention, such as high-risk-track screen-ins, using child welfare administrative data. Additional margins of intended intervention, such as voluntary resources sent to screened-out families, are inferred from team discussion notes. Some interventions are unobserved: services paid for by Medicaid, for example, are not included in administrative data. Workers can recommend opening a prevention case (row 2), can send the referral to law enforcement (row 3), and can connect the family to other resources (row 1) instead or in addition to  screening in (row 4). If workers decide to screen in a family, they also choose whether the screen-in is on a low-risk or high-risk track (rows 5 and 6). Low-risk track investigations are a type of differential response that is less intrusive for the family. Formal investigations are only run for the high-risk track. As a result, having a founded (i.e., substantiated; row 8) investigation requires having a high-risk track investigation. Row 9 shows the change in any other services paid for directly by CPS. The number of minutes it took workers to make a decision (row 7) is measured using time stamp data and is only available for observations where edits to the notes were not made after normal discussion meeting times. The final two rows (rows 10-11) report estimates on whether a child was ever removed to foster care, and the number of re-referrals. We report more details on these last two outcomes in a separate table (Appendix Table \ref{tab:CPSOutcomesTable}). Randomization procedure controls are included throughout. Standard errors are clustered at the household level and are shown at the 90 percent confidence level.\end{spacing}}
\vspace{1em}

\begin{figure}[H]
      \centering
	  \caption{Improvements in Targeting Associated with Reductions in Child Harm} 
	  \label{fig:harm_target_by_random_group}
	  \vspace{-1em}
      \makebox[\textwidth][c]{\includegraphics[width=.8\textwidth]{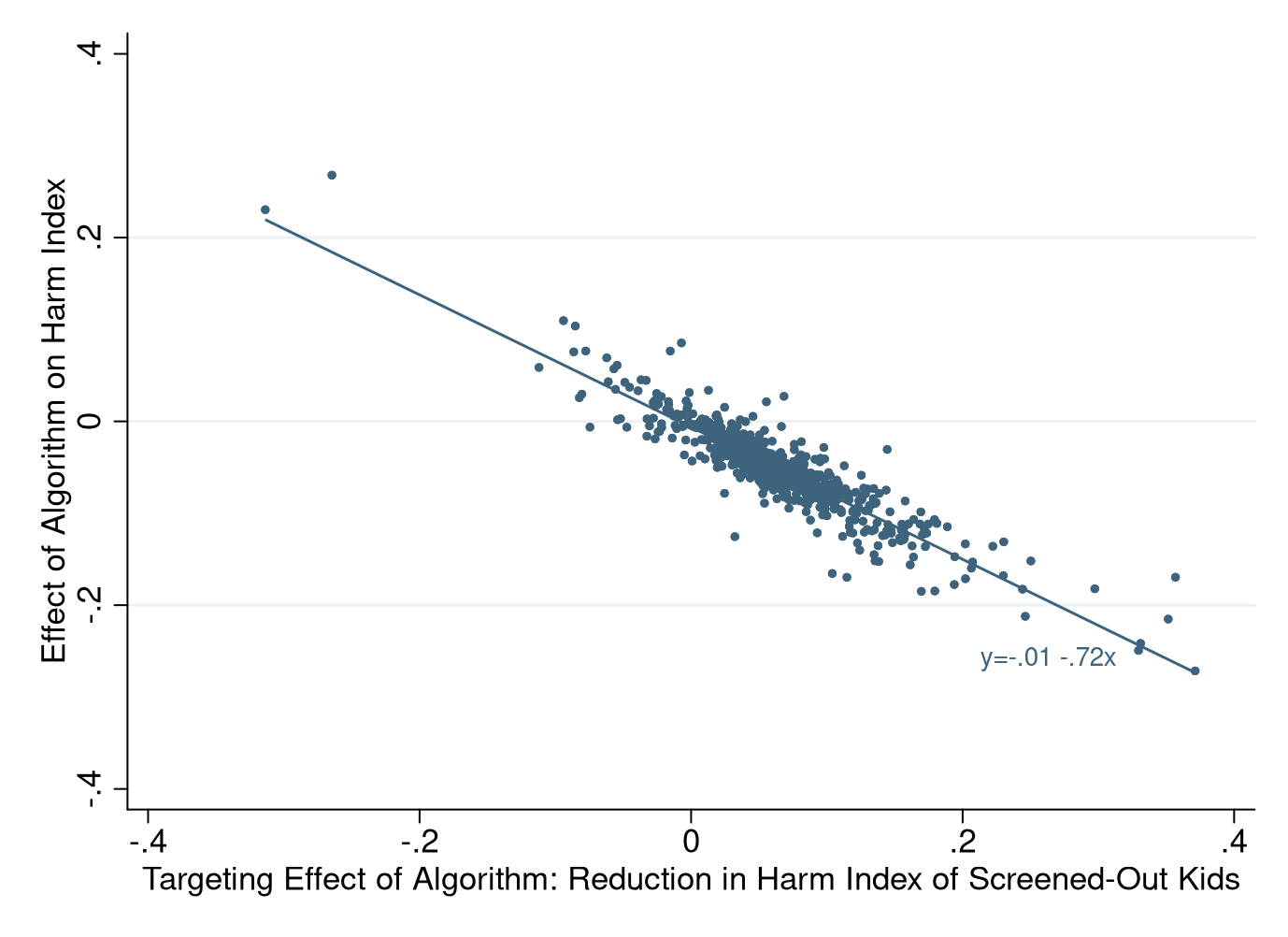}}
	\end{figure}
\vspace{-1em}
{\footnotesize\setlength{\parindent}{0pt}\begin{spacing}{1.0}Notes: We generate a thousand random groups of varying size in our data (minimum size 100 children). Within each group we then estimate effects of tool availability on targeting of screen-outs using the harm index (horizontal axis) and the effect of providing the tool on the harm index in the group (vertical axis). Each dot then represents effects of the tool on harm and targeting for each respective group. The horizontal axis measures targeting by leveraging the difference in harm after the referral among screened-out children with the tool compared to screened-out children without the tool 
(relying on random assignment to treatment and control, implying that the mean harm in treatment and control would have been similar in absence of treatment). This corresponds to the targeting test reported in Column 1 of Table \ref{tab:targeting}. We plot a fitted line and report the coefficient estimates in the bottom right of the figure next to the fitted line.  \end{spacing}}
\vspace{1em}

\begin{figure}[H]
      \centering
	  \caption{Prediction of Algorithm Support from Discussion Text Data}
      
      \textit{Using 100+ Topics Generated from Word Embeddings by BERTopic} 
	  \label{fig:BERT}
      
       \vspace{0.4em} 
      \makebox[\textwidth][c]{\includegraphics[width=0.85\textwidth, trim=10 20 0 70, clip]{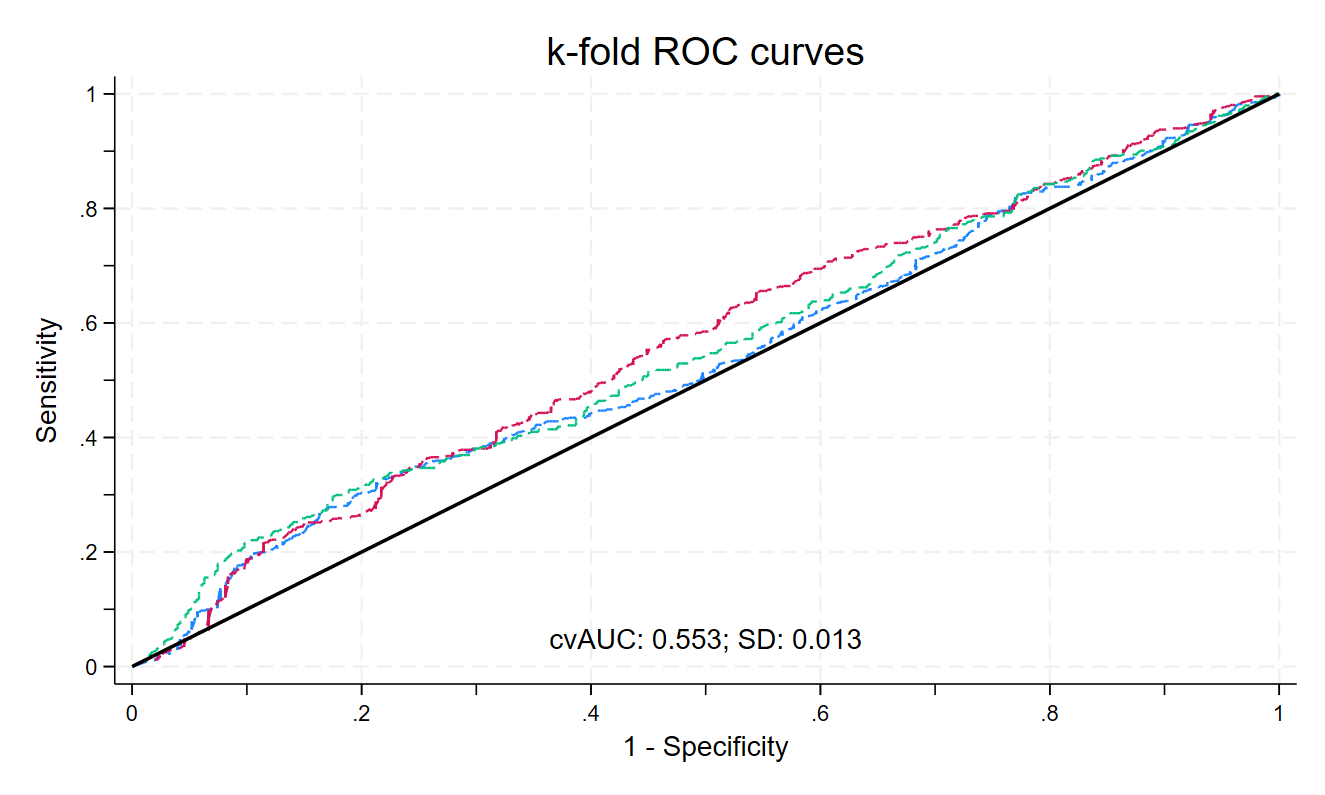}}
	\end{figure}
\vspace{-1em}
{\footnotesize\setlength{\parindent}{0pt}\begin{spacing}{1.0}Notes: This figure shows the three-fold cross-validation ROC curves on the trial sample at the referral level since discussion notes are the same for all children on the same referral ($N=2,668$). A small number of referrals from our sample that have no discussion notes are omitted from the analysis. The topics selected by the BERTopic model are used to predict whether workers had access to the algorithm tool for two sets of folds of the data, and are then validated on the remaining fold. The cross-validated mean AUC and cross-validated standard deviation are reported at the bottom of the figure. The BERTopic model was run on a set of eleven thousand referrals. It converts the text data using sentence transformers into word embeddings, which it then reduces dimensionality of using a state-of-the-art UMAP to avoid the curse of dimensionality when applying the HDBSCAN clustering algorithm for these embeddings. In practice, we used a seeded topic model to guide the model through some of the technical, profession-specific terminology used in the notes and ensure greater replicability across runs. \end{spacing}}

\vspace{8mm}
\begin{figure}[H]
      \centering
	  \caption{Oracle Best-Case Decision Rule: Targeting Screen-Ins by Ex-Post Harm}
	  \label{bestcase_counterfactual_boundAI}
	  \vspace{-1em}
      \makebox[\textwidth][c]{\includegraphics[width=0.7\textwidth, trim=10 20 0 20, clip]{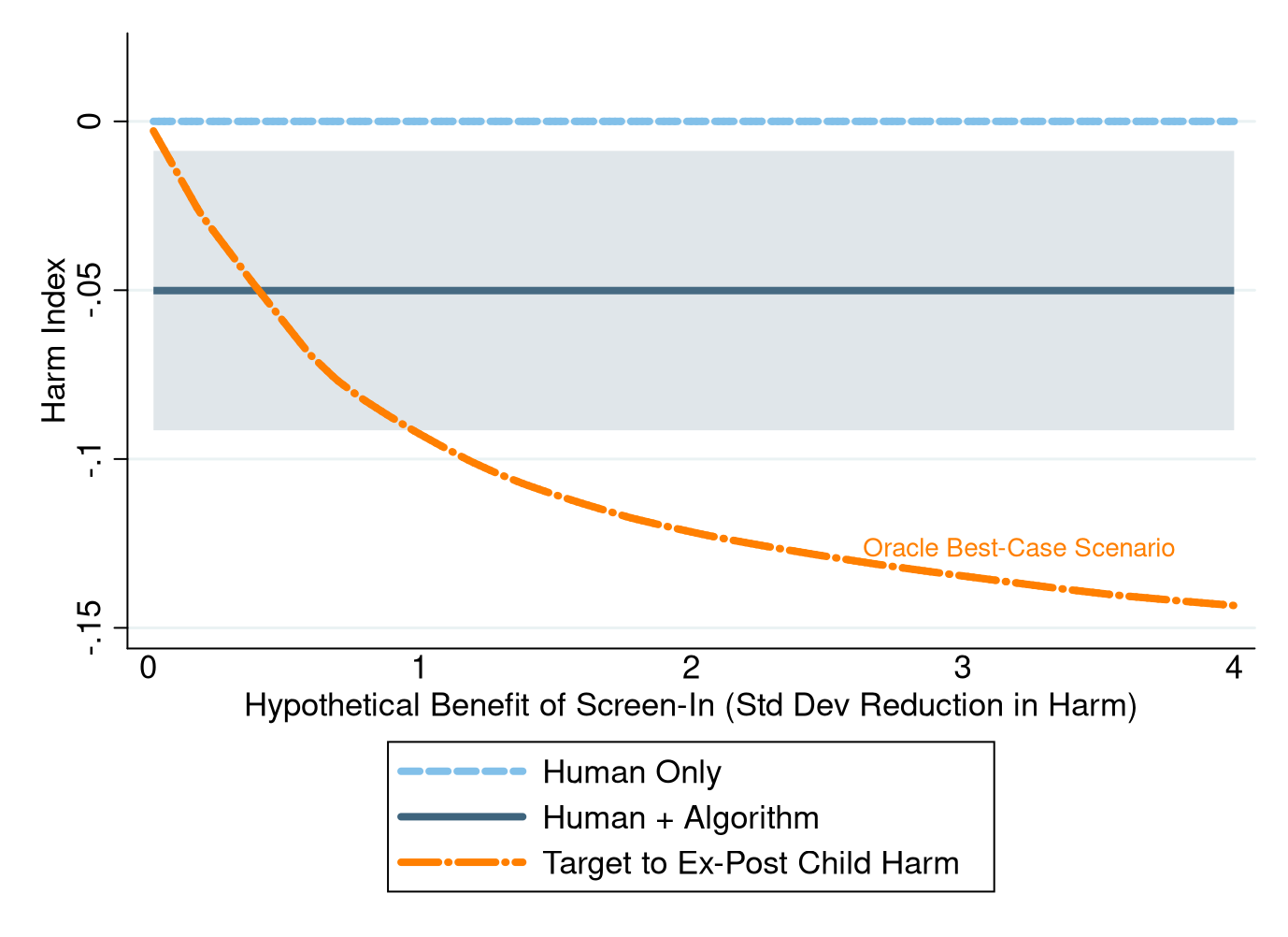}}
	\end{figure}
\vspace{-1em}
{\footnotesize\setlength{\parindent}{0pt}\begin{spacing}{1.0}Notes: This figure shows average child harm under three scenarios. The first scenario is a benchmark control group where humans do not have access to the algorithm (light blue short-dashed line).
The second scenario is when humans have access to the algorithm, which is the experiment's intervention group, shown with a 90\% confidence interval (solid dark blue line with shaded confidence bands; main estimates presented in Table \ref{SERVER_hospitaloutcomes1}). Finally, the figure shows results under a perfect (``oracle'') decision-maker subject to the same screen-in rate, where screen-ins are targeted using ex-post harm under a range of assumptions about the benefits of screen-in (orange dashed-dotted line). The best-case scenario for the oracle assumes that the children who were screened in by humans, but for whom the oracle would have screened out, would have been just as well off under the oracle decision. Effects of this new decision rule depend on hypothetical effects of a screen-in on the child harm index (range of hypothetical screen-in benefits, in terms of standard deviation reductions in harm, shown on horizontal axis).\end{spacing}}

\begin{figure}[H]
	\caption{Health Outcomes for Subgroups of Children with Algorithm-Only Counterfactual}
	    \label{fig:healthoutcomesalgoonly}
	\begin{subfigure}[b]{0.5\textwidth}
		\caption{Black Children}
		\includegraphics[trim=0 20 25 0, clip,width=\textwidth]{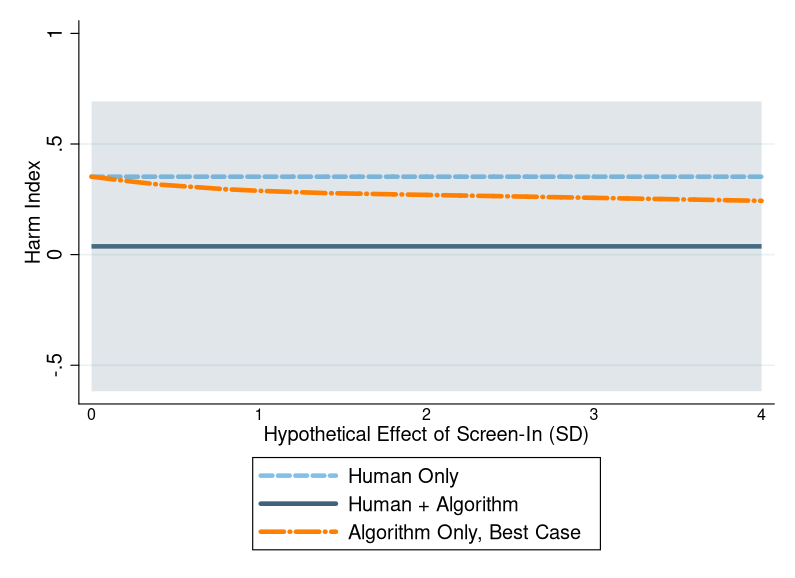}
	\end{subfigure}
	\begin{subfigure}[b]{0.5\textwidth}
		\caption{Hispanic Children}
		\includegraphics[trim=0 20 25 0, clip,width=\textwidth]{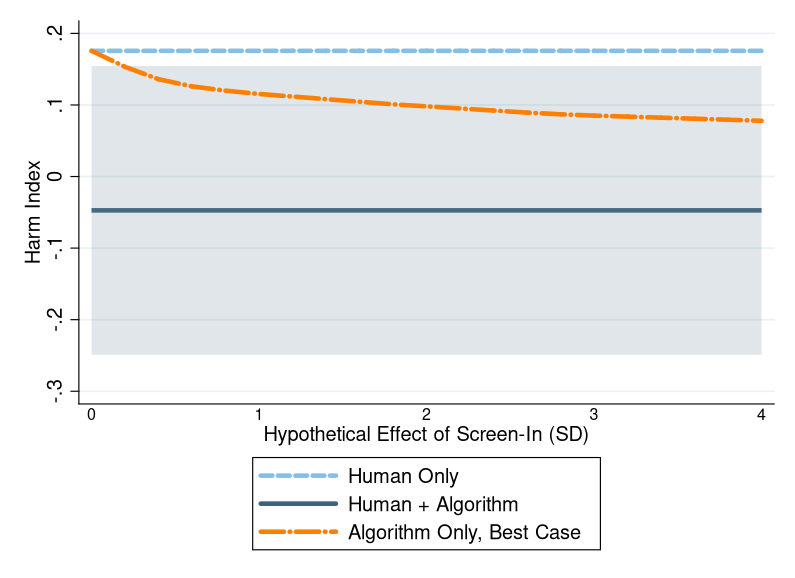}
	\end{subfigure}
 \hfill

	\begin{subfigure}[b]{0.5\textwidth}
		\caption{Girls}
		\includegraphics[trim=0 20 25 0, clip,width=\textwidth]{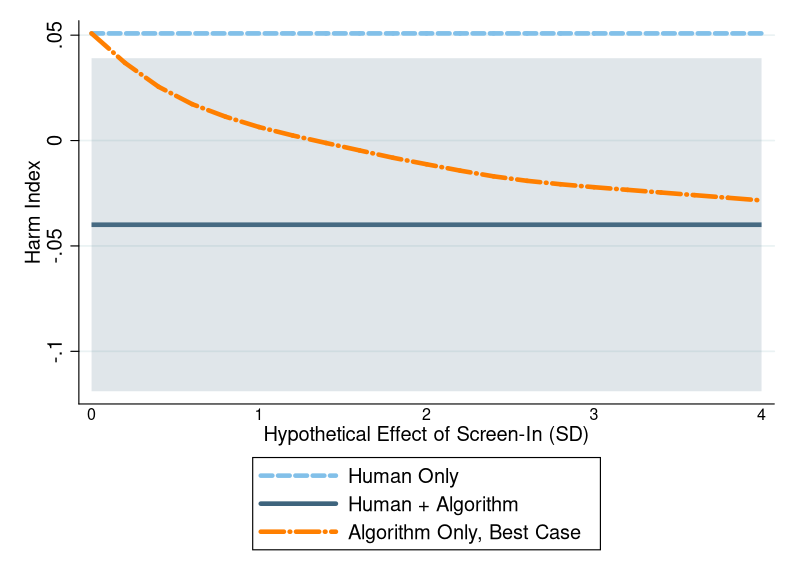}
	\end{subfigure}
	\begin{subfigure}[b]{0.5\textwidth}
		\caption{SNAP Recipients}
		\includegraphics[trim=0 20 25 0, clip,width=\textwidth]{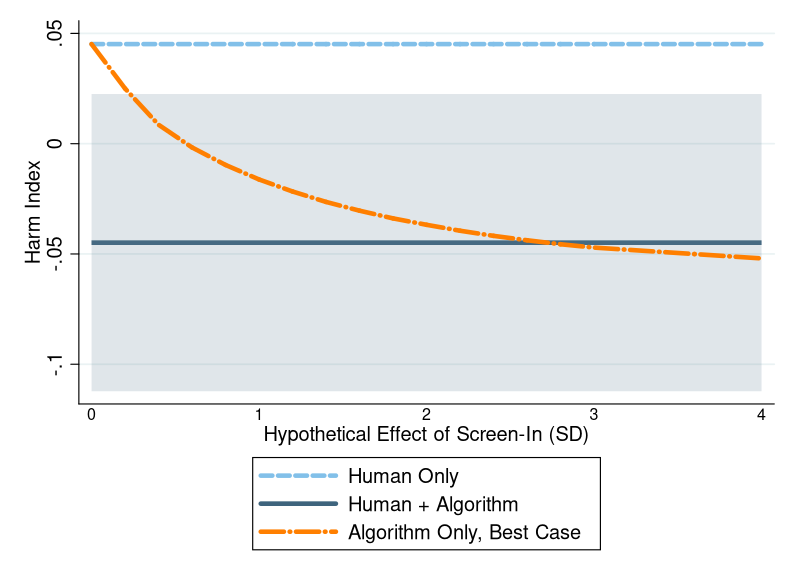}
	\end{subfigure}	
\end{figure}
\vspace{-1em}
{\footnotesize\setlength{\parindent}{0pt}\begin{spacing}{1.0}Notes: For each child subgroup as reported in the title of each figure, we plot the average child harm for three different scenarios: when humans do not have access to the algorithm (light blue short-dashed line) which is our reference/control group, when humans have access to the algorithm (solid dark blue line, with shaded area representing a 90\% confidence interval; main estimates presented in Table \ref{SERVER_hospitaloutcomes1}), and the best-case scenario for the algorithm (orange dash-dot line). The best-case scenario for the algorithm assumes that the children who were screened in by humans, but for whom the algorithm would have screened out, would have been just as well off under the algorithm decision. Effects of the algorithm screening in children whom humans screened out depends on the hypothetical benefits of a screen-in on the child harm index (horizontal axis). SNAP (food stamp) eligibility is a proxy for lower socioeconomic status.\end{spacing}}
\vspace{1em}

\begin{figure}[H]
	\caption{Health Disparities by Child Subgroup with Algorithm-Only Counterfactual}
	    \label{fig:healthdisparityalgoonly}
	\begin{subfigure}[b]{0.5\textwidth}
		\caption{Black Children vs. Others}
		\includegraphics[trim=0 20 25 20, clip,width=\textwidth]{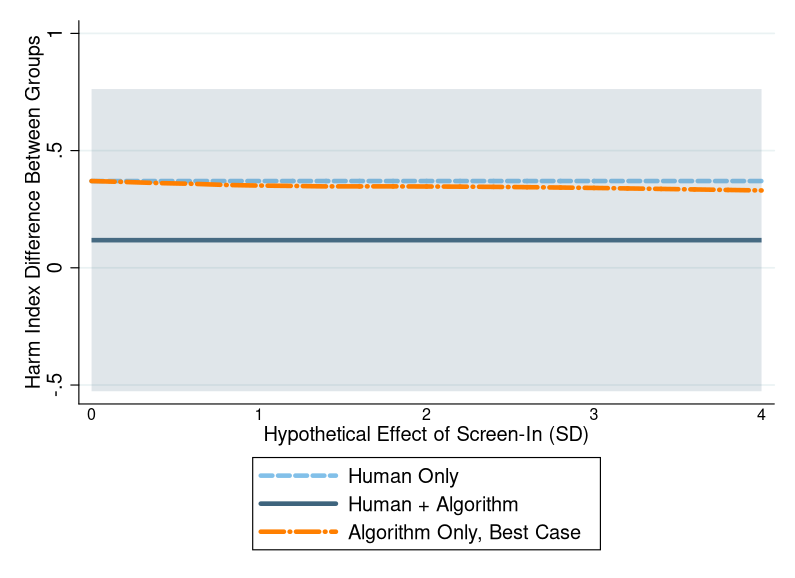}
	\end{subfigure}
	\begin{subfigure}[b]{0.5\textwidth}
		\caption{Hispanic Children vs. Others}
		\includegraphics[trim=0 20 25 20, clip,width=\textwidth]{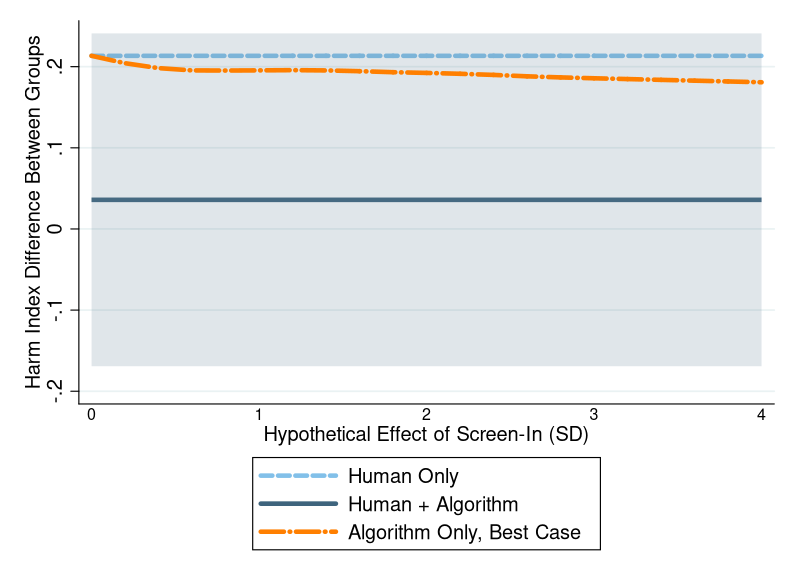}
	\end{subfigure}
 \hfill

	\begin{subfigure}[b]{0.5\textwidth}
		\caption{Girls vs. Boys}
		\includegraphics[trim=0 20 25 20, clip,width=\textwidth]{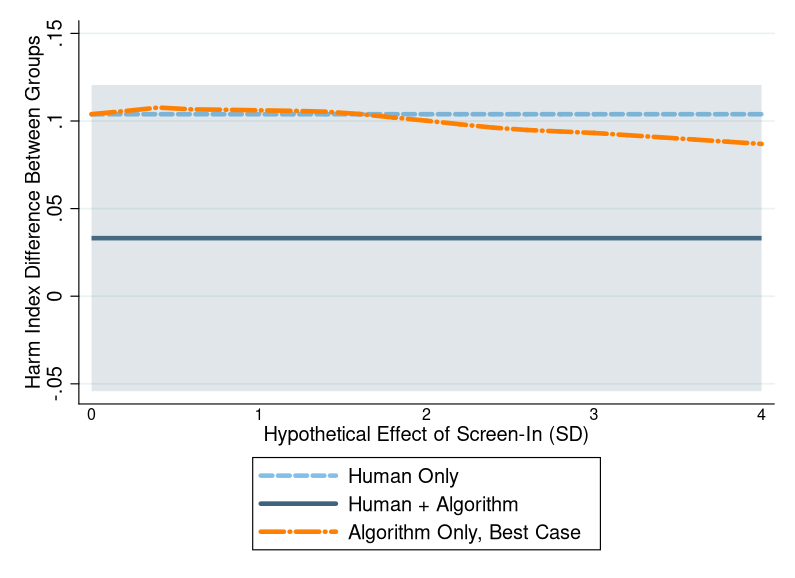}
	\end{subfigure}
	\begin{subfigure}[b]{0.5\textwidth}
		\caption{SNAP Recipients vs. Others}
		\includegraphics[trim=0 20 25 20, clip,width=\textwidth]{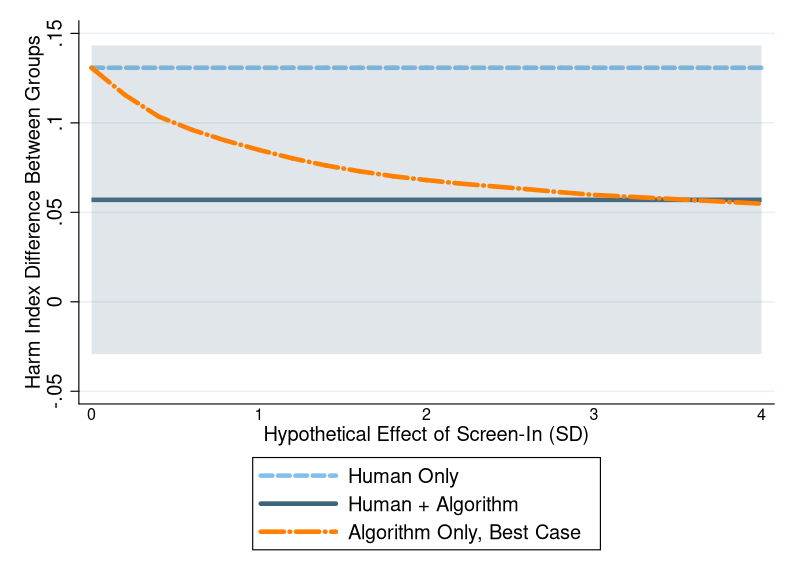}
	\end{subfigure}	
\end{figure}
\vspace{-1em}
{\footnotesize\setlength{\parindent}{0pt}\begin{spacing}{1.0}Notes: For each child subgroup as reported in the title of each figure, we plot the health disparities in child harm for three different scenarios: when humans do not have access to the algorithm (light blue short-dashed line) which is our control group, when humans have access to the algorithm (solid dark blue line, with shaded area representing a 90\% confidence interval; main estimates presented in Table \ref{SERVER_hospitaloutcomes1}), and the best-case scenario for the algorithm (orange dash-dot line). We assume that the effects of a screen-in are homogeneous across groups. The best-case scenario for the algorithm assumes that the children who were screened in by humans, but for whom the algorithm would have screened out, would have been just as well off under the algorithm decision. Effects of the algorithm screening in children who humans screened out depends on the hypothetical benefits of a screen-in on the child harm index (horizontal axis). SNAP (food stamp) eligibility is a proxy for lower socioeconomic status.\end{spacing}}
\vspace{1em}

\begin{figure}[H]
	\caption{Impacts of Algorithm on Harm Index, by Team Size}
        \centering 
	\label{fig:teamsize_indexpost30_std}
	\includegraphics[width=0.6\textwidth]{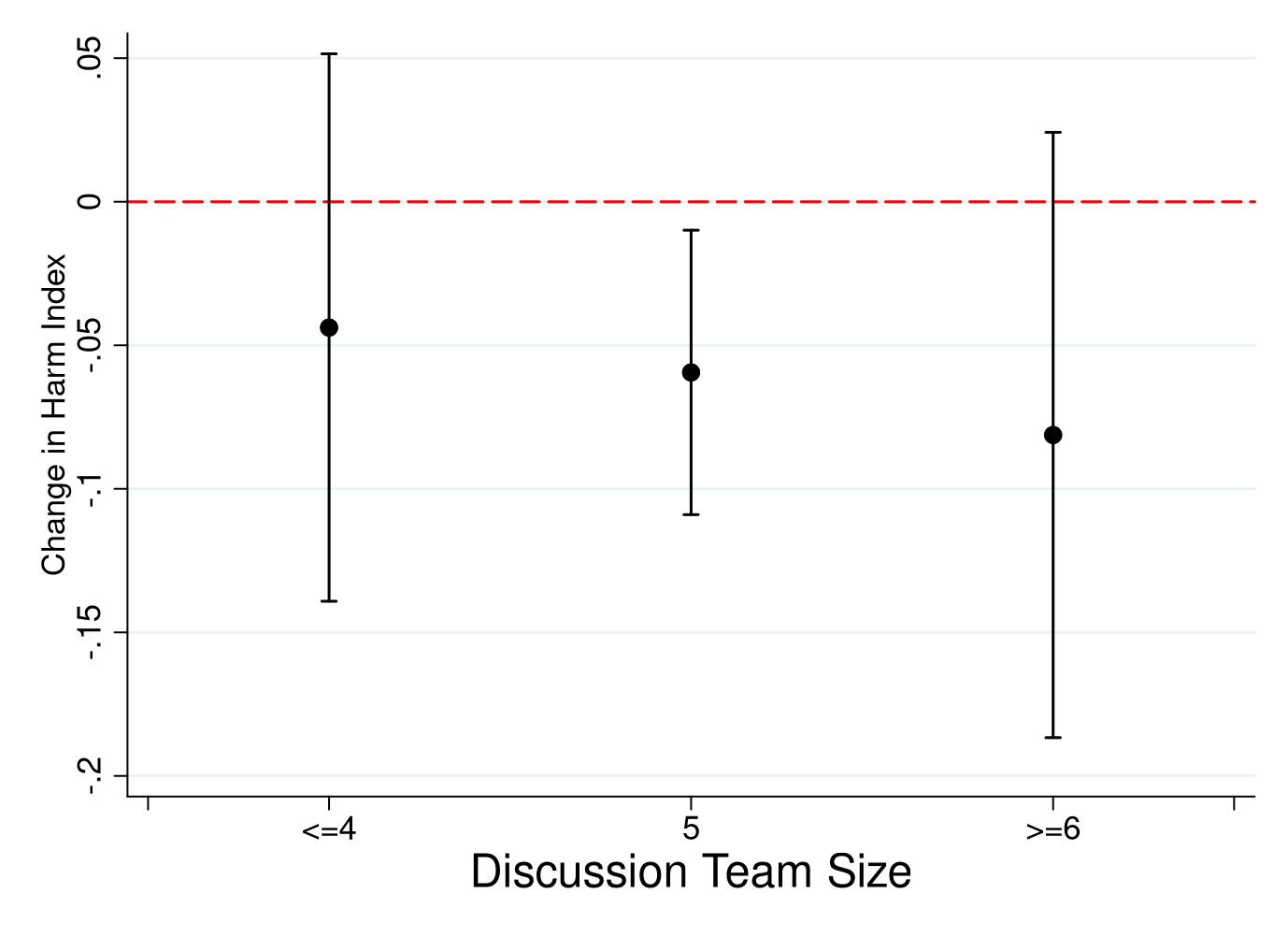}
\end{figure}
\vspace{-1em}
{\footnotesize\setlength{\parindent}{0pt}\begin{spacing}{1.0}Notes: This figure shows estimates of the impact of access to the algorithm tool on the standardized index of child harm (vertical axis), by size of the discussion team making the initial screen-in decision (horizontal axis). Randomization procedure controls are included throughout, as well as unit and month-year fixed effects. Standard errors are clustered at the household level and are shown at the 90 percent confidence level.\end{spacing}}
\vspace{1em}

\begin{figure}[H]
	\caption{Impact of Larger Team on Algorithm's Harm Reduction for Disadvantaged Groups}
        \centering 
	\label{fig:teamsize_equity_indexpost30_std}
	\includegraphics[width=0.6\textwidth]{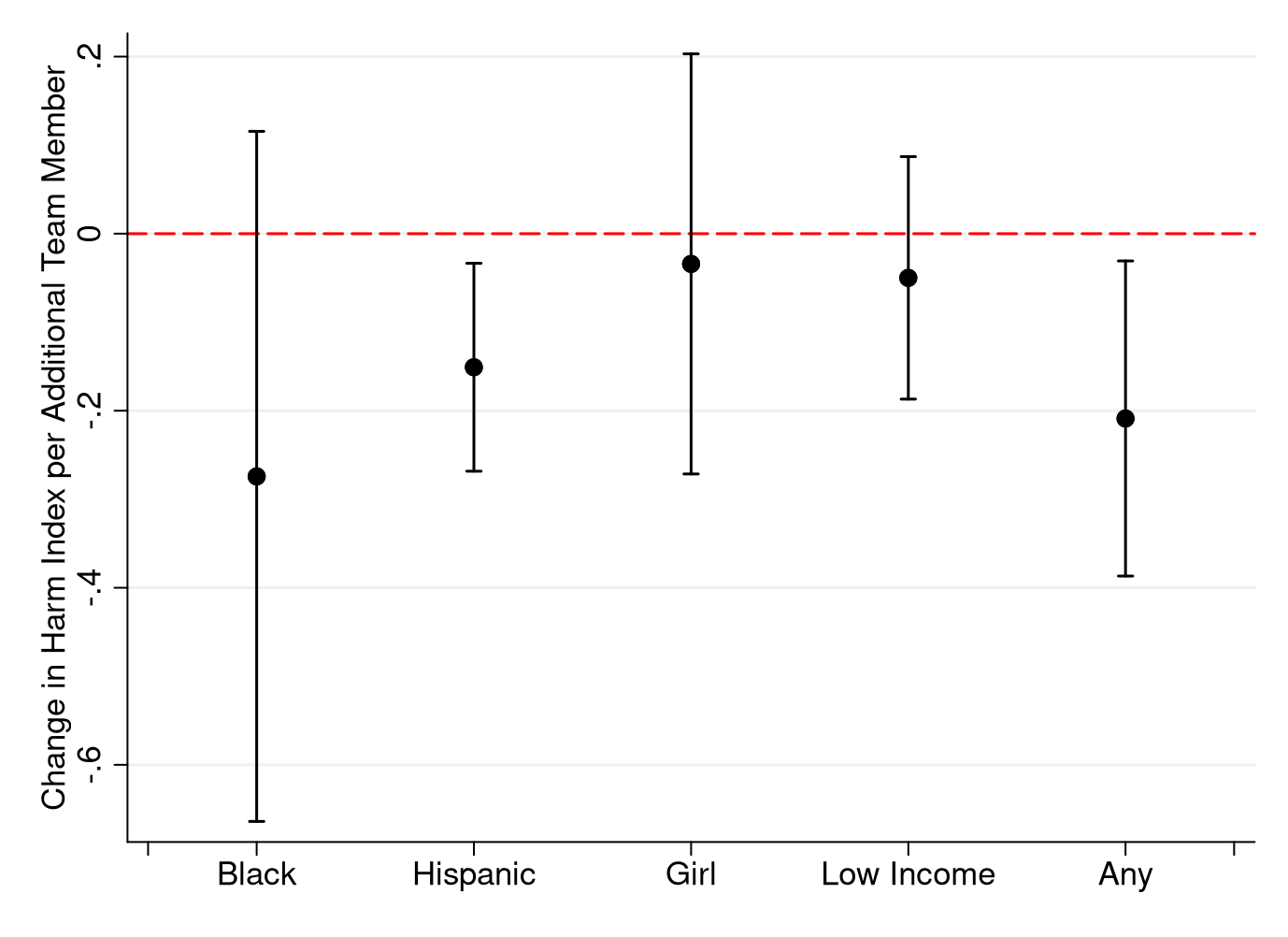}
\end{figure}
\vspace{-1em}
{\footnotesize\setlength{\parindent}{0pt}\begin{spacing}{1.0}Notes: This figure shows a disproportionate improvement in algorithm treatment effects for members of disadvantaged groups with larger teams. The standardized index of child harm is regressed on a triple-interaction of algorithm availability (binary), demographic group (binary), and team size (continuous). The vertical axis indicates the coefficient estimate and standard error bounds on the triple-interaction term, and the horizontal axis indicates the demographic group that is compared to all children not in that group. ``Any'' consists of all children in the sample who fall into at least one of the disadvantaged categories. Randomization procedure controls are included throughout, as well as unit and month-year fixed effects. Standard errors are clustered at the household level and are shown at the 90 percent confidence level.\end{spacing}}
\vspace{1em}

\newpage
\begin{table}[H]
    \centering
    \caption{Sample Restrictions for Trial Analysis Sample, November 2020 -- March 2022}
    \label{SampleSelectionTable}
        \fontsize{9.0pt}{9.0pt}\selectfont
    \begin{tabular}{l|c|c|c} \toprule
\textbf{Restrictions}& \textbf{Observations} & \textbf{Unique Children} & \textbf{Referrals}  \\
\midrule
1. Referral During Trial Period & 15,023 & 7,723 & 8,309  \\
2. Scored & 13,115 & 7,021 & 7,345  \\
\rowcolor{gray!10} 3. First Scored Referral was in Trial & 6,852 & 6,852 & 4,213  \\
4. Excluding Children Seen in Testing Phase (Phase 1) of Trial & 6,601 & 6,601 & 4,073  \\
5. Excluding Children with Household Member already in Trial & 5,998 & 5,998 & 3,631  \\
6. First Call going to Discussion Team & 5,162 & 5,162 & 3,067  \\
7. Omit if Multiple Scored Referrals before First Discussion Team Meeting & 4,919 & 4,919 & 2,935  \\
\rowcolor{gray!10} 8. Final Sample, Excluding Observations with Overwritten Status & 4,681 & 4,681 & 2,832  \\
\bottomrule
\end{tabular}

\end{table}
\vspace{-1em}
{\footnotesize\setlength{\parindent}{0pt}\begin{spacing}{1.0} Notes: This table presents the sample restrictions used to arrive at the main analysis sample. The trial ran from November 1, 2020 until March 29, 2022. Over this time period, the county received $8,309$ referrals involving $15,023$ child-referral observations. The primary unit of analysis is at the child level: children are each randomized once, during their first referral in the trial. $6,852$ unique children were randomized and successfully scored.\footnote{Nine percent of calls could not be scored due to being made during evening (non-standard) hours or being assessed as high emergency by a hotline operator. In order to generate a score for each child, the child's referral needs to be transferred to the state central data system. This happens automatically at the end of each work day: the process begins around 5pm local time and typically takes 12 to 14 hours to complete. Emergency referrals that are immediately screened in will therefore not have had time to be copied over and have a score generated. Furthermore, referrals that arrive approximately after 5pm and before 8am the following morning  will be reviewed by a team before information is copied into the main data system, and therefore will not have a score generated. On certain days, the main data system was down or the transfer process took longer than usual and the team started reviewing referrals before a score could be generated. Therefore scores are not generated for every referral. 
} 
Approximately a third of randomized children are omitted from the analysis sample for various reasons. First, we exclude $N=251$ children who were first scored during a pilot phase prior to the rollout of the full trial, as teams may have remembered prior scores for the subset of these children assigned to the control group. Children with a sibling already randomized during the trial were also excluded from the analysis, as a sibling's prior assignment could have already impacted the newly randomized child prior to their own assignment ($N=603$). We restrict our sample to standard calls assigned to the normal team meeting procedure, and therefore omit $N=836$ non-standard calls that were not seen by a normal discussion team, such as calls listing concerns about a foster family or calls sent to an overflow discussion team where the tool was not available. Given the complications with implementing randomization discussed in Appendix C, we make two further sample restrictions to guarantee random assignment. First, we omit a small subset of referrals where there were multiple calls placed for a family before a team meeting (e.g., a family referred on Friday evening, re-referred on Saturday, and both referrals seen by a team on Monday). In practice, some of these children were re-randomized with multiple treatment statuses before first being discussed, and we therefore omit the entire group ($N=243$). Second, we exclude all children whose original status was overwritten as a result of the implementation issue from Nov 2020-Mar 2021. We thus omit an additional $N=238$ children. Our remaining analysis sample consists of $4,681$ unique children from $2,832$ referrals.\end{spacing}
\vspace{2em}

\begin{table}[H]
\centering
\caption{Sample Characteristics}
\label{Desc_sampfull}
{
\def\sym#1{\ifmmode^{#1}\else\(^{#1}\)\fi}
\begin{tabular}{l*{1}{c}}
\toprule

\emph{Panel A: Child Demographics}&         \\
Female          &     0.52\\
Age at Initial Randomization&     10.5\\
White           &     0.66\\
Black           &    0.036\\
Missing Race    &     0.29\\
Hispanic        &     0.17\\
Missing Ethnicity&     0.34\\
\\\emph{Panel B: Referral Details}&         \\
N Children Listed on Referral&     2.24\\
Neglect Allegation&     0.77\\
Abuse Allegation&     0.43\\
\\\emph{Panel C: Child Experiences}&         \\
Total Referrals in Trial&     2.19\\
Any Re-referral in Trial&     0.42\\
Removed from Home during Trial&    0.022\\
Any Hospital Visit After Referral&     0.43\\
N Hospital Visits After Referral&     1.68\\
\midrule
N Unique Children&    4,681\\
N Referrals     &    2,832\\
N Family Clusters&    2,810\\
\bottomrule
\end{tabular}
}

\end{table}
 \vspace{-.5em}
{\footnotesize\setlength{\parindent}{0pt}\begin{spacing}{1.0} Notes: This table reports the mean characteristics of children who were reported to CPS and successfully randomized during the trial. Panels A and C are at the child level, while Panel B is at the referral level. 
\end{spacing}}
\vspace{2em}

\begin{table}[H]
    \centering
    \caption{First Stage: Workers Report Seeing Algorithm Score when Available}
    \label{tab:firststage}
    {
\def\sym#1{\ifmmode^{#1}\else\(^{#1}\)\fi}
\begin{tabular}{l*{2}{c}}
\toprule
                &\multicolumn{1}{c}{(1)}&\multicolumn{1}{c}{(2)}\\
                &\multicolumn{1}{c}{Wrote Down Score}&\multicolumn{1}{c}{Wrote Down Score}\\
\midrule
Algorithm Available&     0.73\sym{***}&     0.71\sym{***}\\
                &  [0.013]         &  [0.022]         \\
Algorithm Score &                  &-0.000076         \\
                &                  &[0.00039]         \\
Algorithm Available=1 $\times$ Algorithm Score&                  &   0.0014         \\
                &                  & [0.0020]         \\
Randomization Controls &      Yes         &      Yes         \\
\midrule
Control Mean    &    0.005         &    0.005         \\
Observations    &    4,681         &    4,681         \\
Randomization Clusters&    2,810         &    2,810         \\
\bottomrule
\end{tabular}
}

\end{table}
\vspace{-1em}
{\footnotesize\setlength{\parindent}{0pt}\begin{spacing}{1.0} Notes: This table presents the first stage relationship between being assigned to intervention (algorithm tool available) and the worker team recording an algorithm score in their meeting discussion notes. The estimate is likely a lower bound, as some workers may have consulted the tool without writing down the score. Randomization procedure controls are included. Standard errors are clustered at the household level. Significance reported as: * p$<$0.1, ** p$<$0.05, *** p$<$0.01.\end{spacing}} 
\vspace{1em}

\begin{table}[H]
    \centering
    \caption{Balance Test: Randomization of Access to Algorithm}
      \label{tab:balance}
    \fontsize{10pt}{10pt}\selectfont
{
\def\sym#1{\ifmmode^{#1}\else\(^{#1}\)\fi}
\begin{tabular}{l*{1}{c}}
\toprule

Female          &    0.000         \\
                &  [0.014]         \\
Age at Initial Randomization&   -0.002         \\
                &  [0.001]         \\
White           &    0.048         \\
                &  [0.069]         \\
Black           &   -0.048         \\
                &  [0.082]         \\
Missing Race    &   -0.000         \\
                &  [0.077]         \\
Hispanic        &    0.032         \\
                &  [0.029]         \\
Missing Ethnicity&    0.011         \\
                &  [0.037]         \\
Algorithm Score &   -0.003         \\
                &  [0.002]         \\
Mandated Reporter&   -0.011         \\
                &  [0.024]         \\
Child Reported in Danger&    0.038         \\
                &  [0.068]         \\
Team 3          &    0.040         \\
                &  [0.025]         \\
Team 2          &    0.005         \\
                &  [0.026]         \\
Household Size  &    0.001         \\
                &  [0.014]         \\
Number of Alleged Child Victims&    0.000         \\
                &  [0.017]         \\
Active          &   -0.014         \\
                &  [0.034]         \\
Number of Previous Referrals&    0.001         \\
                &  [0.002]         \\
Prior High Priority Hospital Visits&    0.000         \\
                &  [0.018]         \\
Prior Injury Hospital Visits&    0.037         \\
                &  [0.030]         \\
Prior Avoidable ER Hospital Visits&    0.022         \\
                &  [0.036]         \\
Prior Substance Exposure Hospital Visits&    0.007         \\
                &  [0.032]         \\
Prior Maltreatment ICD Hospital Visits&    0.036         \\
                &  [0.135]         \\
Prior Intentional Injury Hospital Visits&   -0.100         \\
                &  [0.070]         \\
Any Prior Hospital Visit&   -0.038\sym{*}  \\
                &  [0.023]         \\
Total Prior Hospital Visits&    0.003         \\
                &  [0.004]         \\
Referred in First Year of Trial&    0.002         \\
                &  [0.023]         \\
Randomization Controls &      Yes         \\
\midrule
Observations    &    4,681         \\
Randomization Clusters&    2,810         \\
F-statistic     &    0.796         \\
\textit{p}-value of F-test&    0.752         \\
\bottomrule
\end{tabular}
}

\end{table}
\vspace{-1em}
{\footnotesize\setlength{\parindent}{0pt}\begin{spacing}{1.0} Notes: This table presents a balance test that suggests stratified random assignment of intervention status to children in the sample. The $p$-value of the joint F-test is presented at the bottom, showing the (lack of) joint significance of the listed observable characteristics with respect to being assigned to the intervention group, conditioning on number of listed children for a subset of referrals with no mother and received after March 2021 (randomization procedure controls; Appendix C). Referrals are assigned to one of three teams for processing. When a case or assessment is already open, the family is said to be ``active.'' All hospitalization variables are calculated on the 180 days prior to randomization to ensure the same time span prior to the first referral in the trial for all individuals. Apart from ``Any Prior Hospital Visit'' which is a binary measure, all other hospitalization variables are counts.
Standard errors are clustered at the household level. Significance reported as: * p$<$0.1, ** p$<$0.05, *** p$<$0.01.\end{spacing}} 
\vspace{1em}

\begin{table}[H]
    \centering
	\caption{Robustness of Harm Reduction Estimates} 
\vspace{-1em}
	\label{tab:HealthRobustness}
        \fontsize{10pt}{10pt}\selectfont
\begin{center}
\begin{tabular}{lcc} \toprule 
& (1) & (2)  \\
& Harm Index & Injury \\
\hline
\addlinespace 1. Main Specification & -0.050** & -0.034* \\
& (-0.091, -0.009 ) & (-0.065, -0.003 ) \\
& [\textit{p} = 0.046] & [\textit{p}=0.068] \\
 \multicolumn{1}{r}{\textit{Permutation Test \textit{p}-value}} & [.041] & [.068]   \\
\addlinespace \multicolumn{1}{r}{\textit{2SLS IV}} & -0.069** & -0.047* \\
& (-0.125, -0.012 ) & (-0.089, -0.005 ) \\
& [\textit{p}=0.045] & [\textit{p}=0.068] \\
\addlinespace \multicolumn{1}{r}{\textit{With Controls}} & -0.046** & -0.033* \\
& (-0.082, -0.009 ) & (-0.063, -0.003 ) \\
& [\textit{p}=0.040] & [\textit{p}=0.068] \\
\addlinespace 2. 60 Day Donut & -0.055** & -0.033* \\
& (-0.096, -0.014 ) & (-0.063, -0.004 ) \\
& [\textit{p} = 0.027] & [\textit{p}=0.065] \\
\addlinespace 3. Including First 30 Days & -0.047* & -0.041** \\
& (-0.088, -0.007 ) & (-0.074, -0.008 ) \\
& [\textit{p} = 0.055] & [\textit{p}=0.042] \\
\addlinespace 4. Count of Harm Visits & -0.167* & \\
& (-0.307, -0.026 ) & \\
& [\textit{p} = 0.051] & \\
\addlinespace 5. Count of Distinct Harm Visits & -0.074 & \\
& (-0.149, 0.000 ) & \\
& [\textit{p} = 0.100] & \\
\addlinespace 6. O'Brien Weighted Index & -0.047* & \\
& (-0.087, -0.008 ) & \\
& [\textit{p} = 0.050] & \\
\addlinespace 7. First Principal Component & -0.094** & \\
& (-0.171, -0.017 ) & \\
& [\textit{p} = 0.045] & \\
\addlinespace 8. Lower Bound - Treatment non-Compliance & -0.041 & -0.029 \\
& (-0.082, 0.000 ) & (-0.059, 0.002 ) \\
& [\textit{p} = 0.101] & [\textit{p}=0.125] \\
\addlinespace 9. Upper Bound - Treatment non-Compliance & -0.140*** & -0.102*** \\
& (-0.188, -0.092 ) & (-0.138, -0.066 ) \\
& [\textit{p} = 0.000] & [\textit{p}=0.000] \\
\addlinespace 10. Clustering at the Referral Level  & -0.050**  & -0.034* \\
& (-0.091, -0.009 ) & (-0.065, -0.003 ) \\
& [\textit{p}=0.045] & [\textit{p}=0.068] \\
\bottomrule
\end{tabular}
\end{center}

\end{table}
\vspace{-1em}
{\footnotesize\setlength{\parindent}{0pt}\begin{spacing}{1.0} Notes: This table presents the robustness of estimated reductions in the child harm index with algorithm support. Column 1 shows the main effect of the algorithm tool on children's harm index, and column 2 shows effects for  injury-related hospitalizations. Row 1 presents benchmark estimates reported previously in the paper. Each subsequent row is a robustness test for these results. 
Below the first row, in brackets, we present a permutation test \textit{p}-value where we randomly permute treatment status and account for randomization controls. We also report instrumental variables (IV) estimates of the effect of the algorithm, using our first-stage estimate of checking the algorithm score (recorded by workers; likely underestimate of true first stage). Finally, we report point estimates controlling for all variables, including prior hospitalizations, listed in the balance test (Appendix Table \ref{tab:balance}). In row 2, instead of excluding index hospitalizations during the pre-specified first 30 days after the referral, we exclude a larger window of 60 days within which most investigations have concluded. Row 3 includes all index hospitalizations after a referral, including within the first 30 days. In row 4, we report effects on the sum of all harm-index-related hospitalizations, where a hospitalization for multiple categories is counted as multiple visits. In row 5, we report effects on a count of harm-index-related hospitalizations where overlapping categories are only counted once (e.g., a hospitalization for both substance exposure and injuries is counted as a single visit). In practice, this outcome is almost identical to our most frequent outcome, the number of high priority visits (see text). In row 6, we present a harm index that is not equally weighted on inputs, but rather uses weights that are optimal if the effect were to be the same on each of the outcome variables included in the index (\citealt{o1984procedures}). In row 7, instead of using the harm index as an outcome, we instead estimate the effects of algorithm access on the first principal component of a principal component analysis using the same six variables included in the harm index. In the final eighth and ninth rows, we present some approximate bounds to address treatment non-compliance (see Appendix C). In row 8, we replace the outcomes of children who ever changed treatment status to the minimum level of harm (no index hospital visits) for an extreme lower bound of treatment effect magnitudes. Estimates of access to the algorithm are still large and negative, suggesting that this small group of children do not drive the effects in our setting. Row 9 presents an extreme upper bound, showing that if these children all would have had the 99th percentile of the harm index, then estimates would be more than double the benchmark findings. 
The final row clusters standard errors at the referral level instead of at the household level. 
All regressions include randomization procedure controls, and standard errors are clustered at the household level unless otherwise specified. Significance reported as: * p$<$0.1, ** p$<$0.05, *** p$<$0.01.\end{spacing}} 
\vspace{1em}

\begin{table}[H]
    \centering
	\caption{Effects on Placebo Outcomes and Other Health Outcomes} 
    \label{tab:otheroutcomes}
    \vspace{-.25em}
\fontsize{7.0pt}{7.0pt}\selectfont
{
\def\sym#1{\ifmmode^{#1}\else\(^{#1}\)\fi}
\begin{tabular}{l*{8}{c}}
\toprule
                    &\multicolumn{5}{c}{Hospital Visits}                                            &\multicolumn{3}{c}{Placebo Specifications}     \\\cmidrule(lr){2-6}\cmidrule(lr){7-9}
                    &\multicolumn{1}{c}{(1)}   &\multicolumn{1}{c}{(2)}   &\multicolumn{1}{c}{(3)}   &\multicolumn{1}{c}{(4)}   &\multicolumn{1}{c}{(5)}   &\multicolumn{1}{c}{(6)}   &\multicolumn{1}{c}{(7)}   &\multicolumn{1}{c}{(8)}   \\
                    &\begin{tabular}{@{}c@{}}N Suggestive \\ Maltreatment \\ (Schnitzer et al.) \end{tabular}   &\begin{tabular}{@{}c@{}}Harm Index\\ Extensive Margin\end{tabular}   &\begin{tabular}{@{}c@{}}Any\\ Visit\end{tabular}   &\begin{tabular}{@{}c@{}}N Visits\end{tabular}   &\begin{tabular}{@{}c@{}}Public+Self \\ Payer Charges \end{tabular}   &\begin{tabular}{@{}c@{}}Cancer\\ Visits\end{tabular}   &\begin{tabular}{@{}c@{}}Harm when\\ Score Not Found\end{tabular}   &\begin{tabular}{@{}c@{}}Lower Priority\\ Visits\end{tabular}   \\
\midrule
Algorithm Available &     -0.0088   &      -0.038   &     -0.0053   &      -0.023   &      -199.8   &      0.0012   &       0.020   &       0.046   \\
                    &    (0.0060)   &     (0.032)   &     (0.017)   &      (0.19)   &     (157.3)   &    (0.0019)   &     (0.042)   &      (0.17)   \\
\addlinespace
Randomization Controls &         Yes   &         Yes   &         Yes   &         Yes   &         Yes   &         Yes   &         Yes   &         Yes   \\
\midrule
Control Mean        &       0.018   &      -0.000   &       0.412   &       1.586   &   1,809.463   &       0.003   &      -0.000   &       1.042   \\
Effect (\%)         &         -50   &               &          -1   &          -1   &         -11   &          44   &               &           4   \\
Design Power               &        0.69   &        0.84   &        0.63   &        0.26   &        0.63   &        0.09   &        0.13   &        0.09   \\
Observations        &       2,445   &       4,681   &       4,681   &       4,681   &       4,681   &       4,681   &       2,393   &       4,681   \\
\bottomrule
\end{tabular}
}

\end{table}
\vspace{-1em}
{\footnotesize\setlength{\parindent}{0pt}\begin{spacing}{1.0} Notes: This table reports the impact of algorithm access (i.e., estimates of equation \ref{eq:eq1}) on the following: the number of ICD codes classified by \cite{schnitzer2011identification} as suggestive of maltreatment for children aged 0-9 years old; an extensive margin harm index where rather than using the number of visits for each component of the harm index we instead use binary indicators for whether there was any such visit after the referral; any hospital visit; a child's number of hospital visits; and public or self payer charges (columns 1-5). Charges are top-coded at the 95th percentile. Placebo specifications where the tool should have no impact are reported in columns 6-8. In column 6 we report the effect on cancer-related hospitalizations, a placebo outcome which was pre-registered but extremely rare. In column 7 we report the effect on treated children's harm index relative to the control group, when workers tried to see the score but were unable (11 percent of treatment group). Finally, column 8 reports impacts on lower-priority hospital visits, which are less likely to be related to maltreatment and should be less affected by the tool. Randomization procedure controls are included throughout. Standard errors are clustered at the household level. Significance reported as: * p$<$0.1, ** p$<$0.05, *** p$<$0.01.\end{spacing}}
\vspace{1em}

\begin{table}[H]
    \centering
	\caption{Effect of Algorithm Tool on CPS Outcomes}
    \label{tab:CPSOutcomesTable}
    \vspace{-0.0em}
    \fontsize{10.0pt}{10.0pt}\selectfont
{
\def\sym#1{\ifmmode^{#1}\else\(^{#1}\)\fi}
\begin{tabular}{l*{4}{c}}
\toprule
                    &\multicolumn{1}{c}{(1)}   &\multicolumn{1}{c}{(2)}   &\multicolumn{1}{c}{(3)}   &\multicolumn{1}{c}{(4)}   \\
                    &Any Re-Referral   &N of Re-Referrals   & Any Removal   &N of Removals   \\
\midrule
Algorithm Available &      -0.018   &       -0.10   &      0.0027   &      0.0079   \\
                    &     (0.020)   &      (0.14)   &    (0.0073)   &     (0.010)   \\
\addlinespace
Randomization Controls &         Yes   &         Yes   &         Yes   &         Yes   \\
\midrule
Control Mean        &       0.612   &       2.162   &       0.035   &       0.051   \\
Observations        &       4,681   &       4,681   &       4,681   &       4,681   \\
\bottomrule
\end{tabular}
}

\end{table}
\vspace{-1em}
{\footnotesize\setlength{\parindent}{0pt}\begin{spacing}{1.0} Notes: This table reports the impact of algorithm access (i.e., estimates of equation \ref{eq:eq1}) on whether a child was ever re-referred, the number of re-referrals, whether a child was ever removed to foster care, and the number of removal episodes. 
Randomization procedure controls are included throughout. Standard errors are clustered at the household level. Significance reported as: * p$<$0.1, ** p$<$0.05, *** p$<$0.01.\end{spacing}}
\vspace{1em}

\begin{table}[H]
    \centering
	\caption{Test of Within-Day Spillovers from Tool Availability}
    \label{spillovers_withinday}
    \fontsize{10pt}{10pt}\selectfont
    {
\def\sym#1{\ifmmode^{#1}\else\(^{#1}\)\fi}
\begin{tabular}{l*{3}{c}}
\toprule
                    &\multicolumn{1}{c}{(1)}   &\multicolumn{1}{c}{(2)}   &\multicolumn{1}{c}{(3)}   \\
                    &   Screen-In   &\begin{tabular}{@{}c@{}}Time to Decision \\ (Minutes) \end{tabular}   &  Harm Index   \\
\midrule
Algorithm Available &       0.001   &      -0.998   &      -0.055** \\
                    &     (0.021)   &     (0.879)   &     (0.026)   \\
\addlinespace
Peer Share with Algorithm Available in Day (Std Dev)&      -0.005   &      -0.387   &       0.001   \\
                    &     (0.015)   &     (0.744)   &     (0.020)   \\
\addlinespace
Algorithm Available $\cdot$ Peer Share (Std Dev)&       0.021   &       0.427   &      -0.018   \\
                    &     (0.020)   &     (0.976)   &     (0.023)   \\
\addlinespace
Randomization Controls &         Yes   &         Yes   &         Yes   \\
\midrule
Mean of Outcome Variable&        0.30   &       20.01   &       -0.04   \\
Observations        &       4,501   &       4,501   &       4,501   \\
\bottomrule
\end{tabular}
}

\end{table}
\vspace{-1em}
{\footnotesize\setlength{\parindent}{0pt}\begin{spacing}{1.0} Notes: This table tests for the presence of within-day spillovers (i.e., experimental contamination from treatment to control) from tool availability. If such within-day spillovers were present, one may expect changes in attention (screen-in or time) to children in the control group -- and similar effects of a smaller magnitude for children in the treatment group -- on days where workers saw a higher percent of children for whom the algorithm tool was available. The share of referrals with the tool available is constructed at the team-day level and excludes a child's own referral (leave-one-out, i.e., jackknife share). Days with only one referral are excluded. Time to decision is constructed using timestamp data which is available for a large subset of referrals and is top-coded at 60 minutes. Randomization procedure controls are included. 
Standard errors are clustered at the household level. Significance reported as: * p$<$0.1, ** p$<$0.05, *** p$<$0.01.\end{spacing}}
\vspace{1em}

\begin{table}[H]
    \centering
	\caption{Effect on Harm Disparities \vspace{0.25em} \\
	{\footnotesize \textit{Outcome: Child Harm Index. Type: Group in Column Title.}}}
	\label{tab:healthdisparity}
	\vspace{-0em}
    {
\def\sym#1{\ifmmode^{#1}\else\(^{#1}\)\fi}
\begin{tabular}{l*{4}{c}}
\toprule
                    &\multicolumn{1}{c}{(1)}   &\multicolumn{1}{c}{(2)}   &\multicolumn{1}{c}{(3)}   &\multicolumn{1}{c}{(4)}   \\
                    &       Black   &    Hispanic   &        Girl   &        SNAP   \\
\midrule
Type                &        0.38   &        0.18*  &       0.092** &        0.11***\\
                    &      (0.38)   &      (0.10)   &     (0.041)   &     (0.040)   \\
\addlinespace
Algorithm Available &      -0.038*  &      -0.025   &      -0.011   &      -0.011   \\
                    &     (0.020)   &     (0.021)   &     (0.024)   &     (0.028)   \\
\addlinespace
Algorithm Available $\cdot$ Type&       -0.23   &       -0.15   &      -0.074   &      -0.061   \\
                    &      (0.38)   &      (0.11)   &     (0.045)   &     (0.045)   \\
\addlinespace
Randomization Controls &         Yes   &         Yes   &         Yes   &         Yes   \\
\midrule
Observations in Group&         170   &         816   &       2,452   &       2,977   \\
Observations        &       4,681   &       4,681   &       4,681   &       4,681   \\
Effect on Disparity (\%)&         -62   &         -82   &         -80   &         -55   \\
\textit{p}-value of Disparity Change&       0.546   &       0.164   &       0.101   &       0.172   \\
\textit{p}-value of Joint Test & \multicolumn{4}{c}{.2136} \\
\bottomrule
\end{tabular}
}

\end{table}
\vspace{-1em}
{\footnotesize\setlength{\parindent}{0pt}\begin{spacing}{1.0} Notes: This table presents estimates of algorithm availability on disparities in the harm index for Black, Hispanic, female, and SNAP-recipient (lower income) children in columns 1-4, respectively, compared to other children in the sample. We report results from regressions of the child harm index on an interacted model of group type (reported in the column titles) with a binary indicator for whether the tool was available.  
For example, the coefficient in row 1 of column 1 indicates that Black children had a 0.38 greater harm index than other children when the tool was not available. Algorithm tool availability reduced harm by 0.038 standard deviations for other children (row 2) and an additional 0.23 for Black children (row 3).  
The number of observations in each demographic group is listed at the bottom of the table. The table also reports the percentage decrease in the harm disparity based on point estimates. The second-to-last row from the very bottom lists the \textit{p}-values from a test of whether health disparities are the same with and without the tool, and the final row reports the \textit{p}-value of a joint test that none of the health disparities changed.   Randomization procedure controls are included throughout. Standard errors are clustered at the household level. Significance reported as: * p$<$0.1, ** p$<$0.05, *** p$<$0.01.\end{spacing}}

\begin{table}[H]
    \centering
	\caption{Control Group Disparities in Screen-Ins (Investigations)} 
    \label{tab:screenintable}
    {
\def\sym#1{\ifmmode^{#1}\else\(^{#1}\)\fi}
\begin{tabular}{l*{4}{c}}
\toprule
                    &\multicolumn{1}{c}{(1)}   &\multicolumn{1}{c}{(2)}   &\multicolumn{1}{c}{(3)}   &\multicolumn{1}{c}{(4)}   \\
                    &       Black   &    Hispanic   &        Girl   &        SNAP   \\
\midrule
Screen-In Disparity &        0.19** &      -0.013   &      -0.026   &       0.049   \\
                    &     (0.077)   &     (0.037)   &     (0.020)   &     (0.032)   \\
\addlinespace
Algorithm Score Control &         Yes   &         Yes   &         Yes   &         Yes   \\
\midrule
Reference Group Control Mean&       0.297   &       0.296   &       0.305   &       0.274   \\
Observations        &       2,098   &       2,098   &       2,098   &       2,098   \\
\bottomrule
\end{tabular}
}

\end{table}
\vspace{-1em}
{\footnotesize\setlength{\parindent}{0pt}\begin{spacing}{1.0} Notes: This table reports control group disparities in screen-ins (investigations) across four demographic categories. Column 1 reports how much more likely Black children are to be screened in relative to non-Black (primarily white) children. Columns 2-4 report disparities for Hispanic children, girls, and children whose families are SNAP recipients (lower income). All specifications include a control for algorithm risk score to compensate for differences in baseline risk across groups. Standard errors are clustered at the household level. Significance reported as: * p$<$0.1, ** p$<$0.05, *** p$<$0.01.\end{spacing}}
\vspace{1em}

\begin{table}[H]
    \centering
	\caption{Robustness Checks for Decrease in Racial Screen-In Disparities \\ Among Children Scored by the Algorithm as Low Risk}
    \label{tab:RaceRobustness}
    \vspace{-0em}
        \fontsize{8.5pt}{8.5pt}\selectfont
    {
\def\sym#1{\ifmmode^{#1}\else\(^{#1}\)\fi}
\begin{tabular}{l*{7}{c}}
\toprule
                    &\multicolumn{1}{c}{(1)}   &\multicolumn{1}{c}{(2)}   &\multicolumn{1}{c}{(3)}   &\multicolumn{1}{c}{(4)}   &\multicolumn{1}{c}{(5)}   &\multicolumn{1}{c}{(6)}   &\multicolumn{1}{c}{(7)}   \\
                    &        Main   &     Overall   &\begin{tabular}{@{}c@{}}No \\ Hispanic\end{tabular}   &\begin{tabular}{@{}c@{}}No Missing \\ Race\end{tabular}   &\begin{tabular}{@{}c@{}}Missing \\ as Black\end{tabular}   &\begin{tabular}{@{}c@{}}Hospital \\ Race\end{tabular}   &\begin{tabular}{@{}c@{}}Score \\ 1-20\end{tabular}   \\
\midrule
Algorithm Available (T)&       0.033   &       0.008   &       0.033   &       0.049   &       0.059   &       0.030   &       0.045   \\
                    &     (0.028)   &     (0.021)   &     (0.029)   &     (0.044)   &     (0.044)   &     (0.028)   &     (0.033)   \\
                    &     [0.238]   &               &               &               &               &               &               \\
Black Child (B)     &       0.426***&       0.201***&       0.487***&       0.240*  &      -0.284***&       0.259** &       0.296** \\
                    &     (0.129)   &     (0.076)   &     (0.131)   &     (0.142)   &     (0.037)   &     (0.124)   &     (0.151)   \\
                    &     [0.008]   &               &               &               &               &               &               \\
Above Median Score (S)&       0.048*  &               &       0.058*  &      -0.071*  &      -0.077** &       0.046   &       0.004*  \\
                    &     (0.029)   &               &     (0.031)   &     (0.039)   &     (0.039)   &     (0.029)   &     (0.002)   \\
                    &     [0.180]   &               &               &               &               &               &               \\
T$\cdot$ S          &      -0.043   &               &      -0.067   &      -0.076   &      -0.083   &      -0.039   &      -0.004   \\
                    &     (0.040)   &               &     (0.043)   &     (0.053)   &     (0.053)   &     (0.040)   &     (0.003)   \\
                    &     [0.282]   &               &               &               &               &               &               \\
B$\cdot$ S          &      -0.314** &               &      -0.371** &      -0.148   &       0.180***&      -0.155   &      -0.011   \\
                    &     (0.154)   &               &     (0.161)   &     (0.166)   &     (0.057)   &     (0.148)   &     (0.010)   \\
                    &     [0.029]   &               &               &               &               &               &               \\
T$\cdot$ B          &      -0.402** &      -0.130   &      -0.426** &      -0.421** &      -0.084*  &      -0.219   &      -0.231   \\
                    &     (0.178)   &     (0.109)   &     (0.186)   &     (0.181)   &     (0.050)   &     (0.164)   &     (0.201)   \\
                    &     [0.030]   &               &               &               &               &               &               \\
T$\cdot$ S $\cdot$ B&       0.399*  &               &       0.431*  &       0.421*  &       0.113   &       0.185   &       0.011   \\
                    &     (0.220)   &               &     (0.229)   &     (0.223)   &     (0.082)   &     (0.203)   &     (0.014)   \\
                    &     [0.066]   &               &               &               &               &               &               \\
Randomization Controls &         Yes   &         Yes   &         Yes   &         Yes   &         Yes   &         Yes   &         Yes   \\
\midrule
Reference Group Control Mean&       0.267   &       0.267   &       0.261   &       0.418   &       0.418   &       0.268   &       0.294   \\
Observations        &       4,681   &       4,681   &       3,865   &       3,321   &       4,681   &       4,681   &       4,681   \\
\bottomrule
\end{tabular}
}

\end{table}
\vspace{-1em}
{\footnotesize\setlength{\parindent}{0pt}\begin{spacing}{1.0} Notes: This table shows estimates of racial disparities in screen-in across a variety of specifications. Columns 1 and 2 provide estimates on the change in disparities for Black vs. non-Black (or unknown race)
screen-in rates when the algorithm tool is made available. Columns 1 and 2 code missing race as white. We report permutation \textit{p}-values in brackets in column 1. Column 3 is the same specification as column 1 but excludes Hispanic children. Column 4 excludes children where information about race was missing. Column 5 recodes children with missing race as Black. Column 6 complements CPS-recorded race records with race information from hospital records. Finally, Column 7 replaces interaction with a binary indicator (above/below median algorithm score) with instead an interaction with a linear algorithm score from 1-20 (pre-registered specification). 
Randomization procedure controls are included. Standard errors are clustered at the household level. Significance reported as: * p$<$0.1, ** p$<$0.05, *** p$<$0.01.\end{spacing}}
\vspace{1em}

\begin{table}[H]
\centering
\fontsize{10pt}{10pt}\selectfont
\caption{Mechanisms: Theories that are Inconsistent with Findings}
\label{table_theories}
\renewcommand{\arraystretch}{1.0} 
\setlength{\tabcolsep}{4pt}      
\begin{tabular}{|p{0.4\linewidth}|p{0.58\linewidth}|}
\specialrule{0.5pt}{0pt}{0pt} 
\rowcolor[rgb]{0.902,0.902,0.902} 
\textbf{Theory} & \textbf{Evidence Against Theory} \\ 
\specialrule{0.5pt}{0pt}{0pt} 
\vspace*{0pt}Workers do not use algorithmic tool, or make arbitrary decisions with the tool. 
& \begin{itemize}[leftmargin=*, topsep=0pt, itemsep=0pt]
    \item Workers record tool use (Table \ref{tab:firststage}).
    \item Reductions in child harm (Table \ref{SERVER_hospitaloutcomes1}).
    \item Reductions in disparities (Figures \ref{fig:HealthDisparity}, \ref{fig:screeninrace}).
    \item Change in discussion content toward topics complementary to tool inputs (Figure \ref{fig:discussioncontent}).
\end{itemize} \\ 
\hline
\vspace*{0pt}Workers err on the side of more intervention. 
& \begin{itemize}[leftmargin=*, topsep=0pt, itemsep=0pt]
    \item No differences in levels or types of CPS interventions (Figure \ref{fig:cpslevel}).
\end{itemize} \\ 
\hline
\vspace*{0pt}Workers mechanically follow tool predictions. 
& \begin{itemize}[leftmargin=*, topsep=0pt, itemsep=0pt]
    \item No increase in CPS interventions for children with high algorithm risk scores (Figures \ref{fig:screeninbyscore}, \ref{fig:cpslevel_score}).
\end{itemize} \\ 
\hline
\vspace*{0pt}Workers exhibit algorithm aversion. 
& \begin{itemize}[leftmargin=*, topsep=0pt, itemsep=0pt]
    \item No reduction in CPS intervention for children with high algorithm scores (Figures \ref{fig:screeninbyscore}, \ref{fig:cpslevel_score}).
    \item No qualitative evidence of tool aversion expressed by workers to researchers (independent from tool developers).
    \item Reductions in child harm (Table \ref{SERVER_hospitaloutcomes1}).
    \item Reductions in disparities (Figures \ref{fig:HealthDisparity}, \ref{fig:screeninrace}).
\end{itemize} \\ 
\hline
\vspace*{0pt}Experimenter demand effects. 
& \begin{itemize}[leftmargin=*, topsep=0pt, itemsep=0pt]
    \item No differential screen-in (or other provision of services) by level of score (Figures \ref{fig:screeninbyscore}, \ref{fig:cpslevel_score}).
\end{itemize} \\ 
\hline
\vspace*{0pt}Workers reallocate CPS-adjacent services to high-risk children. 
& \begin{itemize}[leftmargin=*, topsep=0pt, itemsep=0pt]
    \item Limited evidence of changes in levels/targeting of CPS-adjacent services (family visitor, prevention, law enforcement, community response) (Figures \ref{fig:cpslevel}, \ref{fig:cpslevel_score}).
    \item No evidence that these account for health effects.
\end{itemize} \\ 
\hline
\vspace*{0pt}Workers change behavior (e.g., exert more effort), for instance because they know they are being observed. 
& \begin{itemize}[leftmargin=*, topsep=0pt, itemsep=0pt]
    \item Workers know they are being observed regardless of whether tool is shown.
    \item Tool access is random for each family and randomized within a day, so salience is comparable between intervention and control.
    \item No apparent spillovers (e.g., effort crowd-out) on control group outcomes on days with more treated families (Table \ref{spillovers_withinday}).
\end{itemize} \\ 
\hline
\vspace*{0pt}Investigator behavior (conditional on screen-in) influenced by tool. 
& \begin{itemize}[leftmargin=*, topsep=0pt, itemsep=0pt]
    \item Tool not available to investigators unless they read through group meeting discussion notes.
    \item No evidence of changes in investigator behavior based on level of score.
    \item Improvement in outcomes for screened-out (non-investigated) children: Evidence that gains come at least in part from improved targeting of screen-ins prior to investigation (Table \ref{tab:targeting}; Figures \ref{fig:harm_target_by_group} and \ref{fig:harm_target_by_random_group}).
\end{itemize} \\ 
\hline
\end{tabular}
\end{table}

\begin{table}[H]
    \centering
	\caption{Changes in Worker Discussion Topics}
    \label{tab:text}
    \vspace{-0em}
    {
\def\sym#1{\ifmmode^{#1}\else\(^{#1}\)\fi}
\begin{tabular}{l*{2}{c}}
\toprule
                    &\multicolumn{1}{c}{(1)}   &\multicolumn{1}{c}{(2)}   \\
                    &\shortstack{Proxies for\\ Family Structure}   &\shortstack{Proxies for\\ Attention to Time}   \\
\midrule
Algorithm Available &       0.041*  &       0.043** \\
                    &     (0.022)   &     (0.021)   \\
Randomization Controls &         Yes   &         Yes   \\
\midrule
Control Mean        &       0.592   &       0.651   \\
Observations        &       4,544   &       4,544   \\
\bottomrule
\end{tabular}
}

\end{table}
\vspace{-1em}
{\footnotesize\setlength{\parindent}{0pt}\begin{spacing}{1.0} Notes: This table reports the effects of algorithm availability on worker discussion topics during screen-in decisions, as recorded in discussion notes. The table illustrates changes in worker discussion content toward topics related to the design of the algorithm. Proxies for family structure include mentioning multiple children (terms: sibling, other child, children, kids, sib, brother, sister, older child, younger child) and discussing custody (terms:  custody, co parenting, co-parenting). Proxies for attention to time include referring to the urgency of the call (terms: immediate, emergency, imminent, urgen), mentioning the past (terms: last, hx [history], [name of state record-keeping software], past, prior, previous, histor, cw [past child welfare], dhs [past department of human services], same concern), and referring to a new situation (terms: new, current, now, sober). Randomization procedure controls are included.  Standard errors are clustered at the household level. Significance reported as: * p$<$0.1, ** p$<$0.05, *** p$<$0.01.\end{spacing}}
\vspace{1em}

\begin{table}[H]
\centering
\caption{Counterfactual Exercise: Outcomes under Algorithm-Only Decisions}
\label{tab:counterfactualAIonly}
\begin{tabular}{|c|p{6cm}|p{6cm}|} 
\hline
& \textbf{Algorithm Screen-In}  & \textbf{Algorithm Screen-Out} \\ 
\hline

\textbf{Human Screen-In}                            & Child outcomes observed. &  Assume child outcomes unaffected if screened out. (No additional harm if screened out by algorithm.) \textit{This assumption favors the algorithm.}                                                                               \\ 
 \hline

 \textbf{Human Screen-Out}                            &  Child outcomes depend on marginal benefits of screen-in. \textit{Plug in range of hypothetical values.}        &   Child outcomes observed.                                                                               \\ 
\hline
\end{tabular}
\end{table}
\vspace{-1em}
{\footnotesize\setlength{\parindent}{0pt}\begin{spacing}{1.0} Notes: This table is a 2x2 grid that explains the construction of the algorithm-only counterfactual. We use data from the control group for this exercise. All observations belong to one of four cells: whether or not human decision-makers screened in the child (observed), interacted with whether or not the algorithm would have screened in the child (known under a simple decision rule assumption). \end{spacing}}
\vspace{1em}

\newpage
\section*{Appendix B}
\setcounter{figure}{0} \renewcommand{\thefigure}{B\arabic{figure}}
\setcounter{table}{0} \renewcommand{\thetable}{B\arabic{table}}

\normalsize
\begin{figure}[H]
    \centering
    \caption{Data and Trial Timeline}
    \label{timeline}
  \begin{tikzpicture}[
node distance = -20mm and -20mm,
  start chain = A going below,
   dot/.style = {circle, draw=white, very thick, fill=gray,
        minimum size=3mm},
  box/.style = {rectangle, text width=100mm,
                 inner xsep=6mm, inner ysep=1mm,
            font=\small,
                 on chain},
  every on chain/.style={anchor=north}
                        ]
    \begin{scope}[every node/.append style={box}]
\node[align=left, text height=2mm] {End of Hospital Data} ;
\node[align=left, text height=2mm] {End of Trial: Algorithm support provided for all families.} ; 
\node[align=left, text height=2mm] {Start of Trial} ;
\node[align=left, text height=2mm] {Start of Hospital Data} ;
    \end{scope}
    \draw[very thick, gray, {Circle[length=3pt]}-{Triangle[length=3pt]},
      shorten <=-4mm, shorten >=-8mm]           
    (A-4.south west) -- (A-1.north west);

\foreach \i [ count=\j] in {June 2022, 30 March 2022, November 2020, January 2020}
    \node[dot,label=left:\i] at (A-\j.west) {};
    \end{tikzpicture}
\end{figure}
\vspace{-1em}
\vspace{1em}

\newpage

\begin{figure}[H]
    \centering
    \caption{Agency Decision Protocol}
    \label{track_assignment}
   \makebox[\textwidth][c]{\includegraphics[width=1.00\textwidth]{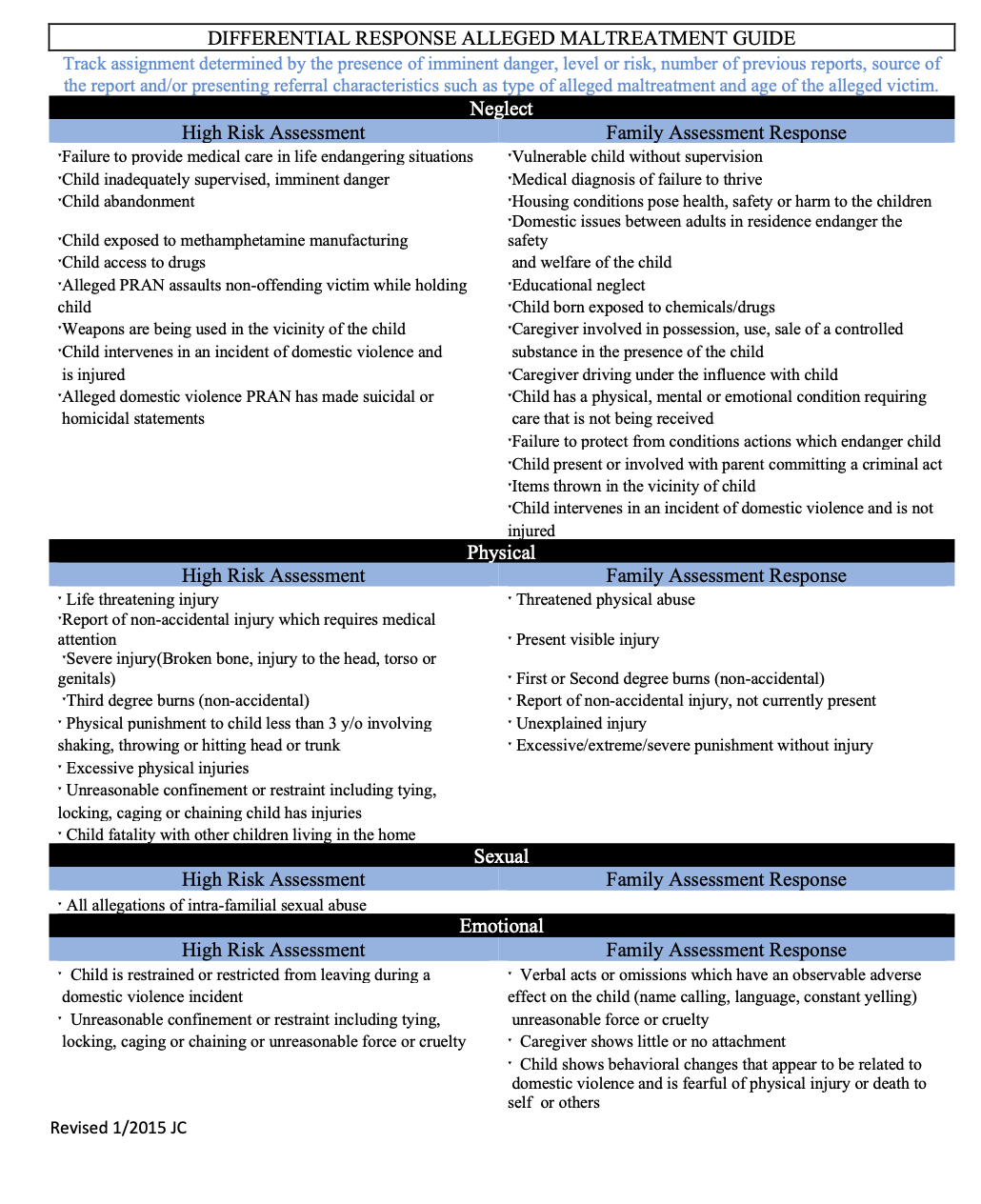}}
\end{figure}
\vspace{-1em}
{\footnotesize\setlength{\parindent}{0pt}\begin{spacing}{0.9}Notes: This image shows the CPS agency's protocol for responding to an allegation of child abuse or neglect. ``High-Risk Assessment'' designates a higher-priority referral. ``Family Assessment Response'' designates a lower-priority referral.\end{spacing}}
\vspace{1em}

{\begin{spacing}{1.0}

{\centering
    \begin{longtable}{| p{11cm}| l |}
    \caption{Redacted Examples of Worker Team Discussions and Corresponding Decisions}  
    \label{table_reasons} \\
    \hline
   \rowcolor{black!10} Discussion Text & Decision \\ \hline
    ``Discussion: [Caregiver] arrested for DUI after getting in an accident with unrestrained child in the car." \newline & Screened in (Low Priority) \\ \hline
    ``Discussion: Firearms in the home and possibly loaded and unlocked.  **** has threatened to kill ****** and reached for a firearm.  **** has hit ****** and tried to strangle him." \newline & Screened in (High Priority) \\ \hline
    ``[Algorithm Tool Scores]:  8, 2 and 1    Discussion: Team feels like the family utilized resources like the crisis center. The family could benefit from case management and prevention resources through CCR [Community Response]. A/N [abuse/neglect] does not meet criteria to assign with CPS. No marks/ bruises.    Dispo[sition]: FFR [File for Reference]. CCR'' \newline & Screened out, voluntary services \\ \hline
    ``discussion: Concerns for being thrown down the stairs.  Cultural situation needs to be assessed and kept in mind. No fear of going home." \newline & Screened out, informal visitor \\ \hline
    ``No current child protection concerns. Past incidents. RP [reporter] heard ****** has public assistance with housing. Worry since she is sleeping at ****** and she is  housing. RP called Alt to violence and they said to call CPS     RP heard ****** has criminal [history] and did heroin in past and now smokes MJ.      Hig.h risk for verbal abuse.      Dad drinks amount [unknown] *** and safe [unknown] if locked.      ***** has dementia.      He is on **d marriage and past wife DV'' \newline & Screened out, no services \\ \hline
    ``No allegation of abuse or neglect. custody issues. home schooled   parents separated for last 3 weeks, kids with dad and his daughter and family ( *** and *****)    in *****, ** ***** was exposed to ****? [...] and worried she will find **** *********?    ******* **** things down and said she doesnt live at moms anymore?    yelling screaming cursing and throwing things in the relationship in the past   dad said mom is bipolar and needs medication. mom admitted to throwing things in the past   dad called mom therapist?    dad family is support to the kids   dad doesnt agree to ******* getting help?" \newline & Screened out, no services
\\ \hline
	\end{longtable}
}
{\footnotesize\setlength{\parindent}{0pt}\begin{spacing}{0.9}Notes: Discussions have been redacted. Brackets have been added or substituted for clarification. Listed text is a subset of the full discussion text in some instances. Stars indicate words that were replaced by a de-identification script to anonymize the text data before it was shared with the researchers. \end{spacing}}
\vspace{1em}

\newpage
\section*{Appendix C: Deviations from Pre-analysis Plan}
\label{sec:preregister}
 The experiment was pre-registered on the AEA RCT Registry under ID AEARCTR-0006311. 

\begin{enumerate}
    \item \textbf{Randomization Implementation Complications.} Two unforeseen implementation challenges with respect to household randomization required us to adapt our design.
    \begin{itemize}
    \item[-] Between November 2020-March 2021, a small group of randomization statuses was unintentionally overwritten by our implementing partner. To address this, we omit the $N=238$ children for whom treatment assignment is unknown (see sample restrictions in Table \ref{SampleSelectionTable}). 
    \item[-] In March 2021, the implementing partner attempted to correct the error, but in doing so inadvertently changed the randomization procedure and introduced two complications. First, when a re-composed family was re-referred with at least one child previously assigned to the intervention group and at least one child previously assigned to control, all children in such conflicting-status referrals were assigned to the intervention group. A small ($N=80$) subset of children in the analysis sample therefore switched from control to intervention status during the trial. We check the robustness of our results to this complication in Table \ref{tab:HealthRobustness}. 
    \item[-] The March 2021 correction introduced a second challenge: children without a listed biological mother were randomly assigned an intervention status at the individual level instead of at the sibling-group level. Given the concurrent change that assigned referrals with conflicting-status children to the intervention group, beginning in March 2021 larger sibling groups were more likely to be assigned to the intervention group. For example, a group of three siblings without a listed mother was assigned to control with a probability of: $\frac{1}{2}\cdot \frac{1}{2} \cdot \frac{1}{2}=\frac{1}{8}$ instead of the intended $\frac{1}{2}$. Most children (80 percent) were referred with a listed mother and therefore unaffected. For the $N=843$ children without a listed mother, treatment assignment remained random conditional on the number of listed children. 
    We therefore include controls for each sibling group size greater than one for children listed without a mother and who were first randomized on or after March 2021. We refer to these controls throughout the paper as ``randomization controls.''
    \end{itemize}
    \item \textbf{Suggestive Maltreatment Outcome.} Among ICD codes suggestive of maltreatment, we had included a vague reference to the trauma mortality prediction model (TMPM). However, the inconsistent level of ICD-code detail in our received data, combined with difficulty of crosswalks between ICD-9 to ICD-10, made it impossible to apply the TMPM to our setting. We thus report effects on an alternative measure of suggestive maltreatment provided by \citet{schnitzer2011identification} in Appendix Table \ref{tab:otheroutcomes}. As would have been the case for the TMPM, this outcome is only available for a subset of child ages. 
    \item \textbf{Aggregating Hospital Outcomes.} Although our pre-analysis plan listed a set of key medical outcomes using hospital data, we had not pre-specified how we would aggregate effects across different outcomes. Acknowledging this, we report various ways one could aggregate results in Table \ref{tab:HealthRobustness} and show that our results do not depend on how we aggregate our hospital outcome variables together. 
    \item \textbf{Gender Disparities.} We had pre-registered examining disparities by race, ethnicity, and socioeconomic status. However, given that we also find gender health disparities in our setting, and these disparities were of interest to the agency partner, we therefore also examine effects of the algorithm tool on gender health disparities. 
\end{enumerate}

\newpage

\section*{Appendix D: Model of Decisions with Algorithm Support}
 \label{sec:model}

 Each child has an unobserved maltreatment risk $r$, assumed in the context of this model to follow a normal distribution.\footnote{The normal distribution provides closed-form solutions in contrast to alternative distributional assumptions.} 
Assume for the purposes of this application that the mean and standard deviation of the normal distribution are such that $r$ is almost always positive. In the absence of a CPS intervention, a child's realized harm is entirely determined by risk $r$. CPS workers can address this risk, however. 

Workers observe a noisy indicator of maltreatment risk, $m=r+\epsilon$, where $\epsilon$ is a normal error term with mean zero that is independent from $r$. Workers are constrained by state legal requirements in how they are instructed to respond to certain allegations (Appendix \ref{track_assignment}), so the model's objective function is to minimize errors in estimating maltreatment risk based on the signal workers receive (i.e., accuracy in discerning risk) rather than solely minimizing child harm. Whereas underestimating risk could harm children, overestimating risk is costly in terms of worker time and effort, agency resources, and increases the risk of litigation. Hence, workers' objective is to accurately assess child maltreatment risk.

\vspace{3mm}

Workers estimate risk using an empirical Bayes estimator: 
	\begin{equation}
		\label{eq:modelbayes}
  \mathbb{E}\left[r|m\right]=\left(1-\gamma\right)\alpha+\gamma m \text{,}
	\end{equation}
	where $\alpha=\mathbb{E}\left[r\right]$ is the mean risk in the population and $\gamma$ is the reliability of the signal as a measure of true risk, defined as $
		\gamma=\frac{\mathbb{V}(r)}{\mathbb{V}(r)+\mathbb{V}(\epsilon)}$. Thus, in the absence of any information $m$, workers fully rely on the group mean to assess risk, and they rely increasingly on $m$ as they become more confident in its signal-to-noise ratio.

\vspace{3mm}
The value $p=r-\mathbb{E}\left[r|m\right]$ describes the extent to which workers misestimate a child's true risk. When workers underestimate child risk, they do not intervene appropriately in the child's family and the child is susceptible to harm.  
The harm that a child experiences is a function of workers' prediction error, denoted as $H(p)$. For simplicity, we assume a functional form of $H(p)=p^2 \mathbb{1}\left(p\geq0\right)$. The function is convex to convey the fact that large underestimation of risk is particularly costly in terms of child harm. In this model, overestimating risk does not harm the child ($H(p)=0$ when $p\leq0$), though it may be a misuse of resources. 

\vspace{3mm}

True maltreatment risk $r$ is more comprehensive than only the algorithm's prediction of foster care placement, which is based predominantly on child and family history. 
The algorithmic tool could still help workers ascertain a more accurate signal of risk, however. The information provided by the algorithm may reduce worker mistakes when checking families' CPS history, prompt a better understanding of family circumstances, allow for more or better discussion of complementary information from the incoming allegation, help assess the allegation in relation to CPS history, or place greater focus on overarching household risk. Each of these potential mechanisms suggest that, in the context of this model, providing an algorithmic tool would reduce the noise in the risk assessment made by workers.

\vspace{3mm}

Denote the noise component in the signal of child maltreatment risk received by social workers by $\epsilon^T$ with algorithm support ($T$ for treatment) and by $\epsilon^C$ in the absence of algorithm support ($C$ for control). Then, if the algorithm reduced noise in workers' perception of child risk, we can rewrite this as for $a>1$:  
\begin{equation}
\label{eq:assumptionnoiseinT }
    \mathbb{V}\left(\epsilon^T\right)=\frac{ \mathbb{V}\left(\epsilon^C\right)}{a^2}
\end{equation}

Note that the model \textit{does not} impose that $\epsilon^T=\frac{ \epsilon^C}{a^2}$. The tool need not necessarily reduce noise for every child by the same ratio ($\frac{1}{a^2}$). In particular, this setup does not exclude the possibility that the tool could reduce noise by different amounts for different groups, or even not at all for some groups.  
The model's assumption is instead weaker: the variance of the noise component in perceived risk will be reduced on average for a random sample by $\frac{1}{a^2}$.  
For Propositions 1.2 and 2.2, we only require similar average effects for groups \textit{B} and \textit{W}: $a^2_B= a^2_W$.

\vspace{6mm}

\begin{spacing}{1.0} \textbf{\textsc{Proposition 1:} Accuracy of Risk Assessment. }
\textit{If providing algorithmic information reduces noise in workers' risk assessment ($\mathbb{V}(\epsilon) \downarrow$),  then: 
\begin{enumerate}
    \item The accuracy of risk assessment with the algorithmic tool will improve: $\mathbb{V}(p) \downarrow$.   
    \item Accuracy will improve more for groups with larger variance in true risk $r$. 
\end{enumerate}}
\end{spacing}

\textsc{Proof:} {\color{gray} \subsubsection*{Proof of Proposition 1.1}
Using the properties of variance, and the independence of $r$ and $\epsilon$, 
	\begin{align*}
		\mathbb{V}\left(p|T\right) =\mathbb{V}\left(r-\mathbb{E}\left[r|m,T=1\right]\right) = & \mathbb{V}\left(r-\left(1-\gamma^T\right)\alpha-\gamma^T m^T\right) =  \mathbb{V}\left(r-\gamma^T \left(r+\epsilon^T\right)\right) \\
		= & \mathbb{V}\left(r\left(1-\gamma^T \right)-\gamma^T\epsilon^T\right) =  \left(1-\gamma^T \right)^2\mathbb{V}\left(r\right)+(\gamma^T)^2\mathbb{V}\left(\epsilon^T\right) \\
		= & \left(1-\gamma^T \right)^2\mathbb{V}\left(r\right)+(\frac{\gamma^T}{a})^2\mathbb{V}\left(\epsilon^C\right) \\
  = & ...\\
  = & \frac{\mathbb{V}\left(\epsilon^C\right)\mathbb{V}\left(r\right)}{a^2 \mathbb{V}\left(r\right)+\mathbb{V}\left(\epsilon^C\right)} 
	\end{align*}
	
Write $f(a)=\mathbb{V}\left(p|T\right) - \mathbb{V}\left(p|C\right) $. Taking the derivative with respect to $a$, since $\mathbb{V}\left(p|C\right)$ does not depend on $a$, $f'(a)=-\frac{2}{a^3}(\gamma^T)^2 \mathbb{V}\left(\epsilon^C\right)<0$ and furthermore $f(1)=0$ since when $a=1, \gamma^T=\gamma^C$. Hence $f(a)<0, \forall a>1$ which is to say 
	$\mathbb{V}\left(r-\mathbb{E}\left[r|m,T=1\right]\right) - \mathbb{V}\left(\mathbb{E}\left[r|m,T=0\right]-r\right) <0$. $\square$

\subsubsection*{Proof of Proposition 1.2 }
Using the example of race, suppose Black children have greater variance in their underlying maltreatment risk compared to white children: $\mathbb{V}(r|B)>\mathbb{V}(r|W)$. Reusing the notation from the previous proof, write the change in the variance of the prediction errors of Black children relative to white children as: $g(a)=f(a|B)-f(a|W)$.  
	
	\vspace{-5mm}
	\begin{align*}
	g'(a) =& f'(a|B)-f'(a|W) =-\frac{2}{a^3}\mathbb{V}(\epsilon^C)(\gamma^T_B)^2+\frac{2}{a^3}(\gamma^T_W)^2\mathbb{V}(\epsilon^C)\\
	 =& \frac{2}{a^3}\mathbb{V}(\epsilon^C)\left[(\gamma^T_W)^2 -(\gamma^T_B)^2\right] \\
	 =& \frac{2}{a^3}\mathbb{V}(\epsilon^C)\left[\gamma^T_W -\gamma^T_B\right]\left[\gamma^T_W+\gamma^T_B\right] 
	\end{align*}
	
$g'(a)$ is of the sign of $\gamma^T_W -\gamma^T_B$. Write $h(x)=\frac{x}{x+d}$. Then,  $h'(x)=\frac{d}{(x+d)^2}>0$.\\

Since $\mathbb{V}(r|B)>\mathbb{V}(r|W)$ and $h$ is increasing, then $h(\mathbb{V}(r|B))>h(\mathbb{V}(r|W))$ which means $\gamma^T_W -\gamma^T_B=h(\mathbb{V}(r|W))-h(\mathbb{V}(r|B))<0$. Hence $g'(a)<0$ and so $g$ is decreasing. Notice that $g(1)=0$ since $f(1)=0$. To conclude: $g(a)<0,$ $a>1$. 
 $\square$} \\

Proposition 1 indicates that if the algorithm helps reduce noise in workers' risk assessment, then the distribution of workers' prediction mistakes will be more compressed, especially in groups whose members have very different levels of risk. The intuition for this result is straightforward: workers will rely less on mean risk $\alpha$ because the signal $m$ they receive is more valuable, and therefore workers' expected maltreatment risk will be closer to a child's true maltreatment risk. A larger variance in true risk implies that the mean will be a worse predictor, and therefore benefits should be larger in groups with greater variance in true risk. Improvements in risk assessment subsequently affect child health as presented in the second proposition, which is 
analogous to the first proposition but for child harm. 

\vspace{6mm} 

\begin{spacing}{1.0} \textbf{\textsc{Proposition 2:} Reductions in Child Harm. }
\textit{If providing algorithmic information reduces noise in workers' risk assessment ($\mathbb{V}(\epsilon) \downarrow$),  then: 
\begin{enumerate}
    \item Child harm will decrease when workers have access to the algorithmic tool: $\mathbb{E}\left[H\left(p\right)\right]\downarrow$.   
    \item Benefits from algorithmic tool use will be larger for groups with more variance in true risk $r$. 
\end{enumerate}}
\end{spacing}

\textsc{Proof:} {\color{gray} 

We know from Proposition 1.1 that $\mathbb{V}\left(p|T\right)<\mathbb{V}\left(p|C\right)$ for $a>1$. Since $p$ is symmetric and centered around 0, then it follows that $\mathbb{V}\left(p|T,p>0\right)<\mathbb{V}\left(p|C,p>0\right)$, which is to say that the algorithmic tool helps reduce workers' errors in risk estimation for the subgroup where risk is underestimated. 

\subsubsection*{Proof of Proposition 2.1}
We want to show that $\mathbb{E}\left[H(p)|T\right]<\mathbb{E}\left[H(p)|C\right]$. Note that: 
\begin{align*}
\mathbb{E}\left[H(p)|T\right]= & \int_{-\infty}^{+\infty}H(p)f^T_p(p)dp = \int_{0}^{+\infty}p^2f^T_p(p)dp\\
= & \mathbb{E}\left[p^2\mathbb{1}(p>0)|T\right]=\mathbb{P}\left(p>0|T\right)\mathbb{E}\left[p^2|T, p>0\right]\\ 
=& \mathbb{P}\left(p>0|T\right) \left( \mathbb{V}(p|T,p>0)+\mathbb{E}\left[p|T, p>0\right]^2\right)
\end{align*}

Define $p^*=\mathbb{E}\left[p|T\right]$,  $\sigma_p=\sqrt{\mathbb{V}(p|T)}$, $Z=\frac{p-p^*}{\sigma_p}$, $\phi$ the pdf of a standard normal distribution, and $\Phi$ its cdf. 

Using the fact that $\phi'(z)=-z\phi(z)$,
\begin{align*}
    \mathbb{E}\left[p|T, p>0\right] = & \mathbb{E}\left[\sigma_pZ+p^*|Z>\frac{-p^*}{\sigma_p}\right]=\sigma_p\mathbb{E}\left[Z|Z>\frac{-p^*}{\sigma_p}\right]+p^*\mathbb{E}\left[1|Z>\frac{-p^*}{\sigma_p}\right]\\ 
    =& \frac{1}{\mathbb{P}\left(Z>\frac{-p^*}{\sigma_p}\right)}\left[\sigma_p\int_{\frac{-p^*}{\sigma_p}}^{+\infty} z \phi(z) dz +p^*\int_{\frac{-p^*}{\sigma_p}}^{+\infty}\phi(z)dz\right] \\
    =& \frac{1}{1-\Phi\left(\frac{-p^*}{\sigma_p}\right)}\left[\sigma_p\int_{\frac{-p^*}{\sigma_p}}^{+\infty} -\phi'(z) dz +p^*\int_{\frac{-p^*}{\sigma_p}}^{+\infty}\Phi'(z)dz\right]\\
    =& \frac{1}{1-\Phi\left(\frac{-p^*}{\sigma_p}\right)}\left[\sigma_p\left[ -\phi(z)\right]_{\frac{-p^*}{\sigma_p}}^{+\infty}+p^*\left[\Phi(z)\right]_{\frac{-p^*}{\sigma_p}}^{+\infty}\right]\\
    =& \frac{1}{1-\Phi\left(\frac{-p^*}{\sigma_p}\right)}\left[\sigma_p \phi(\frac{-p^*}{\sigma_p})+p^*\left[1-\Phi(\frac{-p^*}{\sigma_p})\right]\right]= p^*+\sigma_p\frac{\phi(\frac{-p^*}{\sigma_p})}{1-\Phi\left(\frac{-p^*}{\sigma_p}\right)} 
\end{align*}

Note that $p^*=0$, so 
\begin{align*}
    \mathbb{E}\left[p|T, p>0\right] = & 0+\sigma_p\frac{\phi(0)}{1-\Phi\left(0\right)}=\sigma_p\frac{\frac{1}{\sqrt{2\pi}}}{\frac{1}{2}}=\sigma_p\sqrt{\frac{2}{\pi}}
\end{align*}

Since  $p$ is symmetric around zero in both treatment and control, $\mathbb{P}\left(p>0|T\right)=\mathbb{P}\left(p>0|C\right)=0.5$. Thus, since $\mathbb{E}\left[p|T, p>0\right]\geq 0$, then by using the first set of equations at the top of this proof, if $\mathbb{E}\left[p|T, p>0\right]\leq \mathbb{E}\left[p|C, p>0\right]$, then $\mathbb{E}\left[H(p)|T\right]<\mathbb{E}\left[H(p)|C\right]$.\\

Recall that $\sigma_p^2(a) \downarrow$ for $a\geq1$ from Proposition 1.1. Since control group $C$ can be represented by $a=1$, then $\mathbb{E}\left[p|T, p>0\right]\leq \mathbb{E}\left[p|C, p>0\right]\Longrightarrow \mathbb{E}\left[H(p)|T\right]<\mathbb{E}\left[H(p)|C\right]$.  $\square$

\subsubsection*{Proof of Proposition 2.2 }

We want to show that, if $\mathbb{V}(r|B)>\mathbb{V}(r|W)$, the reduction in harm is larger among group $B$ than group $W$ when the algorithmic tool is available: $\mathbb{E}\left[H(p)|T, B\right]-\mathbb{E}\left[H(p)|C, B\right]<\mathbb{E}\left[H(p)|T, W\right]-\mathbb{E}\left[H(p)|C, W\right]$.\\

From the previous proof we immediately arrive at $\mathbb{E}\left[H(p)|T, B\right]=\mathbb{P}\left(p>0|T,B\right) \left[ \mathbb{V}(p|T,p>0, B)+\frac{2}{\pi}\mathbb{V}\left(p|T,B\right)\right]$. \\

Note that $p$ is symmetric and centered around zero regardless of $T$ and $B$ so long as workers use the right mean for the each subgroup. Hence, $\mathbb{P}\left(p>0|T,B\right)=\mathbb{P}\left(p>0|T,W\right)=\mathbb{P}\left(p>0|C,B\right)=\mathbb{P}\left(p>0|C,W\right)=0.5$.\\

From Proposition 1.2 we know that $\mathbb{V}\left(p|T,B\right)-\mathbb{V}\left(p|C,B\right)<\mathbb{V}\left(p|T,W\right)-\mathbb{V}\left(p|C,W\right)$. 
The only thing left to prove is that $\mathbb{V}\left(p|T,B,p>0\right)-\mathbb{V}\left(p|C,B,p>0\right)<\mathbb{V}\left(p|T,W,p>0\right)-\mathbb{V}\left(p|C,W,p>0\right)$.\\

Note that since $p \sim \mathcal{N}(0,\,\sigma_p^{2})$, then $\frac{p^2}{\sigma_p^2} \sim \chi(1)$.

\begin{align*}
    \mathbb{V}\left(p|T,B,p>0\right)=&\mathbb{E}\left[p^2|T,B,p>0\right]-\mathbb{E}\left[p|T,B,p>0\right]^2\\ 
    =& \frac{1}{\mathbb{P}\left(p>0|T,B\right)}\mathbb{E}\left[p^2\mathbb{1}\left(p>0\right)|T,B\right]-\frac{2}{\pi}\sigma_{p|T,B}^2\\
    =& \frac{\sigma_{p|T,B}^2}{\mathbb{P}\left(p>0|T,B\right)}\mathbb{E}\left[\frac{p^2}{\sigma_p^2}\mathbb{1}\left(p>0\right)|T,B\right]-\frac{2}{\pi}\sigma_{p|T,B}^2
\end{align*}

Using the fact that the mean of a chi-squared distribution is equal to its degrees of freedom, $\mathbb{E}\left[\frac{p^2}{\sigma_p^2}|T,B\right]=1$. Since $p$ is symmetric and centered around zero, $\mathbb{E}\left[\frac{p^2}{\sigma_p^2}\mathbb{1}\left(p>0\right)|T,B\right]=0.5$. Therefore:

\begin{align*}
    \mathbb{V}\left(p|T,B,p>0\right)=&\sigma_{p|T,B}^2 \left(\frac{1}{2\mathbb{P}\left(p>0|T,B\right)}-\frac{2}{\pi}\right)
\end{align*}

Reusing the fact that $\mathbb{P}\left(p>0|T,B\right)=\mathbb{P}\left(p>0|T,W\right)=\mathbb{P}\left(p>0|C,B\right)=\mathbb{P}\left(p>0|C,W\right)=0.5$, we thus conclude that
\begin{align*}
\mathbb{V}\left(p|T,B,p>0\right)-\mathbb{V}\left(p|C,B,p>0\right)-\mathbb{V}\left(p|T,W,p>0\right)+\mathbb{V}\left(p|C,W,p>0\right)=\\ 
(1-\frac{2}{\pi})\left[\mathbb{V}\left(p|T,B\right)-\mathbb{V}\left(p|C,B\right)-\mathbb{V}\left(p|T,W\right)+\mathbb{V}\left(p|C,W\right)\right]<0.\hspace{2mm}  \square
\end{align*} }

Note that until now, all of the above results assume that workers make correct assessments of mean risk: $\alpha=\mathbb{E}\left[r\right]$. One might be interested in relaxing this assumption, for example if workers overestimate risk for certain demographic groups. Suppose now that workers overestimate the risk faced by group $B$ but not group $W=\Bar{B}$: $\alpha_B>\mathbb{E}\left[r|B\right]$ but $\alpha_W=\mathbb{E}\left[r|W\right]$. Disparities in risk estimation is a particularly relevant extension of the model given recent evidence that Child Protective Service workers may overestimate Black children's risk, or underestimate white children's risk \citep{baron2023}. Proposition 3 takes this into account. 

\vspace{6mm}

\begin{spacing}{1.0} \textbf{\textsc{Proposition 3:} Correcting Group-Level Differences in Estimated Risk.} 
\textit{If workers overestimate the risk faced by children in given group B ($\alpha_B>\mathbb{E}\left[r|B\right]$), then providing access to the algorithmic tool will reduce the size of prediction mistakes for those in group B, especially for children with a signal $m$ far from $\alpha_B$.   
}
\end{spacing}
\vspace{2mm}

\textsc{Proof:} {\color{gray} 
First note that, since predictions are unbiased for group W and then replacing $p$ by its expression, we get:
\begin{align*}
    \mathbb{E}[p|T]&= \mathbb{P}(B)\mathbb{E}[p|T,B] +(1-\mathbb{P}(B))\mathbb{E}[p|T,W] \\
    &= \mathbb{P}(B)\mathbb{E}[p|T,B]\\
    &= \mathbb{P}(B)(1-\gamma^T)\left[\mathbb{E}[r|T,B]-\alpha_B\right] \\
    &= \mathbb{P}(B)(1-\gamma^T)\left[\mathbb{E}[r|B]-\alpha_B\right] =p^*<0 
\end{align*}

For $a$>1,
$\gamma^T>\gamma^C \Longrightarrow 1-\gamma^T<1-\gamma^C \Longrightarrow 0> \mathbb{E}[p|T]> \mathbb{E}[p|C]$. 
That is to say, the bias in the estimated prediction error is smaller in the treated sample than in the control sample.\\

Finally, note that the effect of a change in $\gamma$ on predicted risk is larger when $|m-\alpha|$ is larger, which is to say the signal is far from workers' perception of mean risk, 
\begin{equation*}
   \mathbb{E}[r|m]=(1-\gamma) \alpha + \gamma m = \alpha + \gamma (m-\alpha). \hspace{2mm} \square 
\end{equation*}

}

This final proposition indicates that when workers overestimate risk for certain subgroups, then providing the algorithmic tool should decrease this overestimation of risk. Child harm may still fall on average for the group, if for example members of the group with overestimated risk were not concentrated in a part of the risk distribution where children were at risk of being harmed ($r$ sufficiently low). 
\end{spacing} 
\end{document}